\documentclass[12pt,preprint]{aastex}

\newcommand{\kms}{$\rm km~s^{-1}$} 
 
\newcommand{\dd}{$\Delta$} 
 
\newcommand{\mb}{$m_B^{max}$}
\newcommand{\rv}{$R_V$}
\newcommand{\rvth}{$R_V=3.1$}
\newcommand{\rvone}{$R_V=1.7$}
\newcommand{\av}{$A_V$}
\newcommand{\om}{$\Omega_M$}
\newcommand{\ola}{$\Omega_\Lambda$}
\newcommand{\xx}{$\chi^2_\nu$}
\newcommand{\too}{$t_o$}

\begin{document}

\shortauthors{Hicken et al.}
\shorttitle{Improved Dark Energy Constraints}

\title{Improved Dark Energy Constraints from $\sim100$ New CfA Supernova Type Ia Light Curves}

\author{Malcolm Hicken\altaffilmark{1,2}, 
W. Michael Wood-Vasey\altaffilmark{3},
St\'ephane Blondin\altaffilmark{4},
Peter Challis\altaffilmark{1},
Saurabh Jha\altaffilmark{5},
Patrick L. Kelly\altaffilmark{6},
Armin Rest\altaffilmark{2,7},
Robert P. Kirshner\altaffilmark{1} 
}

\altaffiltext{1}{Harvard-Smithsonian Center for
Astrophysics, Cambridge, MA 02138; mhicken, 
kirshner @ cfa.harvard.edu}  
\altaffiltext{2}{Department of Physics, Harvard University, Cambridge,
MA 02138}
\altaffiltext{3}{Department of Physics and Astronomy, University of
Pittsburgh, Pittsburgh, PA 15260}
\altaffiltext{4}{European Southern Observatory, D-85748 Garching, Germany}
\altaffiltext{5}{Department of Physics and Astronomy, Rutgers, the
State University of New Jersey, Piscataway, NJ 08854}
\altaffiltext{6}{Kavli Institute for Particle Astrophysics and Cosmology, Stanford University, 382 Via Pueblo Mall, Stanford, CA 94305}
\altaffiltext{7}{Cerro Tololo Inter-American Observatory (CTIO), Colina el Pino S/N, La Serena, Chile}

\begin{abstract}

We combine the CfA3 supernova Type Ia (SN Ia) sample with samples from the
literature to calculate improved constraints on the dark energy equation of
state parameter, $w$.  The CfA3 sample is added to the Union set of
\citet{kowalski08} to form the Constitution set and, combined with a BAO prior,
produces $1+w=0.013^{+0.066}_{-0.068} (0.11 \rm~syst)$, consistent with the
cosmological constant.  The CfA3 addition makes the cosmologically-useful
sample of nearby SN Ia between 2.6 and 2.9 times larger than before, reducing
the statistical uncertainty to the point where systematics play the largest
role.  We use four light curve fitters to test for systematic differences:
SALT, SALT2, MLCS2k2 (\rvth), and MLCS2k2 (\rvone).  SALT produces
high-redshift Hubble residuals with systematic trends versus color and larger
scatter than MLCS2k2.  MLCS2k2 overestimates the intrinsic luminosity of SN Ia
with $0.7 < \Delta < 1.2$.  MLCS2k2 with \rvth~overestimates host-galaxy
extinction while $R_V\approx1.7$ does not.  Our investigation is consistent
with no Hubble bubble.  We also find that, after light-curve correction, SN Ia
in Scd/Sd/Irr hosts are intrinsically fainter than those in E/S0 hosts by
$2\sigma$, suggesting that they may come from different populations.  We also
find that SN Ia in Scd/Sd/Irr hosts have low scatter (0.1 mag) and reddening.
Current systematic errors can be reduced by improving SN Ia photometric
accuracy, by including the CfA3 sample to retrain light-curve fitters, by
combining optical SN Ia photometry with near-infrared photometry to understand
host-galaxy extinction, and by determining if different environments give rise
to different intrinsic SN Ia luminosity after correction for light-curve shape
and color.

\end{abstract}


\section{Introduction}

One of the limitations of supernova cosmology has been the relatively low
number of cosmologically-useful nearby SN Ia.  The paucity of nearby objects
has caused the statistical uncertainties in measurements of time-independent
dark energy to be on the same order as systematic uncertainties.  For example,
\citet{kowalski08} use 250 SN Ia at high redshift but only 57 at low redshift.
As part of their study, they add eight new nearby SN Ia light curves and find
that their inclusion helps reduce the statistical uncertainty in the
measurement of the cosmological constant.  In this paper, we combine the latest
sample of nearby SN Ia optical photometry from the CfA SN Group (CfA3 sample)
\citep[][hereafter, H09]{hicken09} with samples from the literature and use
multiple light-curve fitters to calculate dark energy values.  The CfA3 sample
consists of 185 objects, compared with 29 from the Calan-Tololo survey
\citep{hamuy96}, 22 from the ``CfA1" sample \citep{riess99}, and 44 from the
``CfA2" sample \cite{jha06}.  As many as 133 are above $z=0.01$ and are useful
for dark energy calculations, depending on what cuts are made.  This increases
the number of cosmologically-useful nearby SN Ia by a factor of roughly 2.6-2.9
and reduces the statistical uncertainties of time-independent dark energy to
the point where the largest uncertainties that remain are systematic.  Because
of these systematic uncertainties, we do not claim than any of our dark energy
values is the ``right" one.  However, as the systematic errors (some of which
are identified and addressed in this paper) are reduced in
future studies, the promise of constraining cosmology from SN Ia with high
precision will become more real.  The task of solving the systematic problems
that limit SN cosmology will be challenging but it is the most important area
to focus on.

There are two main sources of known systematic uncertainty in SN cosmology.
One is in the photometry itself.  For example, good nearby light curves only
have a typical accuracy of 0.03 mag (as opposed to a precision of 0.015 mag)
(see H09) and different groups' nearby samples disagree in their mean Hubble
residuals by about 0.03 mag (see \S3.12 herein).  The second is the method of
SN Ia distance estimation, typically involving corrections for light-curve
shape and color to obtain the absolute intrinsic SN Ia magnitude.  The main
source of uncertainty here is how to treat host-galaxy reddening and
disentangle this from intrinsic SN Ia color variation.  A third source could also be
that there are different populations of SN Ia that are equivalent in
light-curve shape and color but actually have slightly different intrinsic
luminosities.

The CfA3 sample is useful in identifying and reducing systematic errors on the
photometric front because it provides a large, homogeneously-reduced and nearly
homogeneously-observed nearby sample.  The CfA3 sample can be used on its own
as the nearby component and thus reduce the effects of systematic offsets with
other samples.  Also, the same photometric reduction pipeline that was used for
the ESSENCE survey \citep{miknaitis07} was used for the CfA3 sample, reducing
one source of systematic uncertainty between nearby and faraway SN Ia.
Regarding light-curve fitting and distance estimation methods, the CfA3 sample
can be added to existing training sets to improve their accuracy and precision,
especially since the inclusion of the slightly less-common slow and fast
decliners was emphasized.
For the first time we are in a position to examine different
fitting methods on a large sample that was not used to train them.  With the
larger sample, light curve fitters can be trained better and a proper
prediction error can be calculated by excluding individual objects (or groups
of objects) from the training sample one at a time.  \citet{mandel09} have
developed the machinery for this in the near infrared and will be including the
optical bands shortly.  Additionally, many of the CfA3 objects were also
observed spectroscopically \citep{matheson08, blondin09} and photometrically
in the near
infrared \citep{woodvasey08, friedman09}.  The combination of the optical and
near-infrared photometry should help disentangle host reddening from intrinsic
SN Ia color and reduce this source of systematic error.  Finally, the larger sample assists the
search for different SN Ia populations, perhaps by host-galaxy morphology (see
\S4.2) or by host-galaxy color (Kelly et al. 2009). 

Due to the light-curve-shape/luminosity relationship (broader brighter,
narrower fainter), the natural scatter in intrinsic SN Ia luminosity is reduced
by a factor of three.  This makes SN Ia into standardizable candles and the
most useful distance indicators at cosmological scales
\citep[e.g.,][]{phillips93, hamuy96, riess96, perlmutter97, jha99, goldhaber01,
guy05, jha07, guy07, conley08}.  They have been the key element in the
discovery that the universe is accelerating and dominated by dark energy
\citep[e.g.,][]{riess98, perlmutter99, knop03, tonry03, barris04,
astier06, riess07, woodvasey07, davis07, kowalski08}.  Observational efforts
have moved beyond merely establishing the existence of dark energy and are
focused on determining its simplest properties.  This is most often done in
terms of the equation of state, $p = w\rho$, where the equation of state
parameter, $w$, relates the dark energy density, $\rho$, to the dark energy
pressure, $p$.  In a Friedman universe, $\rho$ depends on $1+w$ and the scale
factor of the universe, $a$, as $\rho \sim a^{-3(1+w)}$.  The first question
that arises is whether the dark energy density is constant ($1+w=0$, a
cosmological constant) or not.  We choose to use the notation, $1+w$, since it
is then easier to think about values of $w$ larger than $-1$ ($1+w>0$) or more
negative than $-1$ ($1+w<0$).  In the case of $1+w<0$ the dark energy grows in
density as the universe expands!  The second question is whether the dark
energy properties, as described by $w$, are constant in time or not.  

The first study on the equation of state produced a 95$\%$-confidence limit of
$1+w<0.3$, assuming \om~$\sim0.2$ and zero possibility of $1+w<0$
\citep{garnavich98}.  \citet{knop03} found $1+w=-0.05^{+0.15}_{-0.20}$.
\citet{riess05} reported $1+w=-0.02^{+0.13}_{-0.19}$.  The SNLS and ESSENCE
surveys were designed to narrow the constraints on $1+w$ and their first
reports showed significant improvement in statistical uncertainty over the
previous values, bringing them down to the range where systematic
uncertainties,which they try to reduce as well, are of roughly equal
importance.  \citet[A06, hereafter]{astier06}
found $1+w=-0.02\pm0.09$ while \citet[WV07, hereafter]{woodvasey07} found
$1+w=-0.07\pm0.09$.  Most recently, \citet{kowalski08} (K08, hereafter) made a
compilation of the literature SN Ia, plus several new nearby ones that they
present, and found $1+w=-0.01\pm0.08$ when using the same priors as A06 and
WV07.  All of these studies are consistent with a cosmological constant. 

On the time-evolution of $1+w$, \citet{riess07} rule out rapidly evolving dark
energy.  Below $z\sim0.4$, $1+w$ does not vary.  However, large numbers of new
high-redshift SN Ia ($z\sim1.5$) are needed to provide meaningful constraints
on the long-term time evolution of $1+w$.  Our focus is solely on
time-independent $1+w$ (which we will just refer to as $1+w$), in a flat
universe, where adding nearby SN Ia makes a significant improvement.   We
invite others to use the distances presented in this work and explore a wider
range of dark energy models than we do.

We now summarize the present numbers of cosmologically-useful SN Ia light
curves and some of the significant surveys that will be completed or
operational within seven or eight years from now.   The future estimates are
largely drawn from the \emph{Report of the Dark Energy Task Force}
\citep{albrecht06} and the FoMSWG (Figure of Merit Scientific Working Group)
findings \citep{albrecht09}.  Including the CfA3 sample, there are currently
$\sim$150-200 nearby SN Ia light curves ($z<0.15$) that are useful for dark
energy calculations.  The KAIT sample, with $\sim$100 SN Ia, will soon be
published, as will the Carnegie Supernova Project sample, also with $\sim$100
SN Ia.  Some of these are not in the Hubble flow and a significant number of
these objects were also observed in the CfA2 and CfA3 surveys, so not all of
the KAIT and Carnegie objects will be unique additions.  In total, when these
are included, there will be roughly 300 nearby SN Ia light curves that are
useful for dark energy calculations.  The Nearby Supernova Factory will also be
presenting $\sim$300 spectrophotometric SN Ia light curves ($0.03<z<0.08$) at
some point.  At a more nearby-to-intermediate redshift range, SDSS will soon
publish a total of several hundred SN Ia light curves ($0.05<z<0.35$).  

From the compilation of K08, there are roughly 40 SN Ia from the High-Z team
and 40 from the Supernova Cosmology Project, mostly in the redshift range
($0.3\lesssim z \lesssim 1.2$).  ESSENCE has contributed $\sim$100 SN Ia
($0.2<z<0.8$) and will soon reach a total of $\sim$200 with spectroscopic
identification.  About seventy percent of the ESSENCE SN Ia are useful for dark
energy calculations.  SNLS has published $\sim75$ cosmologically-useful SN Ia
light curves ($0.2<z<0.9$) and will soon have a total of $\sim$500.  There are
also $\sim25$ SN Ia from the Higher-Z team, many above $z=1$.  

All of these soon-to-be-published samples will drive statistical uncertainties
of time-independent dark energy about as low as is possible with the current
state of SN Ia photometry and distance fitting.  Further progress will only be
made by significantly reducing systematic uncertainties.  These additional
samples will also contribute to this.
 
Looking ahead, the Dark Energy Survey should aquire roughly 2000-3000 SN Ia
light curves by 2014 ($0.3<z<0.8$) and Pan-STARRS will observe several thousand
per year ($0.2<z<1$) but only a small percentage will be spectroscopically
identified.  If the Hubble Space Telescope refurbishment is successful then an
additional 50-100 SN Ia with $z>1$ should be obtained.  Farther into the
future, the JDEM will observe 2000+ SN Ia in the range, $0.3\lesssim z
\lesssim 1.7$ with the intention of better constraining the time variation of
dark energy.  The LSST will provide truly staggering numbers of SN Ia, on the
order of $10^5$ per year, most with $z<0.7$ and a smaller portion extending out
to $z=1.2$.  The JWST will be able to study SN Ia beyond $z\approx 2$, giving
further insight into the matter-dominated era of the universe.  In the design
of these surveys, emphasis should be placed on reducing systematic uncertainies,
as opposed to simply acquiring more objects with the same level of accuracy
as in previous surveys.

Returning to the matter at hand, the CfA3 sample was acquired from 2001-2008,
on the F. L. Whipple Observatory 1.2m telescope, mostly using two cameras, the
4Shooter and Keplercam, as described in H09.  A few SN Ia were observed with
the Minicam.  \emph{UBVRI} filters were used on the 4Shooter while
\emph{UBVr'i'} filters were used on the Minicam and Keplercam.  CfA3 comprises
over 11500 observations while CfA1 has 1210 and CfA2 has 2190.  H09 show
relatively good agreement with previous samples of nearby SN Ia in the
distribution of intrinsic color and host-galaxy extinction.  However, CfA3 has
a wider distribution in light-curve shape, in large part due to H09 giving
higher priority to fast and slow decliners in order to fill in the population
sampling of both brighter and fainter SN Ia.  This is of particular value for
retraining light-curve fitters as well as providing light-curve phase
information for spectra of the objects in the CfA3 sample.  The slow decliners
are also a valuable addition since these are found relatively more often at
high redshift.  

As discussed in greater detail in H09, about two thirds of the CfA3 SN Ia were
discovered by professionals and one third by amateurs.  KAIT was the main
single contributor with $46\%$ of the objects.  The median redshift of the CfA3
objects above $z=0.01$ is $z=0.03$.  The discovery limiting magnitude was
typically $\sim19.5$ mag but objects with a peak magnitude fainter than
$\sim18.5$ mag were not observed.  This effective limiting magnitude for the
CfA3 sample captures the whole range of SN Ia intrinsic luminosities out to
$z\approx0.03$.  Beyond that, the fainter SN Ia drop out.  The CfA3 sample is
not representative of the relative numbers of the underlying nearby SN Ia
distribution because of the many selection effects that are part of discovery
and follow-up observations.

Combining the CfA3 sample with high- and low-redshift samples from the
literature improves the constraints on $1+w$.  We use four light curve fitters
to probe for consistency and systematic effects:  SALT \citep{guy05}, SALT2
\citep{guy07}, MLCS2k2 \citep[J07, hereafter]{jha07} with \rvth~(MLCS31) and
MLCS2k2 with \rvone~(MLCS17).  Even though the underlying MLCS2k2 algorithm is
the same for MLCS17 and MLCS31, we refer to them as two ``different" fitters for
the sake of simplicity.  It is important to note that none of these fitters has
been trained on the light curves from the CfA3 sample and yet we find
reasonably consistent results between the CfA3 and literature nearby samples.
We limit ourselves to these fitters so that comparisons with K08, MW07, and A06
can be made.  However, we encourage that other light-curve fitters be used with
the aim of exploring and decreasing systematic errors.  Examples are $\Delta
m_{15}(B)$ \citep{phillips93,phillips99}, stretch \citep{goldhaber01}, CMAGIC \citep{wang03, conley06}, dm15
\citep{prieto06}, and SiFTO \citep{conley08}.  We also encourage 
the addition of appropriate CfA3 objects to the these fitters.

To correctly determine the properties of the dark energy, it is vital to have
distances at high and low redshift that are accurate relative to one another.
The dividing line between high and low redshift in this paper is $z=0.16$ for
SALT, which is consistent with $z=0.20$ in K08 (there are 
no objects with $0.16<z<0.20$,
and only one object, SN 1999ar, with $0.15<z<0.16$), and $z=0.15$ for the other
fitters, consistent with WV07.  In the concordance model, \om$\approx0.3$ and
\ola$\approx0.7$, the SN Ia at $z\approx0.5$ are about $0.25$ mag fainter (and
hence farther) than they would be in a universe with only matter
(\om$\approx0.3$).  If new samples or analysis showed the $z\approx0.5$ SN Ia
to be even fainter (and farther), and the nearby sample remained unchanged,
then this would imply greater \ola (assuming $1+w=0$ in a flat universe) or
lower $1+w$ (with $1+w$ free to vary in a flat universe) than before.  Similar
effects would be seen if the nearby distances decreased (brighter objects), and
the $z\approx0.5$ distances remain unchanged.  




In comparing nearby and faraway SN Ia distances there are two key components:
the underlying nearby and faraway SN Ia populations that nature provides and
the distance estimation methods that we provide.  Ideally, the nearby and
faraway SN Ia populations would be composed of identical objects and the
methods of distance estimation (photometry and light-curve/distance fitting)
would work perfectly across the whole SN Ia range.   

There is good evidence that the faraway population is composed of objects
highly similar to the part of the nearby sample that it overlaps with.
Comparisons of high and low redshift spectra reveal good agreement
\citep[e.g.,][]{matheson05,hook05,howell05,foley05,blondin06,foley08}.
\citet{howell07} mimic there being a difference in faraway and nearby
populations due to evolution by only using SN Ia with light curve stretch
parameter, $s\geq1$ at $z\geq0.4$
and SN Ia with $s<1$ at $z<0.4$ to fit the cosmology and find it is consistent
with the results of the full sample.  We assume that the nearby and faraway
population objects are sufficiently similar for our purposes.

This leaves us with the issue of whether the distance estimation is
sufficiently accurate or not.  It is vital to accurately handle any aspects of
the distance estimation that are unique, or much more heavily weighted, to
either the faraway or nearby samples, such as $K$-corrections for the faraway
SN Ia or including underluminous fast decliners at low redshift.  It is also
key to limit the samples to the range where the distance estimators are
accurate enough and the faraway and nearby subsamples overlap sufficiently in
underlying population and sampling characteristics so that any remaining
inaccuracies in the distance estimators do not give rise to any significant
differences.  One option is to remove any troublesome groups, potentially
gaining systematic safety while losing statistical leverage.  

An example of this is handling host galaxy reddening.  There are more
highly-reddened SN Ia in the nearby sample.  If the host extinction is
estimated too high, perhaps because the physical value of \rv~is less than is
being assumed, then the nearby distances will be too low on average, resulting
in an overestimated dark energy density.  Until the issue of handling host
reddening is improved it is probably advisable to cut out moderately- and
highly-reddened objects, and we follow this course for our ``best"-cut 
samples.  

As SN Ia samples grow in size and quality it may not be statistically necessary
to include less-reliable objects.  However, it is also likely that improved
data sets and distance estimation techniques will result in SN Ia being better
standardizable candles so that a wider range of objects can be safely included.

\subsection{Outline of the Paper}

In $\S$2, we describe the SN Ia light curve samples and what initial ``minimal"
quality cuts we make to ensure that poorly-fit objects are not being used to
calculate cosmological results.  We also describe the four light-curve fitters
we use to calculate distances and how we calculate the cosmological fits.  

In $\S$3, we present our cosmological fits and examine the impact of the CfA3
sample.  It reduces the statistical error on $1+w$ by a factor of 1.2-1.3,
slightly less than rough $\sqrt{N}$ statistics would imply, suggesting that
systematic uncertainties are becoming noticeable.  Our Constitution sample
(Union+CfA3), fit by SALT and using linear color-luminosity and light-curve-shape-luminosity relations, produces $1+w=0.013^{+0.066}_{-0.068} (0.11 \rm~syst)$
when combined with a BAO prior.  This is consistent
with the cosmological constant.  We use the four light-curve fitters and
different SN Ia samples to test for consistency and systematic differences and
find generally good agreement.  An encouraging example is the excellent
agreement of the distances to two SN Ia in the same galaxy (``twins") that was
found using each of the four fitters.  However, there are three areas of concern: SALT,
and to a lesser extent, SALT2, produce a trend in Hubble residuals versus the
color parameter, $c$, at high redshift; MLCS2k2 gives rise to mostly negative
residuals in the moderately underluminous region $0.7<\Delta<1.2$; and
\rvth~seems to overestimate the host galaxy extinction, \av, in MLCS31.  In
\S3.11, we choose to remove these objects for our ``best"-cut samples.  With
and without the best cuts, SALT and SALT2 produce values of $1+w$ that are
statistically consistent with the cosmological constant.  Without the best
cuts, MLCS31 also is consistent with the cosmological constant while MLCS17
produces $1+w\approx0.1$.  After the best cuts, MLCS31 and MLCS17 agree much
better with each other but are slightly more than $1\sigma$ above the
cosmological constant ($1+w=0$).  We believe that when the systematic problems we
discuss are resolved in retrained or future fitters 
that there will be much better agreement amongst different fitters
and a more accurate measurement of the dark energy will exist.

In \S3.12, we break up the nearby sample into its six or seven principal
subsamples, based on observing survey.  We examine the implied consistency of
the photometry and light-curve/distance fitting of the six-to-seven largest
nearby samples by comparing their average Hubble residuals and find typical
agreement to be roughly 0.03 mag.  MLCS17 gives the best agreement between the
samples ($\sim0.02$ mag), MLCS31 and SALT come next ($\sim0.03$ mag), and SALT2
has the most discrepancy ($\sim0.06$ mag).

In $\S$4, We explore the issue of the ``Hubble bubble" and look for trends in
Hubble residuals versus host-galaxy properties.  Our results are largely
consistent with no Hubble bubble.  We find that the SN Ia in Scd/Sd/Irr hosts
are fainter after light-curve correction by $2\sigma$
than those in E/S0 hosts, suggesting that it may be advantageous for
light-curve fitting to divide the SN Ia sample into two or more groups based on
host-galaxy properties.  We briefly comment on the systematic uncertainty in
$1+w$, finding it to be roughly $0.11$ mag and $\sim40-70\%$ larger than the
statistical uncertainty.  The largest systematics seem to be which nearby
sample and which light curve fitter are used.  The treatment of host reddening
is the largest contributor to the uncertainty of a given fitter.
Systematic uncertainties now play the largest role in limiting our
understanding of dark energy.  To reduce these, three main steps should be
taken.  First, future photometry needs to be more accurate, by better
understanding the instrumental passbands and their absolute calibration.
Second, light curve fitters need to be retrained with larger samples,
including the CfA3 and other forthcoming samples, 
treating the intrinsically-red and subluminous SN-1991bg-like objects separately.  And
third, intrinsic SN Ia color and host-galaxy reddening, at both high and low
redshift, need to be disentangled.

\section{Light Curve Fitters}

%
SALT and SALT2 are light curve fitters and require a two-step approach towards
calculating distances:  first, fit each light curve for peak magnitude, stretch
parameter ($s$ for SALT and $x_1$ for SALT2) and color, $c$; second, use
these outputs to calculate 
the best-fit cosmology.  
Typically, the distance modulus is parameterized by a linear dependence on
shape and color parameters and we follow this path.  SALT and SALT2 do not
attempt to disentangle intrinsic color from host reddening.  To be valid for
cosmological calculations, both the high and low redshift samples must obey the
same, combined intrinsic-plus-host-reddening color-magnitude relations.

Using MLCS2k2 also takes two steps to find the best-fit cosmology:  fit for the
distances and then fit for the cosmology.  MLCS2k2 goes farther than SALT and
fits for the distance along with its shape/luminosty parameter, \dd, the
host-galaxy extinction parameter, $A_V$ and the time of \emph{B} maximum, $t_o$.
MLCS2k2 differs from SALT and SALT2 in that it attempts to explicitly calculate
the host reddening by employing a prior on $E(B-V)$ which can then be converted
to the extinction $A_V$ via the reddening law \rv.  We use \rvth and \rvone,
effectively producing two MLCS2k2 fitters.  It also uses a
quadratic dependence on the shape parameter \dd~to describe all intrinsic
variations of the peak magnitude.  No explicit, intrinsic-color parameter is
employed.  Rather, the broader-bluer, narrower-redder relation is incorporated
into \dd~so that the negative-\dd~light curve template is both blue and broad
while the highly positive-\dd~template is red and narrow.  

A weakness of the second step of using SALT and SALT2 is giving the same
color-magnitude relation to both intrinsic SN Ia color variation and
host-galaxy dust.  A weakness of MLCS2k2 is the uncertain nature of the prior
on $E(B-V)$, especially at high redshift, and the reddening law that should be
used.  The use of all four fitters assumes that SN Ia are intrinsically the
same at high and low redshift for a given shape and color, within the observed
scatter.  The veracity of this assumption has been explored and seems to hold
but needs further investigation.

Our main purpose for fitting with SALT is to add the CfA3 sample to the Union
compilation of K08.  We call the combination of the Union and the CfA3 sets the
``Constitution" set, to form a more perfect union.  For the other three
fitters, we do not use all the high-redshift SN Ia samples, specifically
excluding $\sim100$ objects from the SCP and High-Z objects from
\citet{riess98}, \citet{perlmutter99}, \citet{knop03}, \citet{tonry03}, and
\citet{barris04}.  Rather, we use the ESSENCE \citep{miknaitis07}, SNLS (A06) and
Higher-z samples \citep{riess07} where we believe the systematics to be better
controlled.  We will call these three together the ``High-z" sample.  The low
redshift SN Ia compiled in Jha 2006 (referred to hereafter as ``OLD") and the
CfA3 sample constitute our nearby, ``OLD+CfA3," sample.  For each fitter we
separately combine the OLD, CfA3 and OLD+CfA3 samples with the High-z sample.
We will use these three nearby sample names to refer to the three nearby plus
High-z samples since the High-z sample is common to each.  We will also use
``OLD" to refer to the nearby sample and the nearby plus high-redshift sample in
K08.  In the SALT case, OLD is equivalent to Union and OLD+CfA3 is equivalent
to Constitution.

In all our \om-$w$ cosmology fits, we assume a flat universe and combine the
baryonic acoustic oscillations (BAO) constraints on (\om, $1+w$) from
\citet{eisenstein05} with the SN Ia fits to produce our best-fit cosmologies.
The BAO prior provides an effective constraint on the range of \om, primarily,
while the SN Ia data better constrain $1+w$.  Together, they form a complementary
and excellent pair of constraints on \om~and $1+w$.  Additional priors, such as
from the cosmic microwave background, could be applied to increase the
precision of $1+w$ but we prefer to leave the SN Ia effects as unmasked as
possible while still narrowing the range in \om.  The
uncertainty on redshift can be seen in Tables \ref{table_salt},
\ref{table_salt2}, \ref{table_mlcs31}, and \ref{table_mlcs17}.  A peculiar
velocity uncertainty of 400 \kms~is assumed.  A06, J07, K08 and others have
used 300 \kms~while WV07 used 400 \kms.  A06, WV07 and K08 make redshift cuts
at $z=0.015$.  Even though 300 \kms~is closer to the actual peculiar velocity
uncertainty, since we choose to cut at $z=0.01$ and because of potential issues
like a Hubble bubble we would like to deweight the nearest ones slightly and
400 \kms~achieves this.  The peculiar velocity uncertainty of 400 \kms~
produces a difference in distance-modulus uncertainty between $z=0.01$ and
$z=0.015$ that is 0.011 mag greater than if 300 \kms~had been used.  Our use of
`$z$' refers to $z_{cmb}$ throughout.  We also fit for \om~and \ola, assuming
$1+w=0$ and using the BAO prior. 

\subsection{SALT:  Augmenting the Union Compilation}

SALT uses a modified version of the Nugent spectral template that was
developed to reproduce the \emph{UBVR} light curves of 34 nearby SN Ia, 28 of
which are at $z<0.015$ and six others with good \emph{U}-band data at $z>0.015$.  No
SN-1991bg-like objects were included in this training set.  SN-1991bg-like
objects are spectroscopically identifited by strong \ion{Ti}{2} lines. 
Photometrically, they are intrinsically red, fast declining, and subluminous.
H09 show that such objects form more of a separate grouping than a continued
distribution in color and intrinsic luminosity.  SALT fits for
the time of \emph{B}-band maximum light, \too, the flux normalization (which can be
converted to the rest-frame peak magnitude in \emph{B} band, \mb), a time 
stretch factor, $s$, and a color parameter, $c = (B-V)\mid_{t=Bmax} + 0.057$.  
The color, $c$, is a combination of the intrinsic color and reddening due to
host-galaxy dust.   

The next step, taking the SALT output parameters and calculating the best-fit
cosmology uses the following equation:

\begin{equation}
\mu_B = m_B^{max} - M  + \alpha(s-1) - \beta c
\label{eqno1}
\end{equation}

The empirical coefficients $\alpha$, $\beta$, and $M$ are marginalized over as
part of the cosmological fit.  It is remarkably the case that the two different
physical mechanisms (intrinsic color and reddening by dust) that give rise to
the color, $c$, sufficiently obey the same color-magnitude relation that they
can be described by the single parameter, $\beta$.  To calculate the distances
to each SN Ia, we use the 1-d marginalization coefficients, $\alpha$, 
$\beta$, and $M$.

K08 use SALT to fit 307 SN Ia light curves, 57 at $z<0.2$ and 250 at $z>0.2$
to form the Union compilation.  These SN meet the criteria of
having good light curve fits, $z\geq0.015$, time of first observation (divided
by stretch, $s$), $t_{1st}/s$, less than six days after maximum light and each
Hubble residual, scaled by its uncertainty, within $3\sigma$.  The K08 cuts are
different from those that we will use for the other three fitters.  We tested the
K08 cuts on the MLCS31 OLD+CfA3 sample and it only made a difference in $1+w$ of
0.005 so we prefer to use our more permissive cuts, explained below, to allow
for a larger sample.

We take the SALT parameters, \mb, $s$, and $c$, from Table 11 in K08 for the
Union set and calculate the best-fit cosmology from the combination of the 307
SN plus the BAO prior, assuming a flat universe.  There can be slight
differences in the specific procedure to fit for the cosmology so we do not try
to \emph{exactly} reproduce the values for $1+w$ and \om; only get very close so
that we can then confidently add our SALT fits of the CfA3 sample to the K08 
SALT fits of the Union sample and calculate the Constitution set (OLD+CfA3)
cosmology.
 
A peculiar velocity of 400 \kms~is assumed (different than the 300 \kms~in K08
since we want to be more cautious in our assumptions for peculiar velocities).
In order to approximately maintain the same relative weighting for each SN that
was used by K08, including a sample-dependent systematic error and other
uncertainty components, we take the uncertainty of the distance modulus,
$d\mu_B$, from Table 11 in K08 and subtract off the uncertainties of $c$ and
$s$ and use this as a modified uncertainty for \mb~in equation \ref{eqno1}.  We
find that the resulting uncertainty on each SN is too high by 0.007 mag (in
quadrature) to reproduce the same amount of uncertainty in $1+w$ as in K08
and we subtract this from each.  These steps allow us to reproduce
the SN+BAO value of $1+w=-0.011$ from K08.  The statistical uncertainties on
$1+w$ from our calculations are more symmetric:  +0.078 and -0.080 versus
+0.076 and -0.082 for K08 but the difference between the positive and negative
uncertainties is 0.158 for both.

Having sufficiently reproduced the K08 results, the CfA3 sample is added to the
Union set.  To verify that the SALT outputs for the CfA3 sample can be combined
with the Union set we run most of the light curves from the nearby Union sample
through SALT, using the \emph{UBVR} bands only.  We only run SALT on the J07
compilation of nearby SN Ia and so a few of the
Union nearby set are not used to verify that our SALT output is
consistent with K08.  Our calculations of \mb, $s$, and $c$ agree
sufficiently with those of Table 11 in K08, with our peak magnitude fit in \emph{B}
band fainter by $0.01 \pm 0.06$ mag.  Possible reasons for the slight
differences are that K08 perturbs the light curves of the more poorly-fit
objects, there might be slightly different versions of SALT and some other
parameters within SALT may differ.  However, since the typical scatter in the
Hubble diagram for nearby SN Ia is in the range of 0.14-0.20 mag, our SALT
calculations agree well enough with K08 that the SALT fits of the CfA3 sample
can be safely combined with the K08 SALT fits of the whole Union set without
introducing any significant offset.
 
SALT is not suited to fitting 1991bg-like SN so we do not include ones that we
identified as such by their spectra in the CfA3 sample.  We did not remove the
two 1991bg-like SN from the Union set, however, since the number is small and
we are interested in updating the Union set as is.  The CfA3 \emph{UBVRr'} band
light curves were run through SALT.  We wanted to appropriately weight the CfA3
objects so that they can be combined with the Union set for cosmology
calculations.  We use the SALT output from K08 for the Union set and our own
SALT calculations for the CfA3 set.  Although the CfA3 Hubble residuals have a
lower standard deviation than the nearby Union residuals (0.164 mag versus
0.186 mag), we made the more cautious assumption that, on average, a CfA3 SN
Hubble residual should have the same uncertainty as a nearby Union SN Hubble
residual.  To accomplish this, an uncertainty of 0.138 mag is added in
quadrature to the uncertainties of the SALT light-curve fit for each CfA3 SN.
The peculiar velocity uncertainty is set at 400 \kms~and the redshift
uncertainty is 0.001.  Out of 185 CfA3 SN Ia, 90 survive after applying the
Union cuts (these are our minimal cuts for SALT), increasing the nearby sample
with these cuts by a factor of 2.6, from 57 to 147.  We also fit for \om~and
\ola, assuming $1+w=0$ and using the BAO prior.

\subsection{SALT2}

SALT2 differs from SALT by no longer using the Nugent template.
\citet{hsiao07} presents an improved SN Ia spectral template, but rather than
using this, SALT2 has a color- and stretch-dependent model spectrum derived
from a sample of nearby and faraway SN Ia spectra and light curves.  The main
benefit of this is that the model spectrum for a given phase, color, and
stretch is more closely derived from actual SN Ia data and so, in principle,
the range of SN Ia light curves should be better fit than in SALT.  Output
parameters of the fit are the color, $c$, the stretch factor, $x_1$ (analogous
but not equivalent to $s$ in SALT), the \emph{B}-band maximum, \mb, and the
time of \emph{B} maximum, $t_{Bmax}$.  A practical effect of using the high
redshift spectra is that the rest-frame near UV can be modeled out to
2000~$\rm\AA$.  As in SALT, 1991bg-like SN are not included in the training and
should not be fit with SALT2.

We run SALT2 on the OLD, CfA3 and High-z samples, excluding \emph{I} and
\emph{i'} from the nearby samples.  The \emph{U}-band fits were typically poor
but their contribution to the overall light-curve fit was small because the
\emph{U}-band is given significantly less weight in SALT2 than the other
optical bands.  In \S3.9, we explore the effects on $1+w$ of excluding the rest
\emph{U} band and find little difference for SALT2.  
We used the default setting of 3460~$\rm\AA$ for the lower limit on the
wavelength range.  The value used by \citet{guy07} is different:
2900~$\rm\AA$ .  This shorter wavelength limit may improve the precision 
of the color parameter, $c$ for the high-redshift objects, subject to the
uncertainties in calibrating the rest-frame UV.  We did not explore the
effects of of varying the short wavelength limit.
We
also exclude any 1991bg-like objects.  The reduced chi-squared, \xx, of the
light curve fit allows poorly-fit objects to be flagged and removed.  The
precise value of \xx~ that indicates a poor fit, as determined by visual
inspection, varies from fitter to fitter, depending largely on the uncertainty
attributed to the model light curve produced.  If the model light curve has
reasonable uncertainties then a \xx~of one can be expected but if it is too
small then acceptable fits can have higher \xx.  For SALT2 we found that the
poorly-fit light curves had a \xx~of more than 10 while sufficiently-well-fit
light curves were less than this and so we chose this as our cut-off value.  We
also only include SN Ia with $t_{1st}\leq+10d$, ensuring that the fit
parameters can be well constrained.  SN with $z < 0.01$, where peculiar
velocity uncertainties become excessively large, are excluded from our
cosmology fit for SALT2.  These are the minimal cuts for SALT2.  An additional
uncertainty of 0.158 mag is required to give a \xx~of one for the nearby SN Ia
in our SALT2 cosmology fit.  This is added to all objects in the fit, both
nearby and faraway. 

A similar procedure to the SALT cosmology fitting is followed with the SALT2
output.  The distance modulus is 

\begin{equation}
\mu_B = m_B^{max} - M + \alpha x_1 - \beta c
\label{eqno2}
\end{equation}

and $\alpha$, $\beta$ and $M$ are marginalized over to find the best-fit
cosmology.  To calculate the distances to each SN Ia, we use the 1-d 
marginalization values for the coefficients, $\alpha$, $\beta$, and $M$.

\subsection{MLCS2k2 (\rvth)}

MLCS2k2 consists of two main components:  a model light curve parameterized by
\dd~and a physically-inspired attempt at separating instrinsic color variation
from host-galaxy dust extinction.  The intrinsic light curve shapes, color and
luminosity of the model are a function of the single parameter \dd.  Although
MLCS2k2 fits for all of its parameters simultaneously, it is conceptually
useful to think of MLCS2k2 in the following manner.  In calculating distances,
MLCS2k2 assumes that a SN Ia can be corrected to the absolute magnitude of the
fiducial SN Ia by means of the linear and quadratic terms in \dd.  The best-fit
\av~is added and the distance modulus, $\mu$, is the remaining term needed to
reach the ($K$- and Milky Way-reddening-corrected) apparent magnitude.
Obtaining an accurate value of \av~depends in part on using the correct value
of \rv.  If \rv~is too high then \av~will be as well and the distance modulus
will be too small and vice versa.  1991bg-like SN Ia are part of the MLCS2k2
training sample, making a significant contribution to the model colors.  This,
in turn, has an impact on the extinction that is measured.  To run these
objects through MLCS2k2 correctly, $K$-corrections with a 1991bg-like spectral
model are used.  

MLCS2k2 was trained on a large number of high-quality, nearby SN Ia light
curves where the host-galaxy extinction can be accurately estimated from
well-sampled late-time \emph{B} and \emph{V} data and where the distance can be
accurately estimated by only using Hubble flow objects.  For nearby samples,
the prior should be calculated from the sample itself or from a similar one.
MLCS2k2 uses a one-sided exponential of scale-length $\tau_{E(B-V)}=0.138$ mag
as a prior on host reddening.  We refer to this color-excess prior as the
``default" prior and use this for our nearby MLCS2k2 prior.  For faraway
samples, the prior is difficult or impossible to calculate from the data
itself, due to the lack of late-time photometry.  In this case, a nearby or a
theoretical prior can be modified to match the detection effects for a sample
where high-extinction events are less and less likely to be found at higher and
higher redshifts.  For example, the ESSENCE team modifies the Galactic
Line of Sight (glos) prior \citep{riess98,hatano98,commins04,riess05,riello05}
into the redshift-dependent ``glosz" to take into account the
redshift-dependent detection probabilities of their survey (WV07).

We run the SN Ia through MLCS2k2.  We use the 1991bg-appropriate
$K$-corrections for 28 objects that were identified as having 1991bg-like
spectral properties:  SN 1986G, 1991bg, 1992K, 1992bo, 1993H, 1997cn,
1998bp, 1998de, 1999by, 1999da, 1999gh, 2002fb, 2003D, 2005ke, 2005mz, 2006H,
2006bd, 2006bz, 2006cs, 2006em, 2006gt, 2006hb, 2006je, 2006ke, 2007N, 
2007al, 2007ax, and 2007ba.  For the nearby sample we use 
the default host-galaxy reddening prior, assuming that the CfA3 sample is 
close enough to the sample from which the prior was derived.  We use the 
glosz prior for the ESSENCE sample and the modified glosz prior for the SNLS 
light curves, both as described in WV07.  For the Higher-z SN, we also use 
a modified glosz prior as described in \citet{riess05}.  We restrict our nearby 
sample to $z\geq0.01$.  Adding an ``intrinsic" uncertainty of 0.078 mag produces a 
\xx~of one in the cosmological fit for the nearby sample and this is applied
to the high redshift SN as well.  Only those SNe whose MLCS31 fits have a
\xx~of 1.5 or less are included.  Additional criteria for inclusion
are $A_V \leq 1.5$ and $t_{1st} \leq 10d$.  These are the minimal cuts for 
MLCS31.

\subsection{MLCS2k2 (\rvone)}

Previous studies that used stretch or SALT have found the color parameter
$\beta$ to be around 2, significantly lower than the value of 4.1, expected if
$c$ were only measuring the host-galaxy dust and this dust obeys \rvth, the
reddening law seen in the Milky Way \citep[e.g.,][]{astier06,conley07}.
However, it is unlikely that $c$ is only the color excess due to dust.
Therefore, it is not surprising that typical values of $\beta$ are different
than 4.1, rather that they are so much less.  With this in mind, we use MLCS2k2
to find the value of \rv~that minimizes the scatter in the Hubble residuals for
the nearby CfA3 sample.  We run MLCS2k2 with no prior on the color excess
$E(B-V)$, allowing for negative values as well.  We calculate the Hubble
residuals versus $A_V$ for both \rvth~and $2.1$.  With \rvth~and no prior there
is a significant slope while for $R_V=2.1$ it is less severe.  We subtract off
the $A_V$ calculated for each object and fit for the effective \rv~that
minimizes the scatter:  \rvone~is this value.

We fix \rv~to 1.7 and run MLCS2k2 on our 
compilation of SN Ia, using the same priors as in the case of \rvth.  
We used a version of MLCS2k2 that has model \emph{UBVRI} light curves that were 
calculated by training MLCS2k2 with $R_V=1.9$.  We found it made little 
difference to the end results whether we used the version trained with 
\rvth~or $R_V=1.9$ and so we did not bother to train with \rvone, but we did
use the $R_V=1.9$ version to be as consistent as possible. 

We restrict our sample to $z\geq0.01$.  We use the 1991bg-appropriate 
$K$-corrections for the same SN as we did with \rvth.  An intrinsic 
uncertainty of 0.083 mag is added to produce a \xx~of one in the nearby 
sample.  Once again, we cut out objects with $A_V \geq 1.5$ and $t_{1st} > 10d$ days.
The uncertainties in the model light curves trained with $R_V=1.9$ are slightly
different than the \rvth~case and so we choose to cut out SN with MLCS2k2
fits with \xx~greater than 1.6.  This is the only difference in the minimal
cuts between MLCS17 and MLCS31.

\section{Results and Systematics Due to Fitters and Nearby Samples}

This section analyzes the output of the four light curve fitters being employed
for different subsamples, looking for systematic trends and 
errors in both fitters and SN Ia subsamples.  The aim is to find areas where
these fitters work best and provide suggestions for improvement.  We also
want to see how consistent the various nearby samples (from different groups
and instruments) are with each other.

\subsection{SALT--The Constitution Sample}

\begin{figure}
\begin{center}
\scalebox{1.0}[1.0]{
\plotone{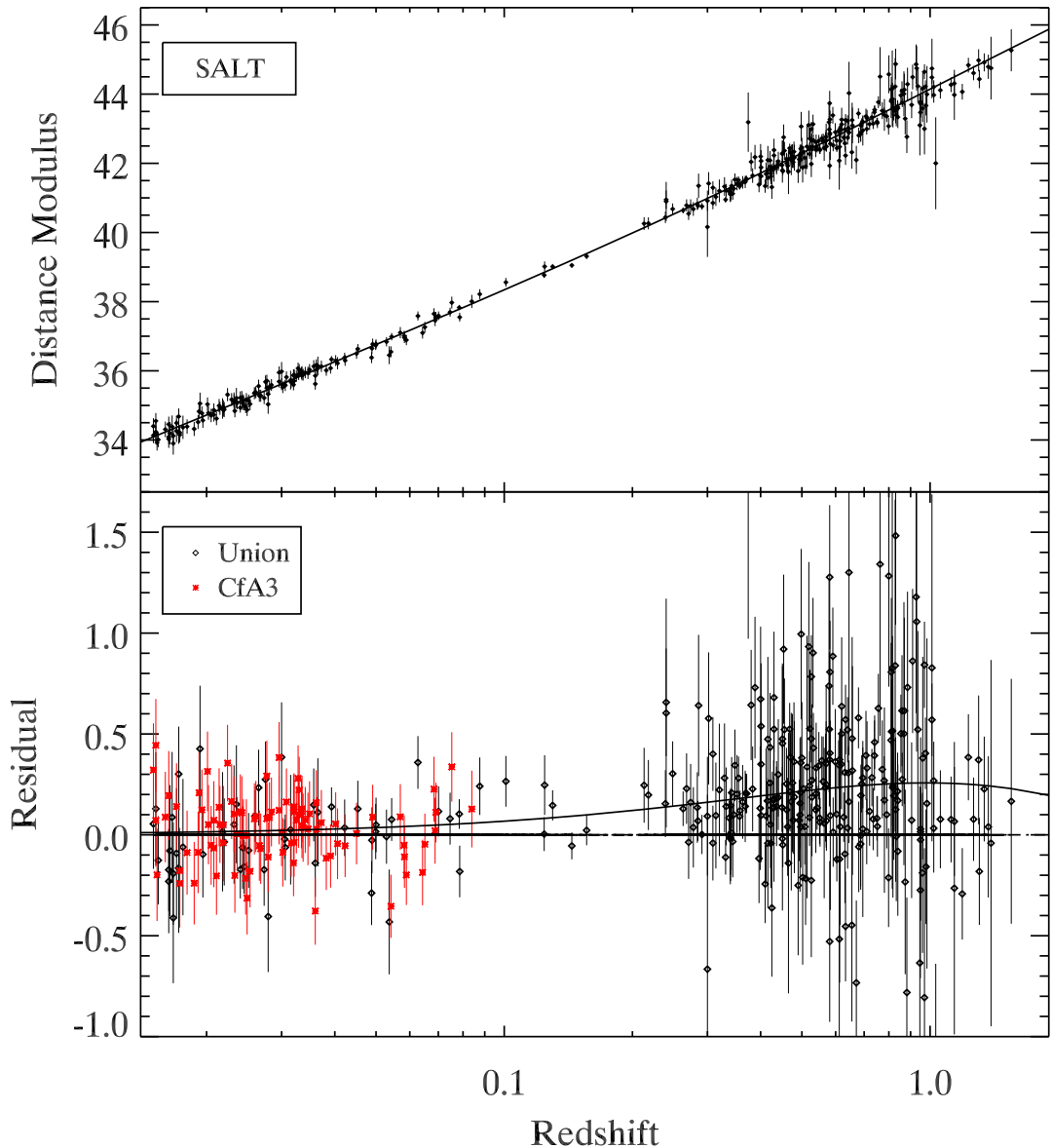}
}
\end{center}
\caption{Hubble diagram and residuals for the Constitution sample fit 
by SALT.  The
bottom panel shows the new CfA3 SN Ia in red and the Union sample
in black.  The residuals are with respect to a universe without dark
energy, \om$=0.27$ and \ola$=0$.  The best-fit cosmology is plotted
in the residuals panels.  The large scatter at high redshift is one of
the main weaknesses of the conventional approach to calculating 
distance moduli from the SALT light-curve fits.
}
\label{fig_hubble_SALT}
\end{figure}

\begin{figure}
\begin{center}
\scalebox{1.0}[1.0]{
\plotone{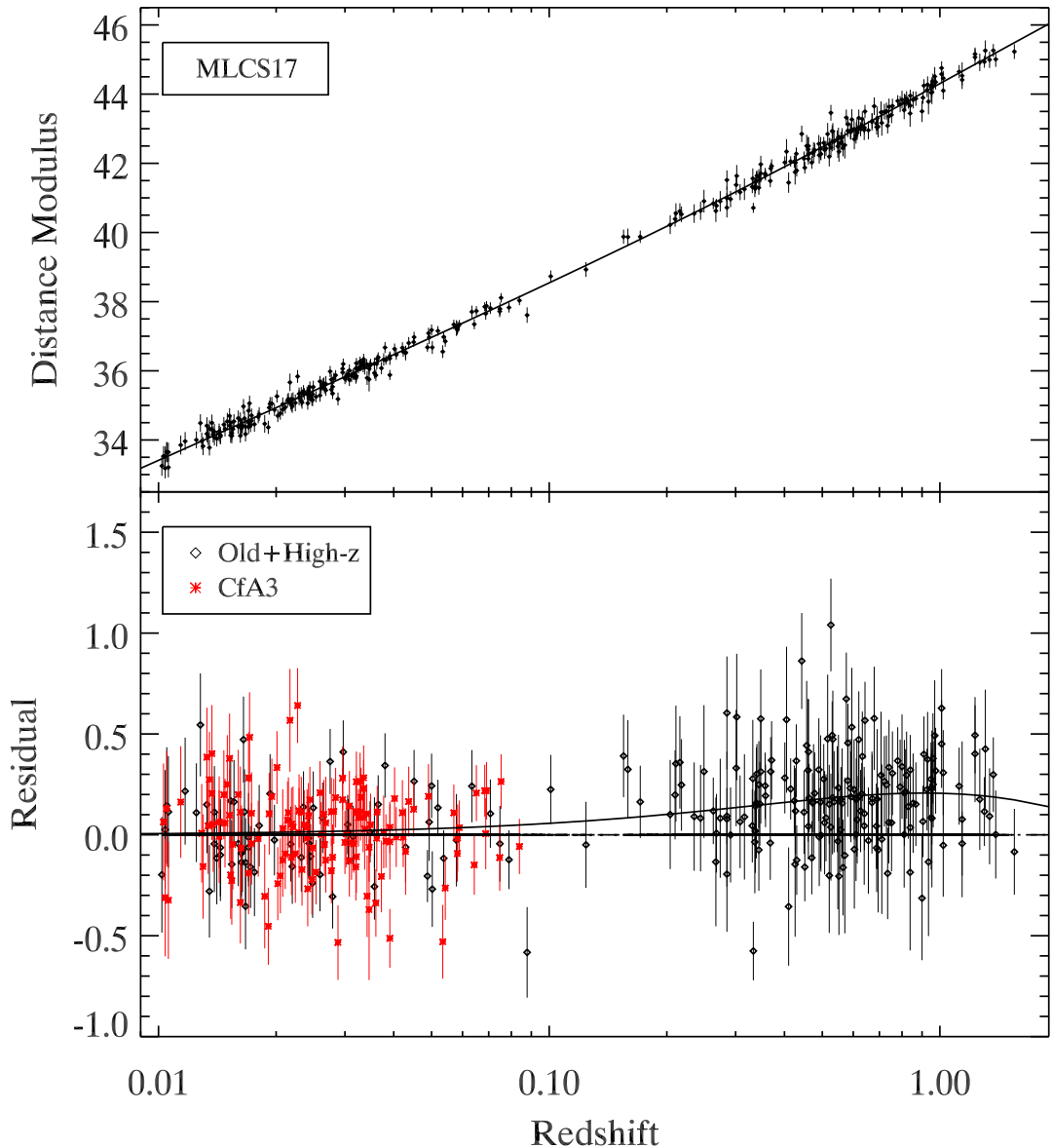}
}
\end{center}
\caption{Hubble diagram and residuals for MLCS17.  The new CfA3 points
are shown in red and the OLD and High-z points are in black.
MLCS17 (and MLCS31) has a smaller dispersion at high 
redshift than SALT (and SALT2).  
The nearby MLCS17 distances are larger than in SALT, making the High-z 
distances smaller relative to a matter-only universe and resulting 
in a greater value of $1+w$.  This effect is seen in how the MLCS17
best-fit cosmology line is closer to the axis.
}
\label{fig_hubble_MLCS17}
\end{figure}

\begin{figure}
\begin{center}
\scalebox{0.60}[0.60]{
\plotone{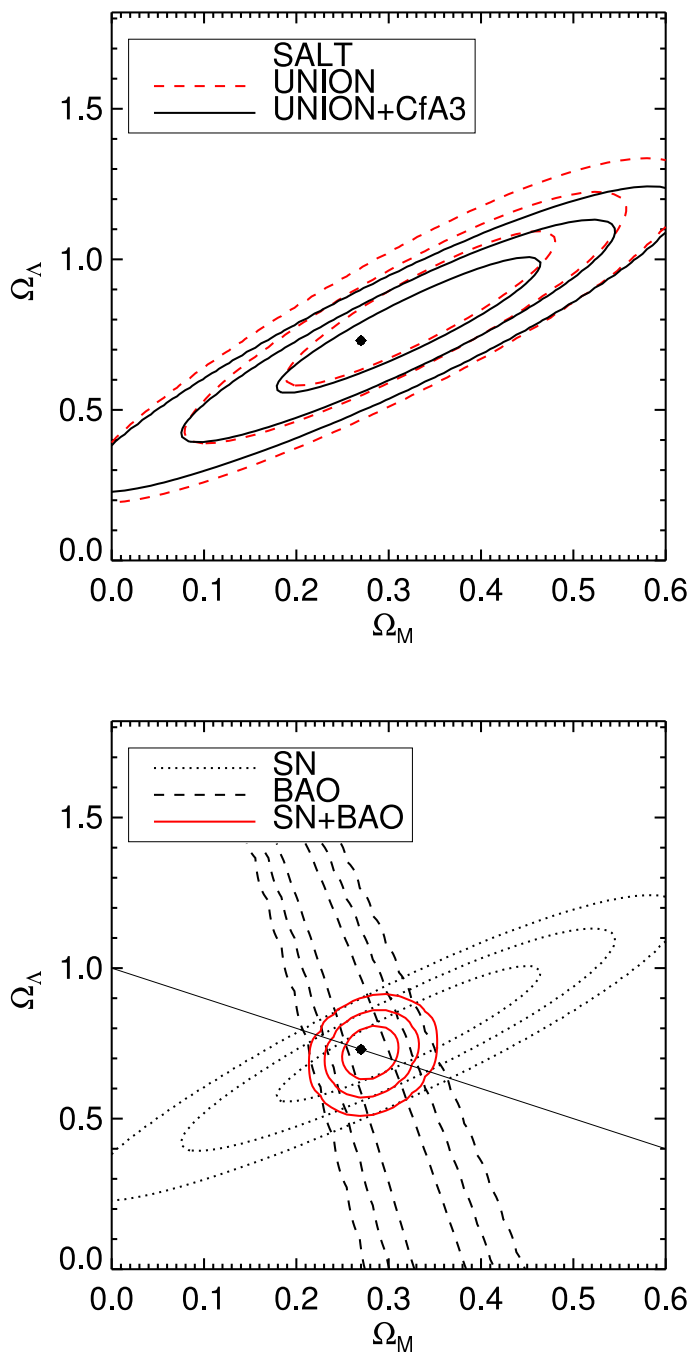}
}
\end{center}
\caption{Contour plots of \ola~vs. \om~for $1+w=0$ for SALT, with no
assumptions about flatness.  The concordance
cosmology (\ola$=0.73$, \om$=0.27$) is shown as a dot.  The
top panel shows how adding the CfA3 sample considerably narrows
the contours along the \ola~axis.  The bottom panel shows the
combination of the SN contours with the BAO prior, with the
flat-universe straight line overplotted for reference.  
} 
\label{fig_ol}
\end{figure}

We show the Hubble diagram and residuals with respect to an (\om$=0.27$,
\ola$=0$) universe in Figure~\ref{fig_hubble_SALT}, with the Constitution
sample best-fit cosmology overplotted.  For comparison, a similar plot, but for
MLCS17, is shown in Figure~\ref{fig_hubble_MLCS17}.  One of the main
differences is the high scatter at high redshift in the SALT Hubble residuals
versus the lower scatter at high redshift in MLCS17 (and MLCS31).  In
Figure~\ref{fig_ol}, we also plot the (\ola, \om) probablity contours from the
(\ola, \om, $1+w=0$) fit for both the SN-only fits and the SN+BAO fits for the
Union and Constitution sets.  The SN-only contours become tighter along the
\ola~axis when the CfA3 sample is added.  Adding the BAO prior gives
\ola$=0.718^{+0.062}_{-0.056}$ and \om$=0.281^{+0.037}_{-0.016}$.  

\begin{figure}
\scalebox{0.90}[0.90]{
\plottwo{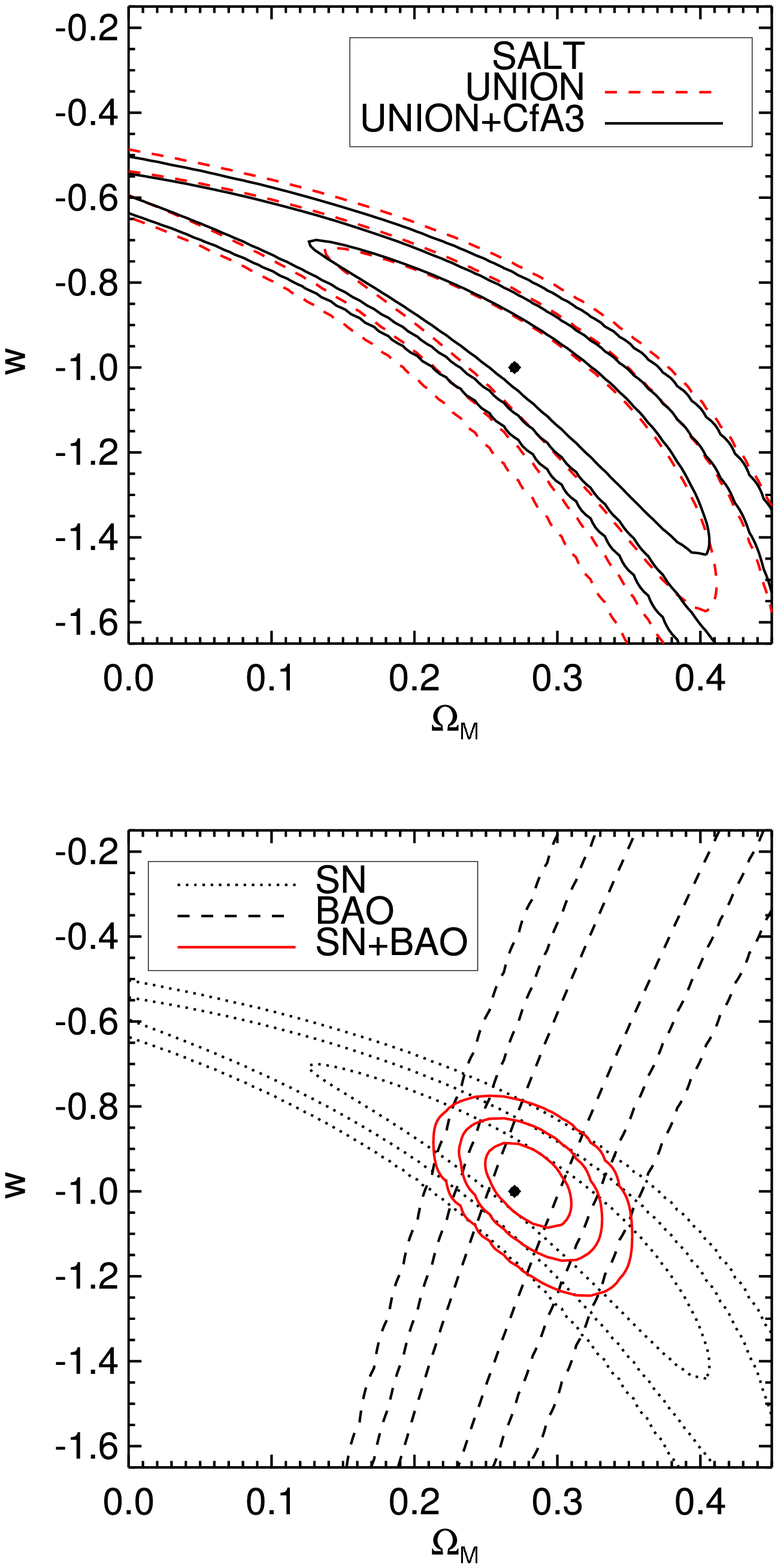}{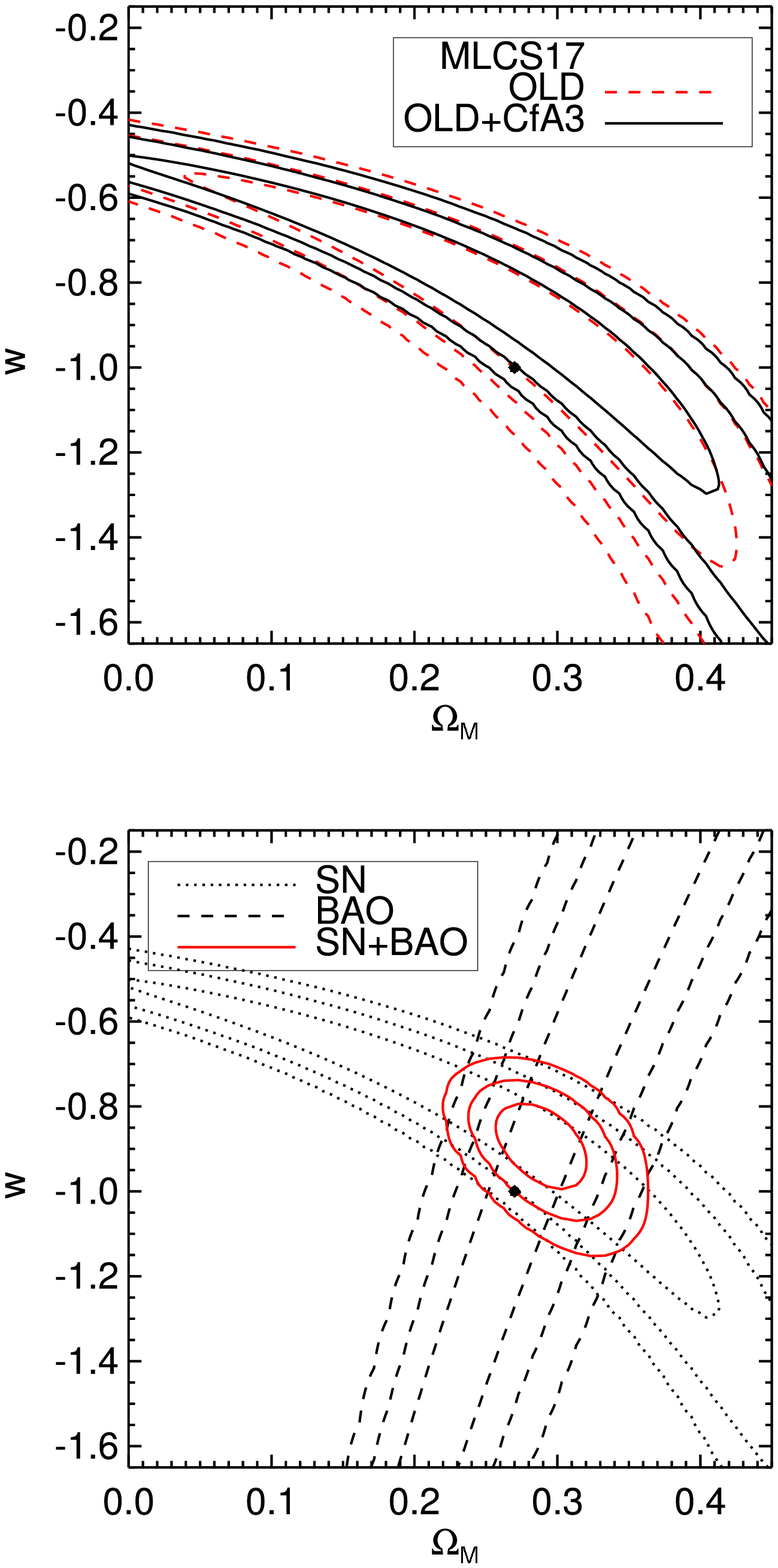}
}
\caption{Contour plots of $w$ vs. \om~in a flat universe.  The
concordance cosmology ($w=-1$, \ola$=0.73$, \om$=0.27$) is shown as a dot in
each plot.  The left two plots are for SALT while the right two plots are for
MLCS17.  The top row of plots show how adding the CfA3 sample
considerably narrows the contours along the $w$~axis.  The second row
of plots show the combination of the SN contours with the BAO prior.
}
\label{fig_w}
\end{figure}

Regarding $1+w$, as described in \S2.1, the SALT fits on the Union sample,
combined with the BAO prior, give $1+w=-0.011^{+0.078}_{-0.080}$, with 57 nearby
and 250 High-z SN Ia used.  When we only use the 90 CfA3 objects that pass the
Union cuts with the 250 High-z SN Ia then we get $1+w=-0.002^{+0.073}_{-0.075}$.
The CfA3 sample, after applying the Union cuts, is 1.58 times larger than the
Union nearby sample.  When we combine the 90 CfA3 SN Ia with the 57 OLD and
250 High-z Union SN Ia sample to form the Constitution sample we get
$1+w=0.013^{+0.066}_{-0.068}$.  This decreases the uncertainty in $1+w$ by a
factor of 1.19, when comparing with the Union-only value.  This effect can be
seen in the reduction in the width of the contours along the $w$-axis in
Figure \ref{fig_w}.  

We now address the reduction in statistical uncertainty on $1+w$ due to adding
new nearby SN Ia.  We compare this with a $\sqrt{N}$
approximation of the expected reduction in statistical uncertainty.
In comparing K08 with WV07, the statistical uncertainty on $1+w$ is
reduced from 0.09 in WV07 to 0.079 (averaging the two error bars) in K08 when
the same SN+BAO fits are compared.  This reduces the 0.09 uncertainty by a
factor of 1.14.  K08 attribute half the improvement to their eight new SN.
This would be a factor of 1.07.  K08 say that if, instead of having zero
systematic or intrinsic uncertainty, their eight SN Ia had an additional
uncertainty of roughly 0.1 mag that the statistical uncertainties on $1+w$
would increase by roughly 10$\%$, going from 0.079 to 0.087.  This would only
be a factor of 1.04 decrease over WV07.  The Constitution sample statistical
uncertainty on $1+w$ would then be smaller than K08 by a factor of 1.3.

\citet{linder06} shows that having no nearby SN Ia increases the uncertainty on
constant $1+w$ by a factor of two (and that a low-systematic-uncertainty sample
of 300 nearby objects is large enough to provide most of the possible nearby
statistical leverage).  In order to get a rough idea of how much the addition
of the CfA3 sample should reduce the statistical uncertainty (without any
thought for systematic uncertaintites), it is reasonable to assume that the
nearby and faraway samples each contribute half of the statistical leverage.
Comparing the Constitution set to the Union set, a simplistic $\sqrt{N}$
analysis would give $\sqrt{147/57}=1.6$ for the nearby contribution and 1.0 for
high-redshift sample since no new objects are added.  Averaging the two gives
1.3.  This is higher than the factor of 1.19 from the comparison of the main
SN+BAO value of K08 but equivalent to the value that is obtained in the
alternative K08 scenario of adding $\sim0.1$ mag.  This suggest that adding the
CfA3 sample achieves much of the expected statistical improvement.  However, we
estimate a systematic uncertainty on $1+w$ of 0.11 in \S4.3.  This is roughly
$65\%$ higher than our statistical uncertainty of $\sim0.067$ and suggests that
the addition of the CfA3 sample has now clearly placed SN Ia cosmology in the
realm where systematic uncertainties are the most dominant.

The standard deviation of the Hubble residuals is 0.186 mag for the Union (OLD)
sample, 0.164 mag for the CfA3 sample and 0.173 mag for the Constitution
(OLD+CfA3) sample.  The difference in the weighted means of the Hubble
residuals of the two nearby samples, OLD minus CfA3, is $-0.017\pm0.027$ mag,
showing that the two samples agree within the expected uncertainty.  On
average, the SALT-based intrinsic luminosities are slightly brighter for the
OLD sample than for the CfA3 sample.  The CMB redshift, distance moduli and
other useful information for the Constitution sample can be found in Table
\ref{table_salt} and can be used by interested parties as an update to the
Union compilation for fitting cosmological models.

\subsection{Three Other Fitters}

We summarize the number of objects that passed our minimal cuts in the other
three fitters here.  The minimal cuts leave samples that have well-fit light
curves so that we can explore for any systematic trends in the Hubble residuals
that would require further cuts.  The minimal-cuts values of $1+w$ for the
three samples (OLD, CfA3 and OLD+CfA3) and four fitters are listed in the
second portion of Table \ref{table_w}.  The results of the best-cuts samples,
described in \S3.11, are in the top portion and values of $1+w$ with other cuts
are listed in the lower portions of the table.  The first column shows $1+w$
for the OLD sample for each of the four fitters, with the mean and standard
deviation of these values below.  It should be noted that this standard
deviation is of the values of $1+w$ from the different fitters, to show how
well they agree, and should not be interpreted as a statistical uncertainty on
$1+w$.  The second column shows the CfA3 results while the third column
contains the CfA3+OLD results.  The fourth column shows the difference in $1+w$
between the OLD and the CfA3 results for each fitter. 

With SALT2, 171 High-z, 65 OLD and 115 CfA3 SN Ia pass the minimal cuts.  The
CfA3 sample is 1.77 times larger than the OLD sample, after these cuts.  In the
MLCS31 case, 72 OLD, 129 CfA3 and 165 High-z SN Ia make it through the minimal
cuts and the CfA3 sample is 1.79 times larger than the OLD sample.  The higher
number of nearby SN Ia, as compared to the SALT2 case, is due mostly to
including the 1991bg-like SN.  The MLCS31 OLD value compares well with the
value of $1+w=-0.069^{+0.091}_{-0.093}$ from WV07 who use an older version of
MLCS2k2 on the OLD, SNLS, and ESSENCE light curves (no Higher-z data) and have
more stringent cuts at z=0.015 and $t_{1st}=+4$ days.  In the MLCS17 case, 70
OLD, 133 CfA3 and 169 High-z SN Ia survive the cuts.  The CfA3 sample is 1.9
times larger than the OLD SN.

\subsection{\dd~and \av~Versus Redshift}

\begin{figure}
\scalebox{0.7}[0.70]{
\plotone{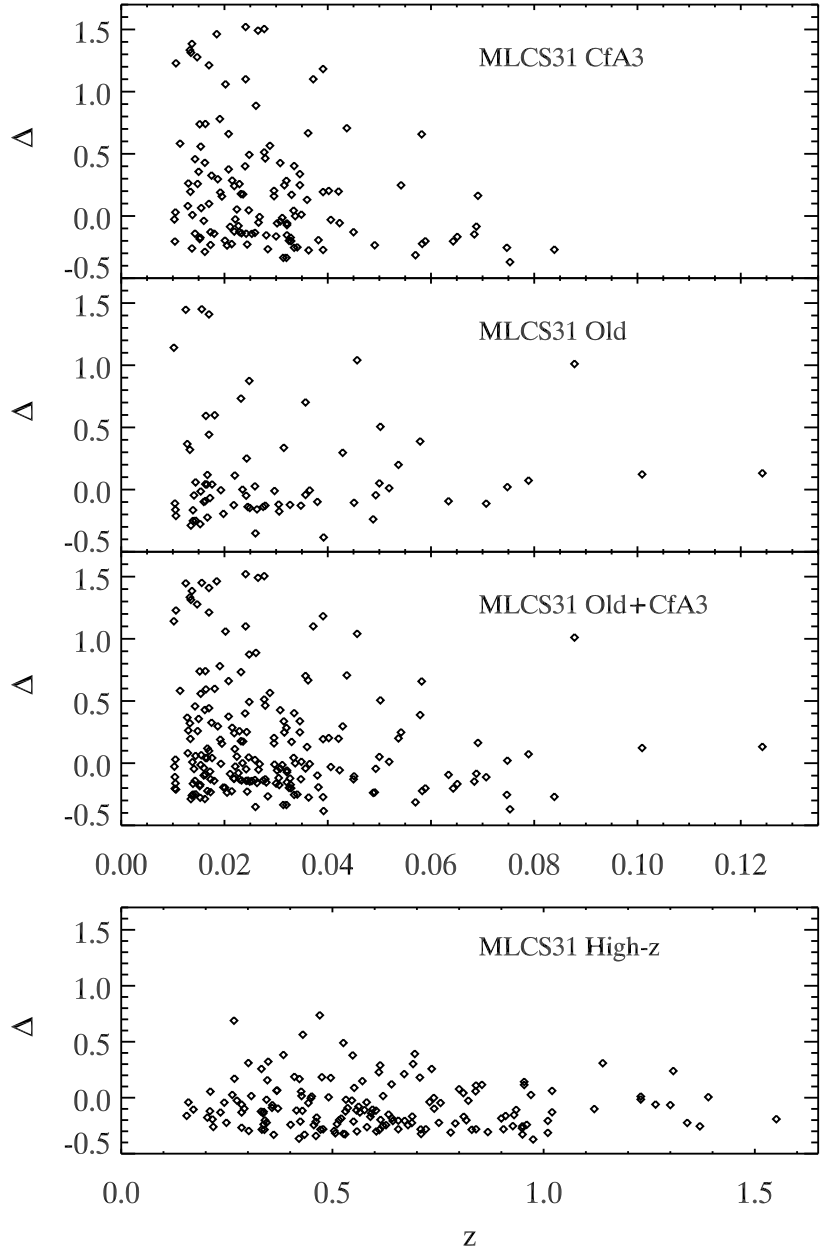}
}
\caption{The distribution of $\Delta$ versus $z$.  Faint SN Ia (high $\Delta$)
are not found at higher redshifts because of magnitude-limited searches. 
The CfA3 sample has an effective limiting magnitude of $\sim$18.5 mag.
}
\label{fig_delta_z}
\end{figure}

\begin{figure}
\scalebox{0.7}[0.70]{
\plotone{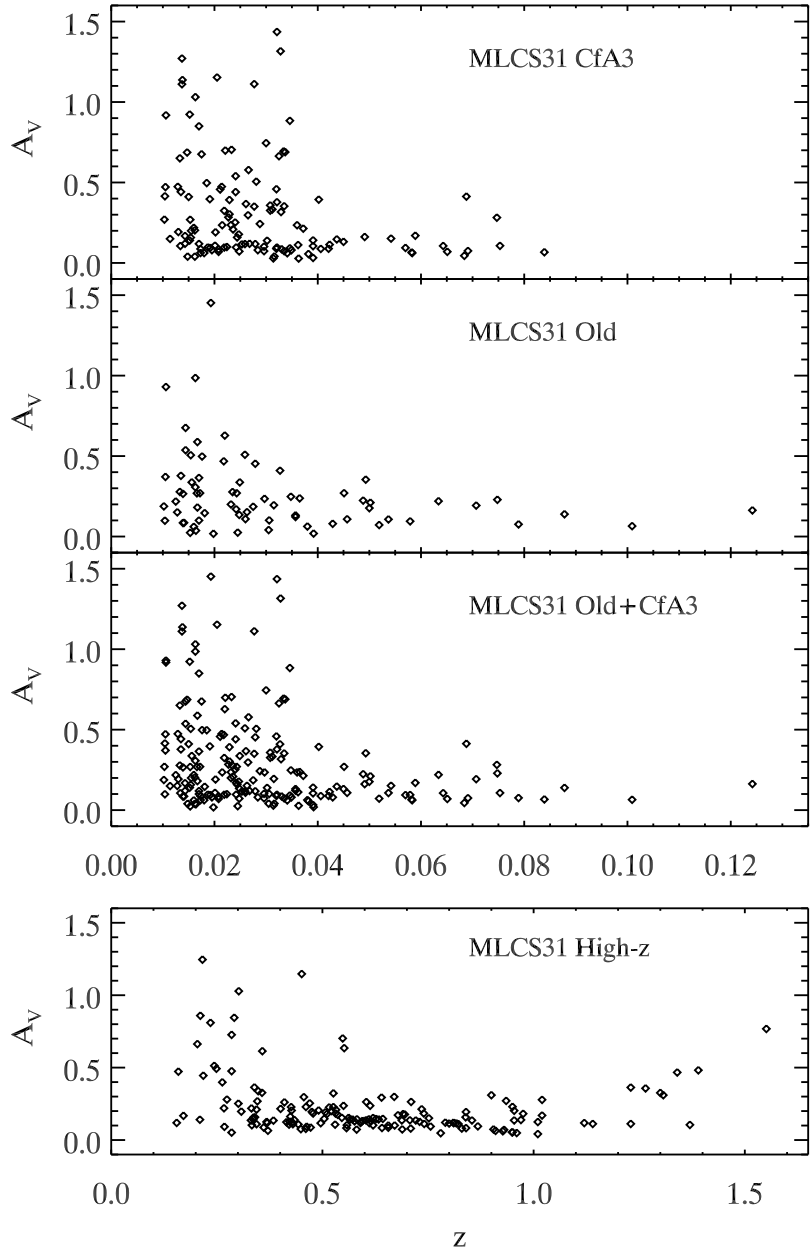}
}
\caption{The distribution of $A_V$ versus $z$.  Limiting-magnitude search 
effects reduce the number of highly extinguished SN Ia found at higher 
redshifts amongst all of the samples except for the Higher-Z, HST sample 
above $z\approx1$, where some moderate-extinction objects are found. 
}
\label{fig_av_z}
\end{figure}

We show the distribution of the MLCS2k2 light-curve shape parameters, $\Delta$
and $A_V$, versus redshift in Figures \ref{fig_delta_z} and \ref{fig_av_z}.
The CfA3 sample does extend the nearby sample with $A_V>0.5$ (and that pass
the minimal cuts) out to $z=0.035$.  Both highly-reddened and low-luminosity
(high-$\Delta$) SN Ia are found increasingly at lower redshifts, mostly due to
the CfA3 limiting peak magnitude of $\sim18.5$ mag.  The different SN Ia
populations being sampled at different redshifts makes it important to ensure
that the light-curve/distance fitters works well across the range chosen for
cosmological calculations so as to not introduce a bias.   

\subsection{Four-Fitter Comparison of Light-Curve Shape and Reddening}

\begin{figure} \scalebox{0.8}[0.80]{
\plotone{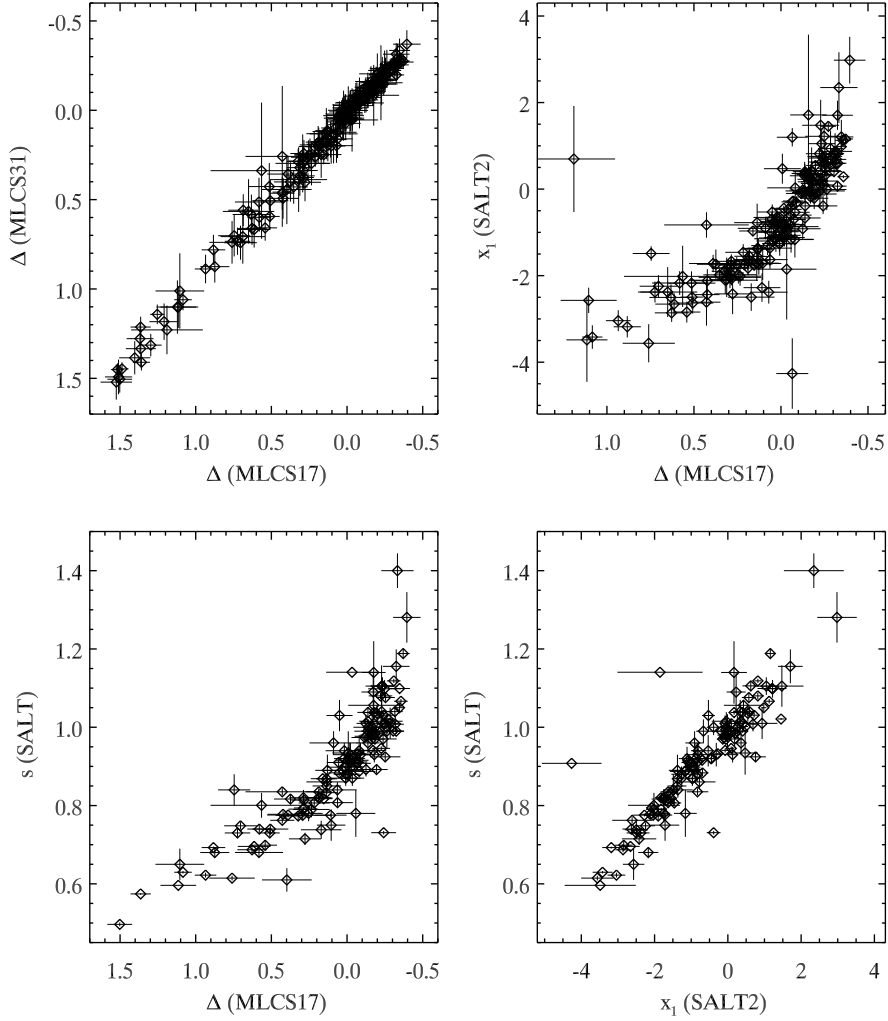} } 
\caption{Relatively good
correlation between light-curve shape parameters.  The two biggest outliers in
each of the upper-right and lower-right panels are due to three objects with
relatively-poor SALT2 fits but good fits with other fitters.  }
\label{fig_comp_lcshape} 
\end{figure}

Next, we want to compare the fitters' output parameters to see that they are
reasonably consistent with each other.  Figure \ref{fig_comp_lcshape} compares
the four fitters' light-curve shape parameters with each other.
$\Delta$(MLCS17) agrees very well with $\Delta$(MLCS31), as expected, and $s$
and $x_1$, from SALT and SALT2, are linearly related with a few outliers.  $s$
and $x_1$ are nonlinearly related to $\Delta$, the most noticeable feature
being that the brightest SN Ia (negative $\Delta$) have a wide range of 
light curve stretch, $s$ or $x_1$ (although still at the slow end).  At face
value, this means that the bright (negative-\dd) SN Ia have a wider range of 
light curve shape than the moderate decliners.

\begin{figure}
\scalebox{0.8}[0.80]{
\plotone{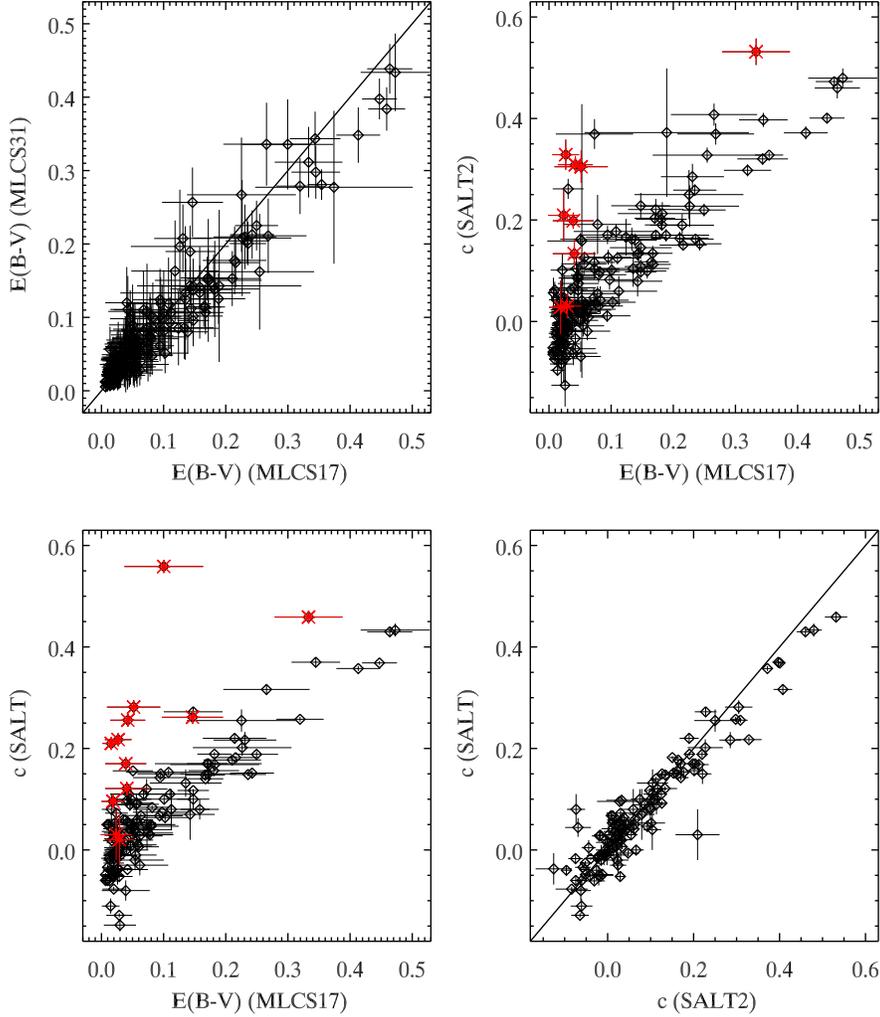}
}
\caption{Good correlation between MLCS17 and MLCS31 host galaxy reddening:
$E(B-V)_{host}=A_V/R_V$.  Diagonal lines plotted in upper-left and lower-right
panels to aid the eye.  In SALT and SALT2, $c$ is a combination of intrinsic
redness and host reddening.  The red asterisks are intrinsically-red objects
with \dd$\geq0.7$.  The lower, diagonal boundary 
in the SALT/2 vs MLCS17 comparisons is the region where the SN Ia
are intrinsically blue but suffer host reddening.  There is 
good correlation between SALT/2 and MLCS17 for the objects that
are intrinsically blue . The points above this
lower edge are intrinsically redder and show that $c$ also measures 
intrinsic redness and cannot be directly compared with $E(B-V)$.
}
\label{fig_comp_reddening}
\end{figure}

In Figure \ref{fig_comp_reddening}, the MLCS2k2 host-galaxy color excess,
$E(B-V)= A_V/R_V$, from MLCS17 and MLCS31, and the color parameter, $c$
(a combination of host reddening and intrinsic SN Ia redness), from 
SALT and SALT2 are compared.  The upper-left panel shows mostly good agreement.
The slope is slightly less than one, meaning that MLCS17 with \rvone~
has a slightly larger best-fit value for $E(B-V)$ than MLCS31 with \rvth.  
However, \av~will usually be larger for \rvth.  The lower-right panel
shows good agreement between $c$ in SALT and SALT2.

In the upper-right and lower-left panels, $c$ should correlate well with
$E(B-V)$ in unreddened cases.  This is evident in the roughly diagonal 
lower bound on the locus of points in these two panels and the roughly
diagonal lower locus of points in general.  The points that
extend significantly above this lower locus are mainly SN Ia with 
higher intrinsic redness that is part of $c$ but not $E(B-V)$.  Points
that have $\Delta\geq0.7$--intrinsically redder SN Ia--are shown in red.

\begin{figure}
\scalebox{0.8}[0.80]{
\plotone{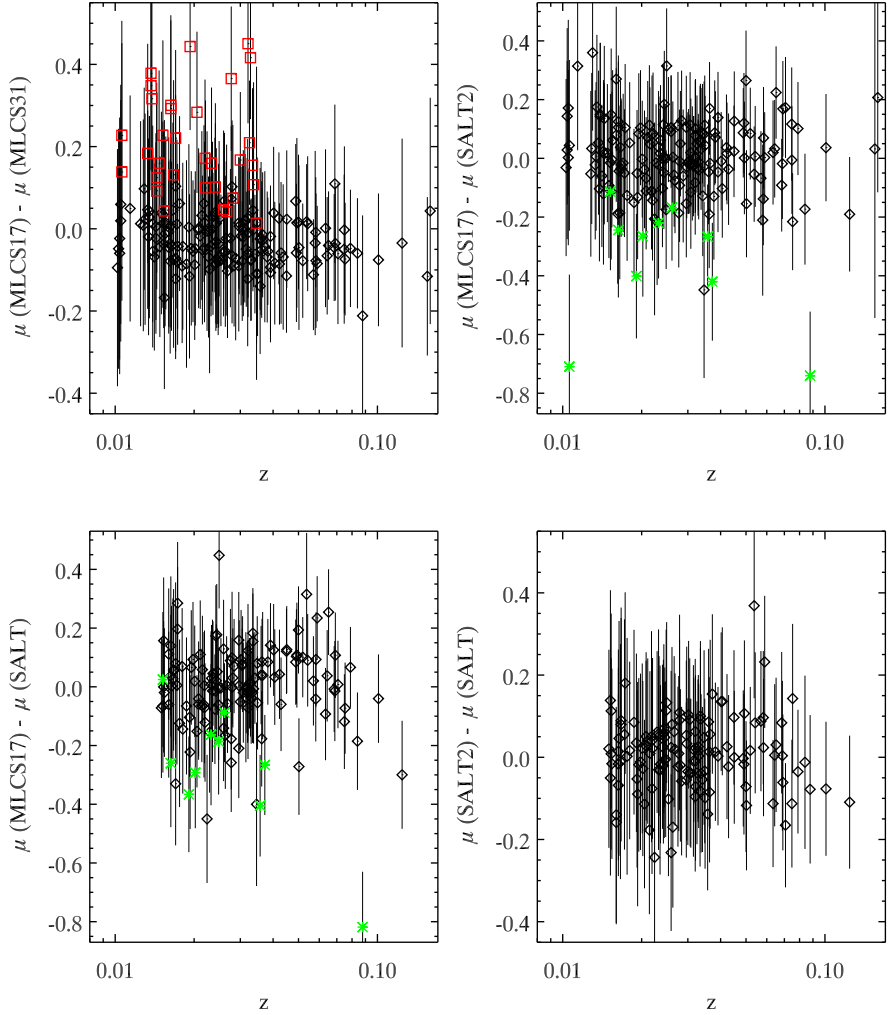}
}
\caption{Difference in distance moduli between fitters versus redshift.  The
upper-left panel compares MLCS31 with MLCS17.  The
red squares in this panel have $A_V>0.5$ in MLCS31 and show a
dramatic difference in distance from the MLCS17 values.  The less-reddened
SN Ia agree well.  There might be a slight offset in the remaining values
since the value of $\mu$ depends on the Hubble constant which was determined
from the full samples.  SALT/2 versus MLCS17 comparisons are good, especially
when the problematic MLCS2k2 points with $0.7\leq \Delta \leq1.2$ in green
are ignored.  SALT and SALT2 agree well.
}
\label{fig_comp_diffmu_z}
\end{figure}

Finally, in Figure \ref{fig_comp_diffmu_z}, we plot the difference between the
fitters' distance moduli versus redshift.  In order to put them on the same
scale, the value of $\rm H_o$ was calculated for the four fitters' samples with
the minimal cuts and a correction was added to make them all consistent with
$\rm H_o=65 km s^{-1}$ and $M_V=-19.504$ mag for MLCS2k2, $M_B=-19.46$ mag for
SALT and $M_B=-19.44$ for SALT2.  The actual additive correction is sensitive
to the exact subsample used.  The upper-left panel shows significant
disagreement below $z=0.04$, where the highly-reddened MLCS31 points (shown in
red) have significantly shorter distances.  This is most likely due to
\rvth~being too high and is discussed further in \S3.8.  The low-\av~points do
show excellent agreement though.  The slight average offset from zero of the
slightly-reddened points is mainly due to the whole MLCS31 sample being adjusted
to be consistent with $\rm H_o=65 km s^{-1}$, as opposed to only the
lowly-reddened portion.

The lower-right panel shows excellent agreement between SALT and SALT2 while
the upper-right and lower-left panels show good agreement across redshift,
especially when the green points are ignored.  These points are in a range
($0.7\leq \Delta \leq1.2$), where MLCS2k2 seems to produce distances that are
too short (see H09 and \S3.7).  This will be discussed further below.

Overall, the four fitters produce light-curve shape and reddening/color
parameters that agree well with each other.  This is an important result
because it shows that they are all similarly characterizing the SN Ia light
curves but that determining the distance modulus is an area where
different methods can introduce systematic offsets.   It is thus a good sign
that the distance moduli do not exhibit any trends versus redshift when the
areas of low reliability are removed.

\subsection{Twins}

As another test of the light-curve and distance fitting, SN Ia in the same host
galaxy--``twins"--should produce the same distances.  This is also a test of
the underlying photometry.  One example where this has been discussed in the
literature is NGC 1316, the host of SN 1980N and SN 1981D.  \citet{hamuy91}
find that the peak apparent magnitudes agree well but mention that 1981D has
significant host contamination.  This makes the photometry of 1981D less
reliable.  \citet{krisc00} find that MLCS optical
light-curve fits of 1980N and 1981D give almost identical values of $\Delta$,
implying similar absolute magnitudes, but due to the estimated host reddening
the distance of 1981D is smaller by about 0.4 mag than that of 1980N.  They
find no evidence for host reddening from \emph{VJHK} analysis but do from the \emph{BV}
data.  Since the peak \emph{V} magnitudes agree well, they conclude that the
\emph{B}-band photometry of SN 1981D may have a hidden systematic error.  This could
well come from the host contamination.  A recent addition to this conflicted
family--once twins, now triplets--is SN 2006dd \citep{monard06}.  
Hopefully it will settle the score.

On a brighter note, SN 1999cp \citep{krisc00} and SN 2002cr occured in the same
galaxy, NGC 5468.  Optical photometry of
SN 2002cr was obtained as part of the CfA3 sample.  We fit both SN with all
four fitters and the respective agreement in MLCS31, MLCS17, SALT and SALT2
distance moduli, $\mu_{2002cr}-\mu_{1999cp}$, are:  $0.006\pm0.138$,
$0.017\pm0.137$, $0.027\pm0.195$, and $0.077\pm0.233$ mag.  The uncertainties
include both the uncertainty from the light-curve/distance fit and the
additional ``intrinsic" uncertainty for each fitter mentioned in \S2.

It should be emphasized that the photometry was aquired and reduced by
separate groups and instruments.  The excellent agreement in MLCS17,
MLCS31 and SALT, and the adequate agreement in SALT2 imply that the 
photometry from the two groups is accurate and the fitters are working well.

\subsection{Comparing the Results from the Minimal-Cut Samples}

Table~\ref{table_meanres} shows the standard deviation and the weighted mean of
the Hubble residuals of each fitter for each SN Ia sample.  We want to know if
the two nearby samples are consistent internally and with each other.  The
similarity in the standard deviations across all four fitters and in the OLD,
CfA3 and OLD+CfA3 samples suggests both.  With all four fitters, the OLD sample
has a more negative mean of its Hubble residuals than the CfA3 sample by about
0.03 mag when all four fitters' values are averaged.  This means that the OLD
sample is being interpreted as intrinsically brighter than the CfA3 sample.
Whether this is due to photometry errors or intrinsic differences in the SN Ia
is hard to say.  SALT and MLCS17 give differences in mean residuals between the
two samples that are within the uncertainty while SALT2 and MLCS31 are slightly
outside the quoted uncertainty.  The reality of the situation is probably a
combination of both photometric offsets and intrinsic SN Ia differences.  We
examine this further in \S3.12 by calculating the mean residuals for the six or
seven main sources of the nearby sample.  We find a scatter of $\sim0.03$ mag
amongst these subsamples.  We conclude that this level of offset between
nearby samples is representative of the current state of nearby photometry and
light-curve fitting.  Since the difference in the mean residuals of the CfA3
and OLD samples, across all four fitters, is roughly at the $1\sigma$ level it
is not too worrisome.  Whatever the reason for the difference in the mean
residuals, it is consistent with the OLD sample implying larger \ola~(with
$1+w=0$) and lesser $1+w$ than the CfA3 sample.  At high redshift, SALT and
SALT2 have large standard deviations of their Hubble residuals (see \S3.8
for one possible cause of this) while MLCS17 and MLCS31 have values that are
roughly the same as at low redshift.  Using the lower wavelength range of 
2900~$\rm\AA$ should reduce the scatter of the SALT2
high-redshift objects but we do not explore that in this work.

We now turn to looking for trends in Hubble residuals versus light-curve
fitter parameters.  In \S3.7, we focus on the light-curve-shape parameter
and look at the effects on $1+w$ of removing $\Delta>0.7$.  In \S3.8, we
focus on the color/extinction parameters and look at how only including
$-0.1\leq c \leq 0.2$ for SALT/2 and $A_V<0.5$ for MLCS31 and MLCS17 affect
$1+w$.  In the end, we adopt all of these as the ``best" cuts and recalcute
$1+w$ in \S3.11.

\subsection{Residuals versus Light Curve Shape}

\begin{figure}
\scalebox{0.85}[0.80]{
\plotone{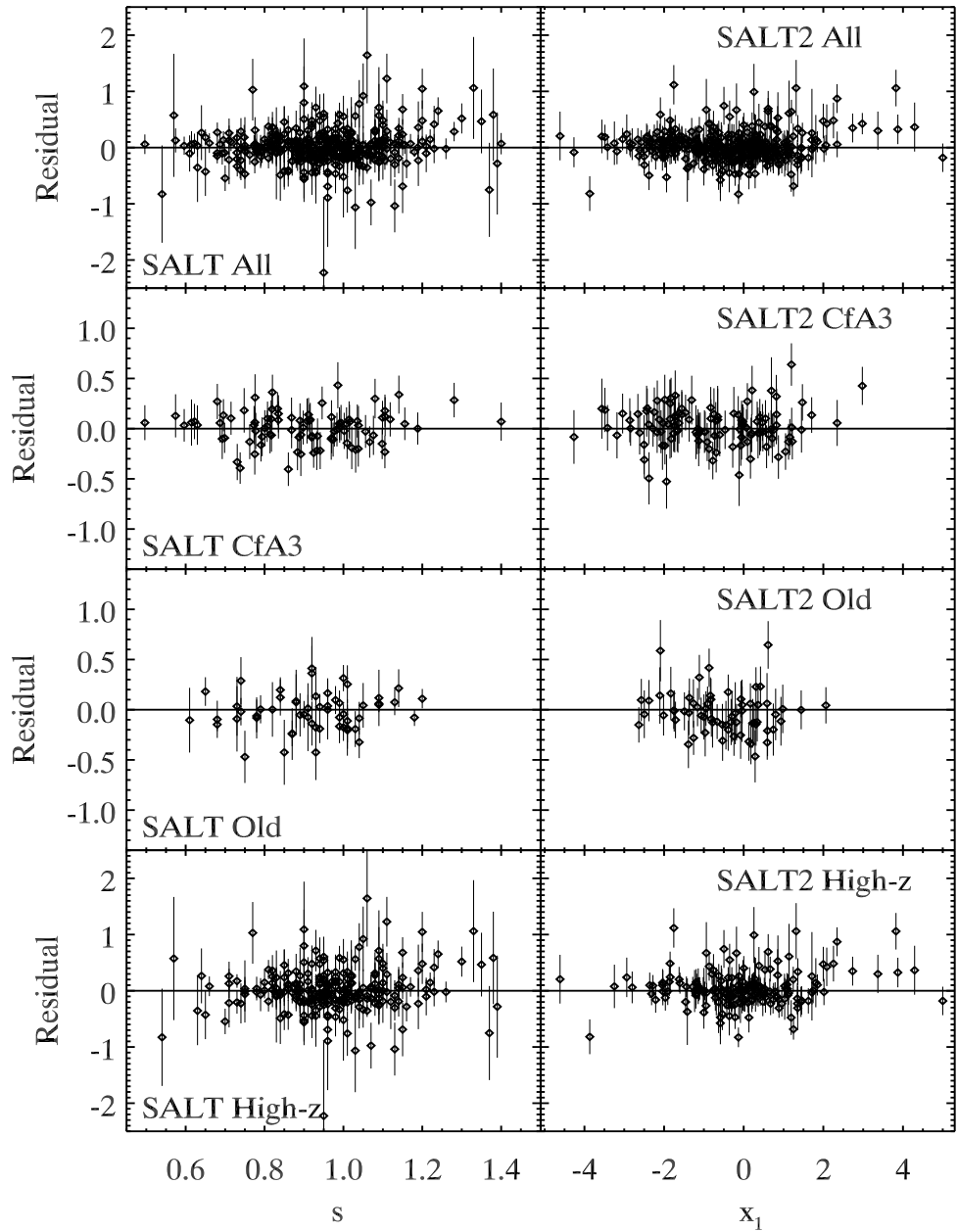}
}
\caption{Hubble residuals relative to the best-fit cosmologies versus $s$ 
and $x_1$ for SALT and SALT2, respectively.
}
\label{fig_res_s_x1}
\end{figure}

We use the definition of Hubble residual, $\rm HR\equiv \mu_{SN}-\mu_z$, so
that a positive residual means that the SN-derived distance is greater than the
redshift-derived distance with a given set of cosmological parameters.  A
positive residual can also be interpreted as the SN being fainter than expected
for the reference cosmology, and vice versa for a negative residual.  Figure
\ref{fig_res_s_x1} shows the Hubble residuals versus the light curve shape
parameters, $s$ and $x_1$, for SALT and SALT2, respectively.  No significant
trends exist in any of the subsamples but it should be noted that there are
more broad-shaped SN Ia (large $s$ and $x_1$) at high redshift and that for
SALT2 these have mostly positive residuals, meaning that they are being
interpreted as intrinsically fainter than predicted by the best-fit cosmology.  
Some of the largest outliers in the high-redshift SALT and SALT2 samples are
due to a trend in residuals versus color parameter, $c$.  Cuts on $c$ that will
remove many of these outliers will be explored in \S3.8.

\begin{figure}
\scalebox{0.85}[0.80]{
\plotone{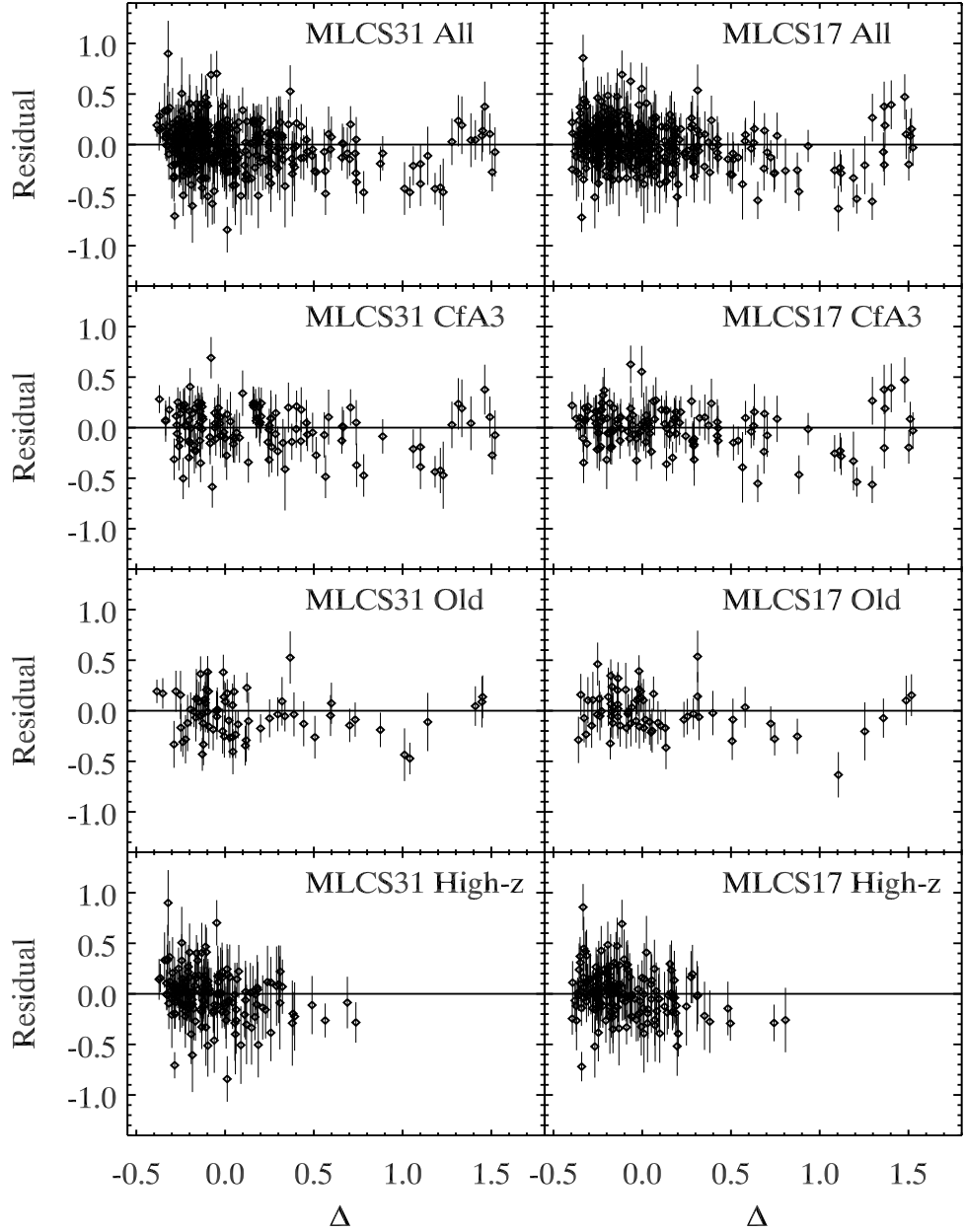}
}
\caption{Hubble residuals relative to the best-fit cosmologies versus \dd~
for MLCS31 and MLCS17, respectively.  There is a noticeable region of
negative residuals between $0.7<\Delta<1.2$.
}
\label{fig_res_delta}
\end{figure}

Figure \ref{fig_res_delta} displays the Hubble residuals for MLCS31 and MLCS17
versus \dd.  With both fitters, and in both the OLD and the CfA3 samples, there
is a peculiar region between \dd~$=0.7$ and \dd~$=1.2$ where the residuals are
mostly negative.  At high redshift there are only two objects at the edge of
this range (due to selection effects) but they also have negative residuals.
This ``dip" in the residuals is most likely due to a sharp change in the
intrinsic absolute magnitude of 1991bg-like SN Ia compared to their neighboring
non-1991bg-like SN Ia with slightly lower values of \dd.  H09 find that, at
peak, the 1991bg-like SN Ia are about 0.5 to 1.0 mag fainter in \emph{B} and
0.3 to 0.5 mag fainter in \emph{V} than the non-1991bg-like SN Ia that are
nearest to them in \dd.  In the nearby samples, the residuals beyond \dd~$=1.2$
are fairly well centered on zero, with a slight positive offset in the CfA3
sample.  The inclusion of the 1991bg-like SN Ia in the MLCS2k2 training set
gives rise to the positive coefficient for the quadratic term in \dd.  The
MLCS2k2 model light curves are accurately calibrated for the range of \dd~that
the 1991bg-like SN Ia occupy but this comes at the expense of the model light
curves being too faint in the region $0.7<\Delta<1.2$.  This, in turn, produces
negative residuals in this region.  MLCS2k2 should be retrained by treating the
1991bg-like objects (either identified spectroscopically or by making a cut
above \dd$\gtrsim1.2$) separately.  This will likely improve the performance of
MLCS2k2, which already produces smaller scatter than SALT and SALT2 at high
redshift.

\begin{figure}
\scalebox{0.8}[0.80]{
\plotone{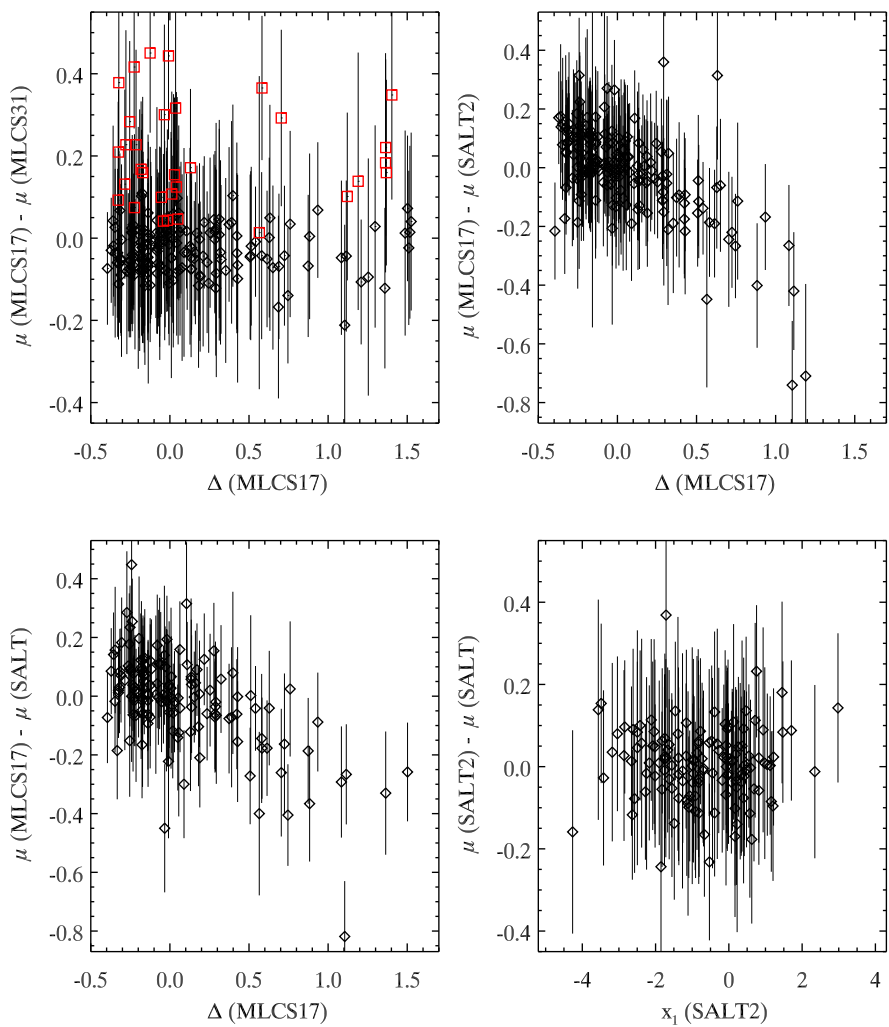}
}
\caption{Comparison of distance moduli between the four fitters versus
light-curve shape parameter.  MLCS17 and
MLCS31 agree well when the highly-reddened SN Ia (red squares) are ignored. 
SALT and SALT2 agree well.  There is a systematic trend between MLCS2k2 and
SALT/2, where fainter objects have relatively smaller distances, especially
beyond $\Delta=0.7$.
}
\label{fig_comp_diffmu_lcshape}
\end{figure}

Another check on this region of \dd~is to compare the differences in the
MLCS2k2 and SALT/2 distances versus \dd.  SALT and SALT2 were not trained using
1991bg-like SN.  Figure \ref{fig_comp_diffmu_lcshape} shows that the MLCS2k2
distances get smaller as compared to SALT/2 as \dd~increases and there does
seem to be an even steeper drop starting around $\Delta\approx0.5$.  There are
no 1991bg-like SN Ia in the MLCS17-versus-SALT2 plot while in the
MLCS17-versus-SALT plot there are two SN Ia with $\Delta>1.2$ because K08 did
not exclude 1991bg-like SN Ia and we kept all the K08 SN in our minimal cuts.
In spite of the trend in differences in distances versus \dd, it is reassuring
that there is no trend versus redshift (see Figure \ref{fig_comp_diffmu_z}).
It can be expected that there would be some systematic differences in distances
produced by different fitters but we believe that a retraining of MLCS2k2 and
SALT2 with larger data sets, avoiding the 1991bg-like SN Ia in MLCS2k2 and
better treatment of the \emph{U} band in SALT2 will bring them into better
agreement.  Figure \ref{fig_comp_diffmu_lcshape} also shows
that the MLCS17 and MLCS31 distances agree well
versus \dd~when the reddened ones are ignored (upper-left panel) 
and the SALT and SALT2 distances agree well versus $x_1$ (lower-right panel). 

To avoid the problems of the MLCS2k2 fits with $0.7<\Delta<1.2$, and since the
High-z SN only extend out to $\Delta\approx0.7$, we make a cut on \dd~$>0.7$ and
rerun the MLCS31 and MLCS17 cosmology fits.  By making this cut we are removing
more negative than positive residual objects and so we would expect an increase
in $1+w$.  For MLCS31, $1+w$ increases by about 0.03 to 0.035 for each of the
OLD, CfA3 and OLD+CfA3 subsamples, and a similar effect is seen for MLCS17.
The specific results are presented in Table \ref{table_w}.  In $\S$4.2, we will
look for any trends in residuals and \dd~versus host galaxy morphology.

As SN Ia samples increase in size it may prove worthwhile to only include
objects that are within the light-curve-parameter regions that are common to
both high- and low-redshift samples.  It is uncertain if our cut on \dd~moves
our results on $1+w$ closer to physical reality since we are removing a region
of the parameter space that was used in the training.  Ideally, the training of
the light-curve/distance fitter would be performed over the same range of
parameters that is used in the cosmological calculations.  The cut at
$\Delta=0.7$ will be one part of our best cuts.  The other part wil be to only 
include objects with $-0.1 \leq c \leq 0.2$ and $A_V\leq 0.5$, which we discuss
next.

\subsection{Residuals versus $c$ and \av;  Discussion of $\beta$}

\begin{figure}
\scalebox{0.8}[0.80]{
\plotone{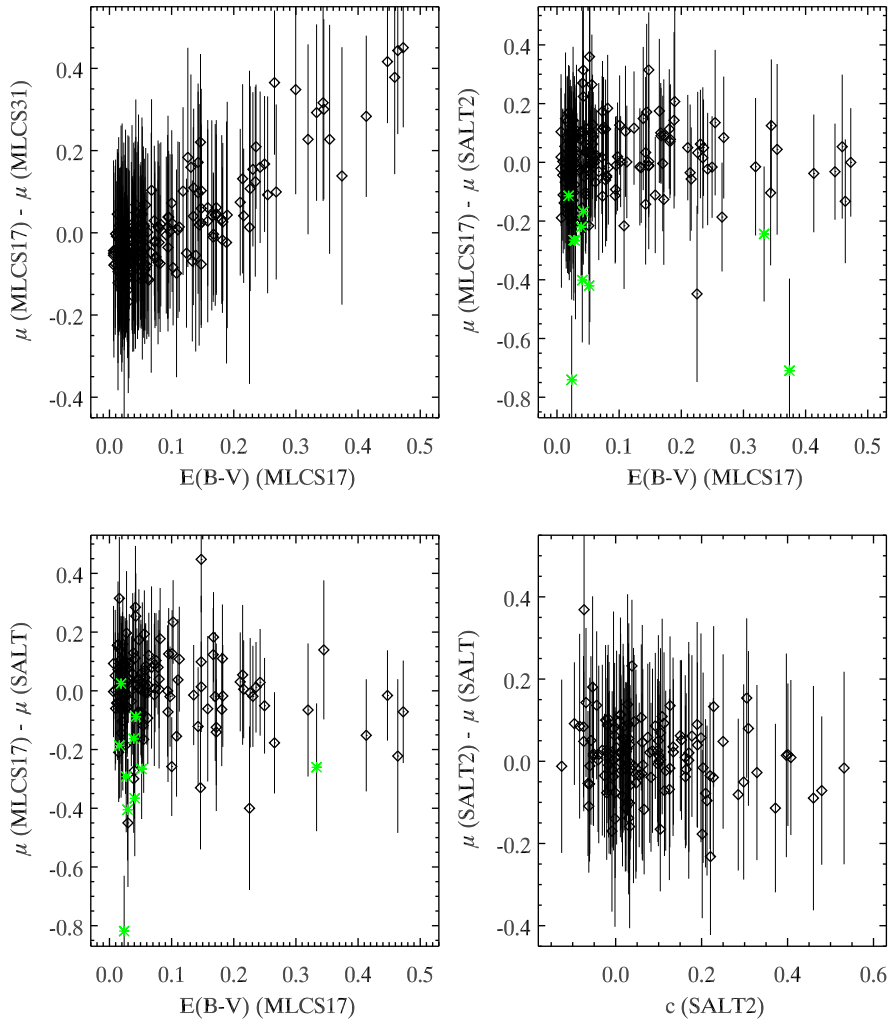}
}
\caption{Comparison of distance moduli between the four fitters versus
host-galaxy reddening in three panels and SALT2 $c$ in the fourth.
The upper-left panel shows that MLCS31 (with $R_V=3.1$) 
probably overestimates the extinction, $A_V$, for highly-reddened
objects while MLCS17 agrees well with SALT/2, especially when the
green points ($0.7\leq \Delta \leq1.2$) are ignored.  SALT and SALT2 agree
well.
}
\label{fig_comp_diffmu_reddening}
\end{figure}

Figure \ref{fig_comp_diffmu_reddening} shows the difference between distance
moduli for the four fitters versus color parameter.  The comparisons between
MLCS17, SALT and SALT2 all show no trend while it is very clear that the higher
value of \rvth~makes the MLCS31 distances increasingly smaller than the MLCS17
distances with higher host reddening.

\begin{figure}
\scalebox{0.85}[0.80]{
\plotone{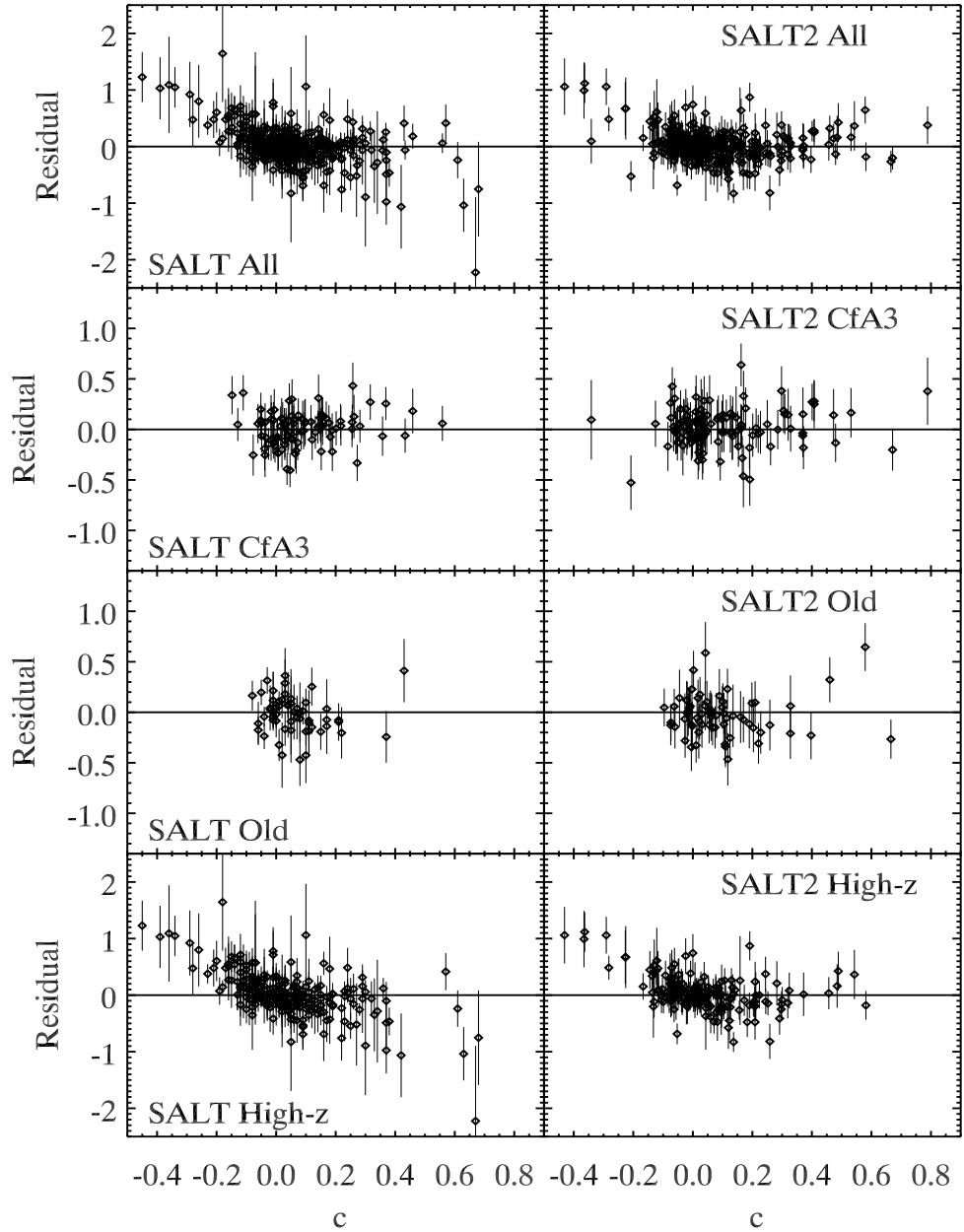}
}
\caption{Hubble residuals relative to the best-fit cosmologies 
versus $c$ for SALT and SALT2, respectively.  The nearby samples do not
show any significant trends versus $c$ but the High-z sample does and
we explore making a cut on $c$. 
}
\label{fig_res_c}
\end{figure}

In Figure \ref{fig_res_c}, we plot the Hubble residuals versus $c$ for SALT and
SALT2, with the minimal cuts (for SALT, these are the Union cuts from K08).  
Neither of the nearby samples show any significant trend but the
high-redshift sample has highly positive residuals at the blue end with both
SALT and SALT2 and several highly negative residuals at the reddest end with
SALT.  The High-z SALT sample shows a clear negative slope with increasing
color.  We explore the effects on $1+w$ by making a sensible cut on $c$.  

Most of the OLD SALT SN Ia fall within the range $-0.1 \leq c \leq 0.2$.
Cutting all objects outside this range would remove most of the extreme
outliers in the SALT High-$z$ sample.  We choose this as our cut on $c$ for
SALT.  The High-$z$ SALT2 sample does not have the same problem at the red end
but we make the same cut on $c$ to be consistent.  
Also, $c\approx0.2$ corresponds well with this
value.  If a typical unreddened $s\approx1$ SN Ia has $c\approx0$ then this
same SN with $c\approx0.2$ would experience $0.2$ mag of host reddening.  If we
assume that \rv$\approx2.5$, a rough average of the two values in this paper,
then $c\approx0.2$ corresponds to \av$\approx0.5$ and these cuts on $c$ and 
\av~are fairly consistent.   

Without the cuts on $c$, SALT and SALT2 have respective values of $\beta$ of
$2.59^{+0.12}_{-0.08}$ and $2.48^{+0.10}_{-0.12}$ for the OLD+CfA3 samples
(including High-z) and $2.87^{+0.17}_{-0.17}$ and $2.69^{+0.22}_{-0.21}$ with the
cuts on $c$.  The corresponding values of $\alpha$ are $1.34^{+0.08}_{-0.08}$ and
$1.26^{+0.09}_{-0.09}$ for SALT, before and after the cut on $c$, and
$0.104^{+0.018}_{-0.018}$ and $0.109^{+0.026}_{-0.021}$ for SALT2.  We also fit
after only cutting out the objects bluer than $c=-0.1$ and find
$\beta=2.55^{+0.10}_{-0.10}$ for SALT2 ($\alpha=0.106^{+0.019}_{-0.019}$).
This cut does not remove the redder ones.  It is only $0.14\pm0.24$ lower than
the SALT2 case where the additional cut above $c=0.2$ is also made.  This
difference in $\beta$ of 0.14 is small but implies that the reddening law for
dust is smaller for highly-extinguished SN Ia.  When $\beta$ is calculated in
SALT for the minimal-cuts High-z and nearby objects separately, they are
virtually the same, consistent with the findings in K08.  In SALT2 they differ
by $0.14\pm0.22$, within the $1\sigma$ uncertainty.  This suggests that the
single color correction of the nearby and faraway SN Ia is adequate.  However,
there is a large difference in SALT between $\alpha=1.13^{+0.09}_{-0.08}$ in
the low-redshift sample and $\alpha=1.75^{+0.15}_{-0.13}$ in the high-redshift
sample:  $0.62\pm0.16$.  In SALT2, $\alpha$ is $0.037^{+0.029}_{-0.039}$ larger
in the high-redshift sample than in the nearby.  Since the nearby and High-z
values of $\beta$ are consistent with each other, but the values of $\alpha$
are not (more in SALT than SALT2), it is interesting that the trend in High-z
residuals is seen versus color, $c$, and not stretch, $s$ (see Figures
\ref{fig_res_c} and \ref{fig_res_s_x1}).

If there were no differences in intrinsic SN Ia color then $c$ would just be
the host reddening and $\beta$ would be equivalent to $R_B=R_V+1$.  Even though
$\beta-1$ is not equivalent to \rv, it is interesting to see that the values of
$\beta-1$ are about 1.55 without the cuts on $c$ and 1.78 with the cuts, both
very close to the value of \rvone~ derived to minimize the dependence of the
nearby Hubble residuals on \av. 

\begin{figure}
\scalebox{0.85}[0.80]{
\plotone{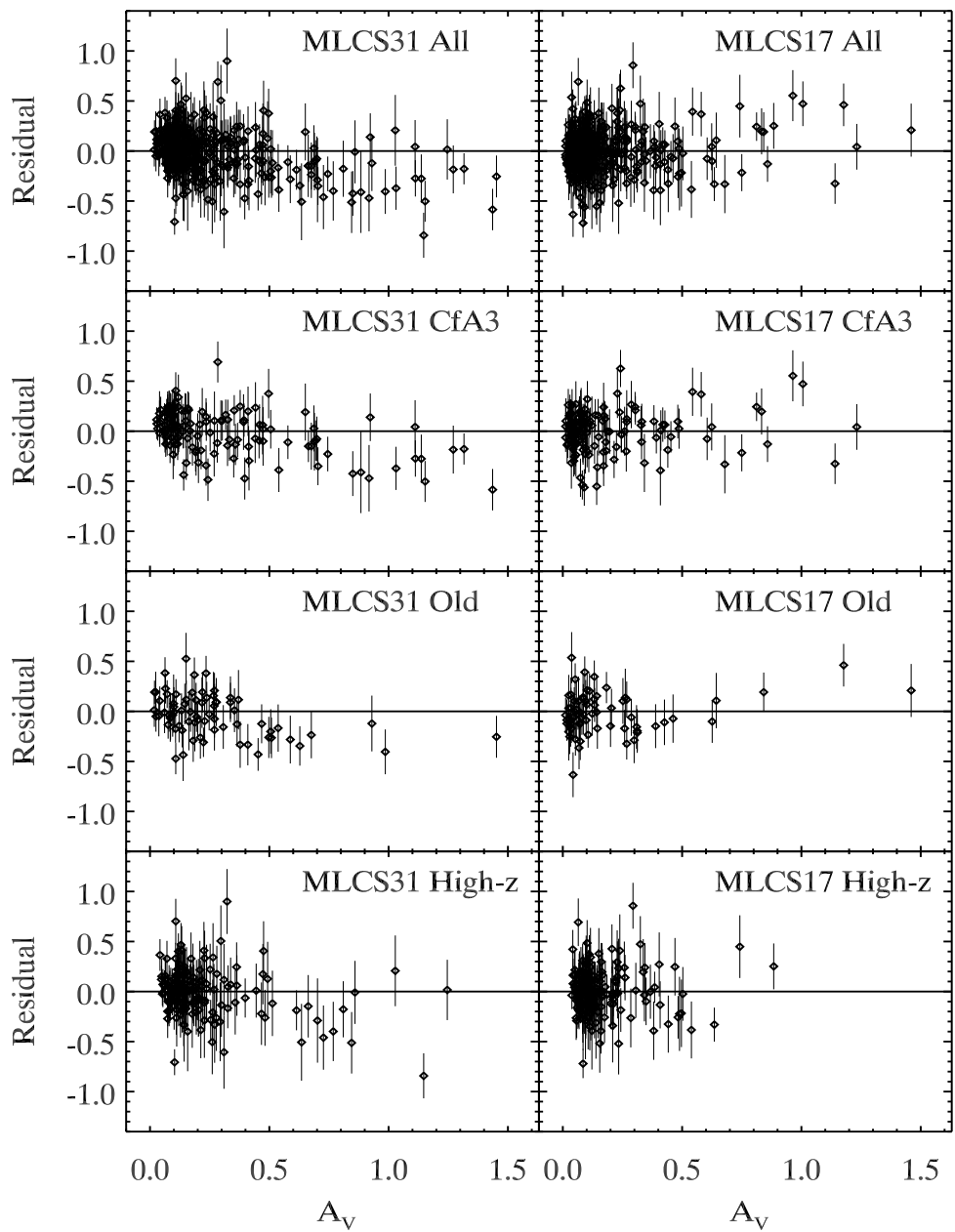}
}
\caption{Hubble residuals relative to the best-fit cosmologies 
versus \av~for MLCS31 and MLCS17, respectively.  The residuals are 
noticeably negative beyond $A_V\approx0.5$ for \rvth, suggesting that
$A_V$ is being overestimated by MLCS31.
}
\label{fig_res_av}
\end{figure}

In Figure \ref{fig_res_av}, we plot the Hubble residuals versus \av~for MLCS31
and MLCS17.  All three subsamples (CfA3, OLD and High-z) have mostly negative
residuals beyond \av$=0.5$ with MLCS31.  The slope in residuals versus \av~for
MLCS31 is $-0.34$ both with or without the cut on \av~while for MLCS17 it is
$+0.01$ (nearly zero) with the cut at $A_V=0.5$ and $+0.12$ without it.
Although a slightly higher value in \rv~would eliminate the trend in residuals
versus \av, \rvone~succeeds in eliminating the trend in residuals versus \av~
when the cut at $A_V=0.5$ is made.  $A_V=0.5$ is adopted as a cut-off
value for our best cuts in \S3.11 and so there is no point in searching for a 
higher value of \rv~that would eliminate the trend at greater \av.  
The fact that, without the cut on \av, MLCS17
has a slightly positive slope while it was practically zero for the low-\av~
objects implies that the high-extinction objects are interacting with dust with
larger \rv~than the dust of the low-extinction objects.  This is the opposite
of what was seen in SALT2 and suggest that neither implication about the dust
in the highly-extinguished objects should be taken seriously. 

If we assume that highly reddened SN Ia are intrinsically similar to their
less-reddened counterparts with similar \dd~then this means that \av~(with
\rvth) is overestimated, giving rise to smaller MLCS31 distance moduli than
those from the best-fit cosmology.  This seems more likely than assuming that
the highly-reddened SN Ia are intrinsically brighter and that the MLCS31 \av~is
accurate.  In contrast, the MLCS17 Hubble residuals show little trend with
respect to \av, especially when cut at $A_V=0.5$.  

Table \ref{table_w} shows the values of $1+w$ after making the cuts on $c$ and
\av.  SALT, SALT2 and MLCS31 all have greater values of $1+w$ for the OLD+CfA3
samples while MLCS17 has a smaller value of $1+w$, bringing them into better
agreement with each other.  The standard deviation of $1+w$ from the four
fitters drops from 0.051 to 0.025 for the OLD+CfA3 samples, from 0.051 to 0.046
for the CfA3 sample and from 0.075 to 0.059 for the OLD sample, and the average
difference in $1+w$ between the OLD and the CfA3 samples drops from -0.068 to
-0.060, suggesting an improvement in our calculation of $1+w$.  While the four
fitters agree better now for a given sample, and the mean value of $1+w$ from
each sample agrees slightly better, the CfA3 and OLD samples do not improve in
agreement for any single fitter. 

Until the intrinsic color variation of SN Ia and the host
dust properties are better understood, and distance fitters are better able
to incorporate this information, it is safer to cut out the overly 
blue and red objects in SALT and SALT2 and the moderate to high extinction
objects in MLCS2k2.  The improved agreement in $1+w$ between the four fitters
and the relative constancy in the uncertainties on $1+w$ suggest that the 
cuts are worth making.  The cuts on $c$ and \av~will be combined
with the cut on \dd~to form our best cuts.

\subsection{No \emph{U} band}

\citet{jha06} present the first large body of nearby \emph{U}-band light
curves.  H09, as part of the CfA3 sample, present an even larger body.
\citet{ellis08} compile a large body of rest-frame UV spectra of SN Ia at
$z\approx0.5$ and \citet{foley08} do something similar.  The \emph{U} band is
the most poorly-constrained optical band for SN Ia.  It is also poorly fit by
SALT2, where the model spectra in the rest-frame \emph{U} region are highly
influenced by the high-$z$ \emph{U}-band photometric calibration, which might
be discrepant from that at low $z$.  It is worth briefly examining the effects
of excluding the rest-frame \emph{U} band from the light curve and cosmology
fits.  We refit the nearby SN Ia samples with MLCS31 and SALT2, excluding the
\emph{U} photometry.  Many of the high-redshift SN Ia have strong contributions
from rest-frame \emph{U}.  To check if this matters we fit the cosmology for
the no-\emph{U} nearby samples plus the ESSENCE sample, since ESSENCE is mostly
composed of rest-frame \emph{B} and \emph{V} data.  There is virtually no
difference between the values of $1+w$ found using just the ESSENCE sample as
the High-z sample and using the complete High-z sample (the uncertainty is
larger though when using only ESSENCE as the High-z sample, due to fewer objects) so we choose to focus
on the cosmology fits of the no-\emph{U} nearby samples combined with the
complete High-z samples.  All values of $1+w$ increase by about 0.03 for the
three MLCS31 samples.  The OLD SALT2 sample has its value of $1+w$ increase by
about 0.02, the CfA3 SALT2 $1+w$ is virtually unchanged and the OLD+CfA3 SALT2
$1+w$ increases by an insignificant 0.005.  SALT2 was designed partially to
use the rest-frame \emph{U}-band data.  The MLCS17 CfA3 value of $1+w$ is
virtually unchanged as well.  See the exact values in Table \ref{table_w}.  The
change in the MLCS31 values may be due to the importance that the \emph{U} band
plays in determining the color excess (and subsequent extinction, \av),
especially for SN Ia in dusty host environments.  Because of the negligble
effect on MLCS17 and SALT2 and the 0.03-level effect on MLCS31 we do not
explore this further beyond combining the no-\emph{U} cuts with cuts on \av~and
\dd~in \S3.10.

\subsection{Combining Multiple Cuts}

We briefly explore making a cut at \av$=0.5$ on the no-\emph{U} MLCS31 sample
(see Table \ref{table_w}).  The results are quite similar to the cut at
\av$=0.5$ on the MLCS31 sample (with \emph{U}).  This is consistent with the
possibility that the \emph{U} band has a strong impact on the extinction estimate
for the more highly-reddened objects.  If we remove these objects with the cut
at \av$=0.5$ then it is not so important whether the \emph{U} band is kept or not
in the nearby samples.  The MLCS17 CfA3 sample, which we expect to be not as
sensitive to reddening effects, has its value of $1+w$ decrease by 0.009,
compared to the no-\emph{U} case. 

We also cut out SN Ia with \dd$\geq0.7$ and \av$>0.5$ for the 
no-\emph{U}-band MLCS31 samples (see Table \ref{table_w}).  Just as we saw an
increase in $1+w$ by about 0.03 to 0.035 in making only a \dd$>0.7$ cut on
the MLCS31 sample, we also see a similar increase in $1+w$ compared to the 
no-\emph{U} plus \av$\leq0.5$ sample from the previous paragraph while the
MLCS17 CfA3 sample increases by 0.05 relative to the value in the previous
paragraph.  This is to be expected since the \dd$>0.7$ cut removes more 
negative than positive nearby residuals while leaving the High-z sample
virtually unchanged.  

\subsection{Best Cuts Adopted}

Finally, for MLCS17 and MLCS31, we combine the cuts of the two areas of
most-obvious difficulties, namely, cutting above $A_V=0.5$ and above
$\Delta=0.7$.  We adopt these cuts (on top of the minimal cuts) as our 
best cuts for the two MLCS2k2 fitters.

Both of these cuts remove more negative-residual objects than positive and
this causes $1+w$ to rise to $0.074^{+0.064}_{-0.065}$ for MLCS31 and
$0.118^{+0.063}_{-0.065}$ for MLCS17, both with the OLD+CfA3 sample (see Table
\ref{table_w}).  The statistical uncertainties on $1+w$ are lower than for
MLCS31 and MLCS17 with the minimal cuts, in spite of fewer objects.

For SALT and SALT2, the best cut is to include objects with $-0.1 \leq c
\leq 0.2$.  The OLD+CfA3 values of $1+w$ for SALT and SALT2 are
$0.026^{+0.069}_{-0.071}$ and $0.026^{+0.078}_{-0.082}$, respectively,
in excellent agreement with each other.

Amongst the four fitters, SALT and MLCS17 give the lowest differences in
weighted-mean residuals and have the best agreement between the CfA3 and OLD
values of $1+w$.  With the best cuts, MLCS17 and MLCS31 have the smallest
statistical uncertainties on $1+w$.  SALT2 and MLCS31 produce the highest
difference in weighted-mean residuals between the OLD and the CfA3 samples and
they produce the greatest differences between these two samples' values for
$1+w$.  

In all three samples, the MLCS17 values of \ola~($1+w=0$) are the lowest and
the MLCS17 values of $1+w$ are the largest.  With its lower value of \rv,
MLCS17 attributes less host extinction to each SN Ia.  This produces a larger
distance compared to MLCS31 for the more-reddened nearby SN Ia, and from
that, a larger value of $1+w$.  Although there are some highly-reddened SN Ia in
the High-z sample, most of them are below $z\approx0.35$ and do not affect
$1+w$ as much as if they were at $z\approx0.5$.  The High-z sample is
relatively unreddened in the redshift region that has most influence on $1+w$.

It is also important to look at the standard deviations of $1+w$ for the four
fitters in the OLD and the CfA3 samples (Table \ref{table_w}).  They are
roughly the same:  0.066 and 0.064, respectively.  The fitters were all trained
on many objects from the OLD sample and none from the CfA3 sample, yet the CfA3
sample gives a similar standard deviation in $1+w$.  This suggests that the
CfA3 sample is internally consistent.  Also, as seen in Table
\ref{table_meanres} and discussed in \S3.6, the scatter in the residuals of the
OLD and the CfA3 samples is similar.

Repeating the simplistic $\sqrt{N}$ analysis of \S3.1, of attributing half the
statistical leverage to the faraway sample and half to the nearby sample, we
expect a decrease in the  statistical uncertainty of $1+w$ to be roughly 1.3 in
the four fitters.  We find a decrease of roughly 1.2-1.3.  It is likely that we
are beginning to run up against the effects of systematic differences
in different light curve samples and in different
light-curve/distance fitting assumptions and methods.   

The fact that our best cuts on MLCS31 and MLCS17 give values of $1+w$ about
0.05 and 0.09 higher than the SALT/2 values is somewhat troubling.  In the case
of MLCS2k2, although these seem like the correct regions to cut out, it may
introduce a bias to the remaining SN, since it was trained on the larger range
of values.  We expect that if the CfA3 sample is added to the training sets,
the 1991bg-like SN Ia are treated separately (in MLCS2k2), improved treatement
of the \emph{U} band is achieved, and the correct reddening treatment is
discovered and used, that the retrained fitters (SALT/2 and MLCS2k2) will then
produce values of $1+w$ that agree well with each other.

\subsection{Mean Hubble Residuals of Six Different Nearby Samples}

We have already looked at how the weighted mean of the Hubble residuals of the
OLD and CfA3 nearby samples agrees in the four fitters.  In MLCS17 and SALT
we found better than $1\sigma$ agreement while in MLCS31 and SALT2 it was
slightly larger than $1\sigma$.  We now want to compare the main sources of
nearby data more closely to see if subsamples of the OLD and CfA3 samples are
causing the slight discrepancy.  We break up the OLD sample into the
Calan-Tololo (CT) \citep{hamuy96}, CfA1 \citep{riess99}, CfA2
\citep{jha06} and Other subsamples 
(and K08 in the case of SALT where the eight SN Ia that
K08 present are included in the Union and Constitution samples).  We also break
up CfA3 into CfA3(4Sh) and CfA3(Kep) for the two cameras (4Shooter and
Keplercam, see H09 for details) that observed the vast majority of the CfA3
sample.  

For SALT and SALT2, since there are no obvious trends in residuals versus
stretch or color at low redshift (it was the trend versus color at high
redshift that necessitated the cut on $c$), and we are only looking at nearby
objects for this comparison, we use the minimal-cuts nearby sample.  For
MLCS17 and MLCS31, we look at both the minimal-cuts sample and the best-cuts
(\av~and \dd) sample.  

For each subsample we calculate the weighted mean of the Hubble residuals and
the uncertainty.  In the best-cuts cases of MLCS17 and MLCS31, we are still
using the residuals from the minimal-cuts cosmological fits and so we subtract
off the weighted mean of the entire best-cuts sample and recalculate the
difference between the OLD and the CfA3 mean residuals.  The standard deviation
of the subsamples' weighted means is calculated both with and without the
most-discrepant OLD subsample.  We also remove the most discrepant OLD
subsample and recalculate the modified OLD*$-$CfA3 differences.  The CfA3(Kep)
sample has the largest influence on the mean of the whole sample due to its
size.  The results are displayed in Table \ref{table_res_6samples}.  We
reiterate that the CfA3 sample played no role in training the light-curve
fitters and so the good agreement between the two CfA3 subsamples and most of
the others is a good sign that the training samples are representative of the
nearby SN Ia population and working well for the most part. 

In both MLCS17 and MLCS31, for both the minimal and the best cuts, the CfA1
sample is the most discrepant.  In MLCS17, the other five subsamples all
agree to better than 1$\sigma$.  In MLCS31, the best cuts actually make
the two CfA3 subsamples disagree more but this may be due to different
distributions in \av~where the slope in residuals versus \av~of $-0.34$
is present.  

SALT2 has the largest discrepancies between all the subsamples, with CT being
the most discrepant from the mean of the whole sample.  SALT has seven 
subsamples and Other is the most discrepant but it only has five objects.

Some of the differences between subsamples is likely due to systematic offsets
in photometry but a good portion is also due to the selection effects that
different cuts have on which objects remain.  The agreement between OLD and
CfA3 mean residuals improves with the best cuts for MLCS17 and stays about the
same for MLCS31.  When the most-discrepant subsample is removed from the OLD
sample, the resulting OLD*$-$CfA3 differences are all at the $1\sigma$ level
or less.  As a crude way of measuring the scatter in the subsamples in the
different fitters, we ignore the most discrepant subsample and take the
standard deviation (StdDev*) of the weighted means.  MLCS17 has 0.02, MLCS31
and SALT have 0.03 and SALT2 has 0.06.  If we take 0.03 mag as representative,
then that is about how good any subsample's weighted mean of its Hubble
residuals can be expected to agree with another's.  Hopefully, the inclusion
of several subsamples reduces the offsets from the true value of the
individual subsamples.  

As we have seen in this work, in comparing the OLD and the CfA3 samples, the
difference in $1+w$ is about twice as large as the difference in the weighted
mean of the residuals.  Going forward, to not only measure static $1+w$ better
but to measure for time variation as well, SN Ia photometry needs to be
improved by lowering systematic errors in calibration and understanding
passbands better.  Improved signal-to-noise measurements will also be helpful.
The light curve and distance fitting needs to be more robust, including how
it deals with host-galaxy reddening, so that the distances do not have
systematic errors.

\section{Looking for Physical Systematics}

\subsection{Hubble Bubble}

We address the issue of whether we live inside a Hubble bubble, possibly due
to a local void, where space is expanding faster (or slower) within a given
region than outside.  \citet{zehavi} find an apparent bubble.  J07 confirms
its existence, finding the inner expansion rate to be $6.5\pm1.8\%$ higher
than the outer rate at the partition redshift of $z\approx0.025$.
\citet{conley07} shows that MLCS2k2, with \rvth, and SALT, SALT2, and SiFTO,
with $\beta=4.1$, all show evidence of a Hubble Bubble but that SALT, SALT2
and SiFTO do not produce significant bubbles when the best-fit value of
$\beta\approx2$ is used.

\begin{figure}
\scalebox{0.8}[0.80]{
\plotone{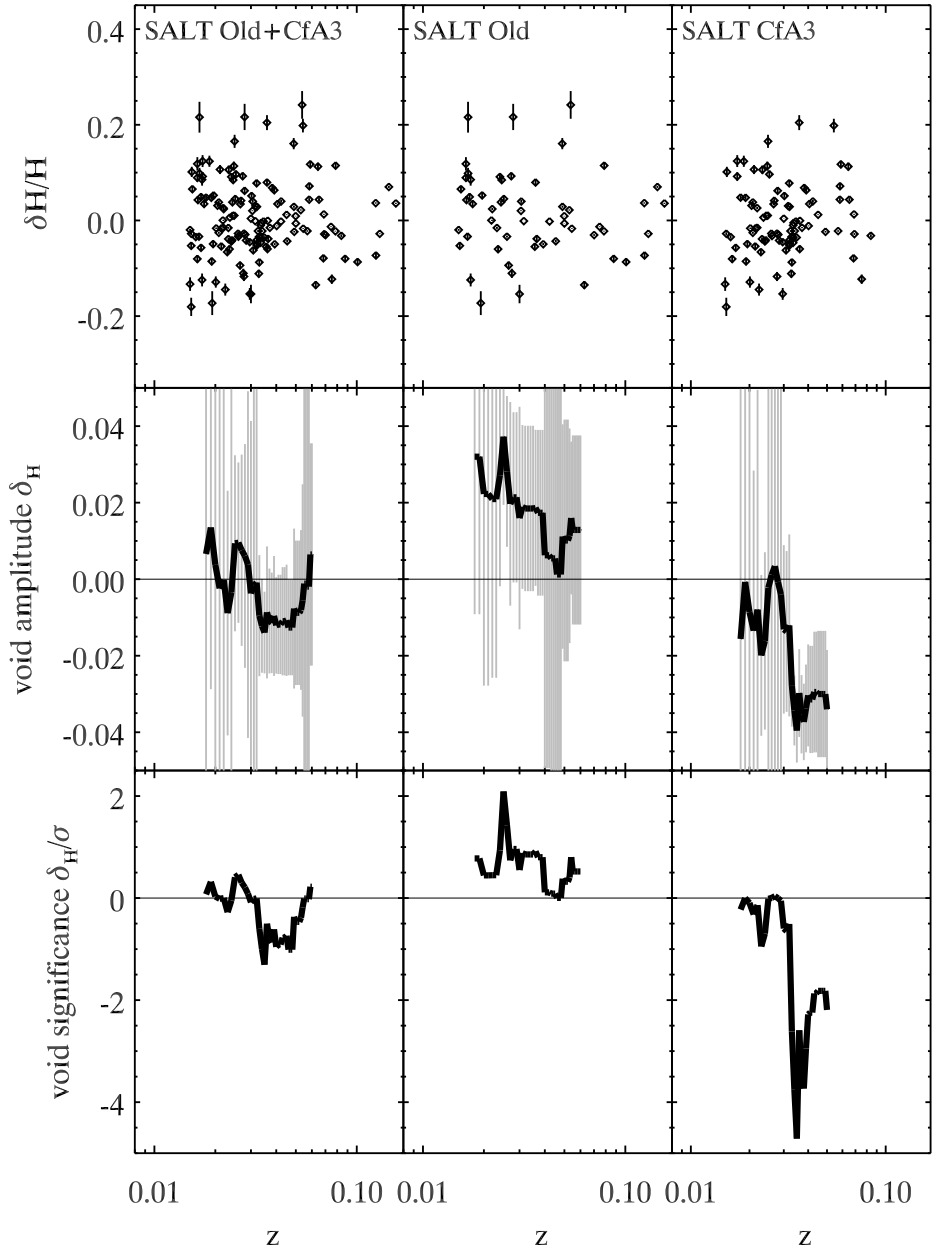}
}
\caption{
Hubble bubble for SALT.  A negative but insignificant Hubble 
bubble is present in the OLD+CfA3 sample.
}
\label{fig_bubsalt}
\end{figure}

We follow the same approach as J07 and \citet{zehavi}, converting our SN Ia
light-curve-fitter-derived distance modulus, $\mu_{LC}$, into a luminosity
distance, $d_{LC}$ and multiplying by $H_o$ to get the light-curve-based
expansion velocity.  We also take the luminosity distance predicted by the best
fit cosmology, $d_{cosm}(z)$ and multiply by $H_o$ to get the
best-fit-cosmology velocity.  We subtract the two to get the host galaxy
peculiar velocity, $u=H_od_{cosm}(z)-H_od_{lc}$.  The deviation from the Hubble
law is equivalent to the fractional peculiar velocity and can be expressed as
$\delta H/H = u/(H_o d_{lc})$.  This is displayed for the SALT results in the
top panels of Figure \ref{fig_bubsalt}.  We partition the nearby samples in
increments of $\Delta z=0.001$, starting at $z=0.015$ for MLCS31, MLCS17, and
SALT2, and $z=0.018$ for SALT and ending at $z=0.06$ for all fitters for the
OLD and OLD+CfA3 samples and at $z=0.05$ for the CfA3 sample.  These ranges on
the redshift partition ensure adequate numbers of SN Ia both above and below
each partition.  For each partition, we calculate the best-fit Hubble constants
$H_{inner}$ and $H_{outer}$ and their uncertainties as in \citet{zehavi}.  We
calculate the void amplitude, $\delta_H=(H_{inner}-H_{outer})/H_{outer}$, and
its uncertainty.  These are displayed in the middle panels of Figure
\ref{fig_bubsalt}.  Finally, in the bottom panels, we show the void
significance, $\delta_H/\sigma$.

For the minimal-cuts samples, we look for the largest absolute value of the
void significance to identify the most likely redshift for a Hubble Bubble.
Table \ref{table_hubble_bubble} shows the results from the various fitters and
samples.  The minimal-cuts samples are used for all four fitters as well as the
$A_V\leq0.5$ samples for MLCS31 and MLCS17.  We find a Hubble bubble at
$z\sim0.025$ in the OLD samples with both MLCS31 and SALT, with significances
of 3.0 and 2.0, respectively, while MLCS17 and SALT2 do not yield significant
Hubble bubbles in the OLD sample.  The SALT and MLCS17 CfA3 samples show
negative Hubble Bubbles at $z\approx0.035$ with absolute-value void
significances greater than 3.0, and the SALT2 CfA3 sample shows one there with
signficance of 1.6.  The MLCS31 CfA3 sample has its most significant void at
$z=0.04$.  The difference in void location between the CfA3 and OLD samples
illustrates the sample-dependent nature of Hubble bubble calculations.  

Adding the samples should amplify any real voids.  However, we see the
opposite.  Each of the most-significant voids in the OLD and CfA3 samples
becomes weaker and, in some cases, replaced by a different
most-significant-void redshift.  The SALT and SALT2 voids have a significance
on the order of 1.  MLCS17 has a $2.75\sigma$ void at $z=0.034$ with amplitude
-0.020 while MLCS31 has a $5.56\sigma$ void at $z=0.028$ with amplitude 0.029. 

By adding the CfA3 sample we have shown that the MLCS31 Hubble bubble is
smaller and at a slightly different redshift than was seen in Jha and Zehavi.
The negative bubble in the case of MLCS17 shows that the existence of a Hubble
bubble in any data set may be influenced by the reddening law, \rv, used.
Whereas in Conley they show that by using a larger $\beta$ they can find
positive Hubble bubbles in SALT, SALT2, and SiFTO, we show that by using a
smaller value of \rv~in MLCS2k2 that the positive Hubble bubble is removed and
even becomes negative.

As was discussed above, if the MLCS31 values of \av~are indeed being
overestimated then the light-curve-derived distance modulus will be
underestimated.  The distance modulus calculated from the redshift and the
best-fit cosmology will be larger than the light-curve-derived one, giving rise
to a positive peculiar velocity.  This seems to be why MLCS31 is finding
a Hubble bubble. 

\begin{figure}
\scalebox{0.8}[0.80]{
\plotone{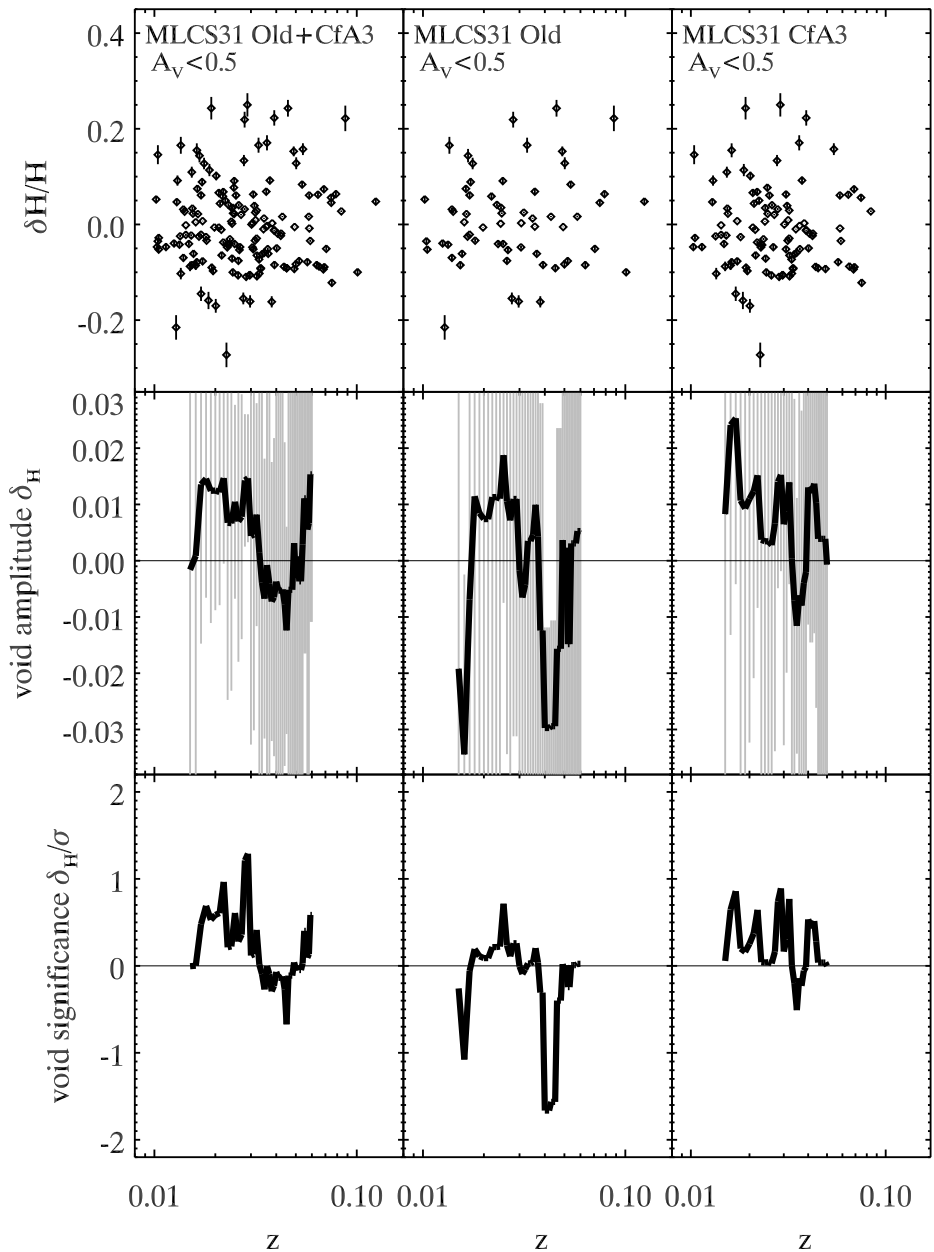}
}
\caption{
After a cut at $A_V=0.5$, the positive Hubble bubble for MLCS31 is now
insignificant.
}
\label{fig_bub31avcut}
\end{figure}

We return to the \av-versus-redshift plot for the three MLCS31 samples in
Figure \ref{fig_av_z}.  It is interesting to note that the \av$>0.5$ SN Ia
are at $z<0.035$ for the CfA3 sample and at $z<0.025$ for the OLD sample, very
close to the values at which the most significant Hubble bubbles are seen.  We
redo the MLCS31 and MLCS17 Hubble bubble calculations, removing all objects
with \av$>0.5$.  We show the results for MLCS31 in Figure
\ref{fig_bub31avcut}.  The most-significant void amplitudes of the OLD+CfA3
samples are now just slightly greater than one, similar to those of SALT and
SALT2.  We conclude that any Hubble bubble with redshift partition in the range
$0.015\leq z \leq0.060$ is likely to be less than $2\%$ and could be positive
or negative.  Our analysis is consistent with no Hubble bubble.  This
has important implication for SN cosmology since nearby SN Ia down to
$z\approx0.01$ can be safely included.  
We make an estimate of the systematic
uncertainty due to a potential Hubble bubble in \S4.3.

\subsection{Host Galaxy Morphology and Projected Galactocentric Distance}

\begin{figure}
\scalebox{0.8}[0.80]{
\plotone{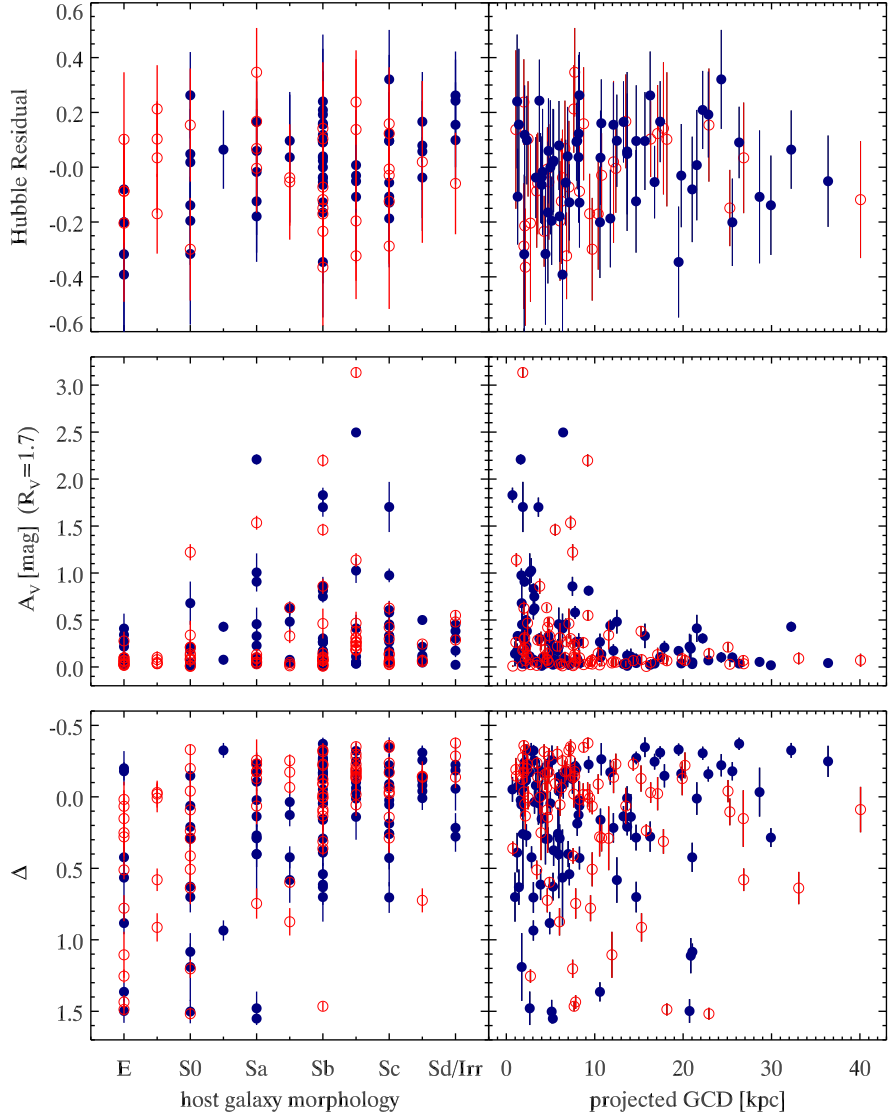}
}
\caption{
MLCS17 properties versus morphology and PGCD.  
Red empty circles are OLD and blue filled circles are CfA3.  
The negative 
residual SN Ia stand out in the E hosts.  The same qualitative pattern
is seen in MLCS17 as in MLCS31.  Low
host reddening for E/S0 and Scd/Sd/Irr SN Ia is seen.
Highly reddened SN Ia do not occur beyond $\rm\sim10kpc$.
Objects with $0.7\leq\Delta\leq1.2$ and \av$>0.5$ have been removed from the
Hubble residuals panels, so as 
to not influence possible trends.  Two elliptical-host SN Ia have 
$\Delta\approx -0.2$.  
}
\label{fig_morph_mlcs17}
\end{figure}

In Figures \ref{fig_morph_salt2} and \ref{fig_morph_mlcs17}, we plot the Hubble
residuals, color or host-reddening, and light-curve shape parameter versus the
host galaxy morphology and projected galactocentric distance (PGCD) for SALT2
and MLCS17, using the best cuts for the Hubble residuals panels.  The SALT2
plot is very similar to SALT and the MLCS17 plot is very similar to MLCS31 so
we only show these two.  It should be reiterated that our samples are not
complete, especially for the nearby samples where field galaxies tend to be
searched less than clusters.  We are only reporting the SN Ia that were
discovered, have good light curves and have morphological typing.
Morphological type is drawn from NED\footnote{http://nedwww.ipac.caltech.edu/}.
About $70-75\%$ of the nearby SN Ia host galaxies have morphological typing
listed in NED.  For the color/reddening and light-curve-shape plots we want to
focus on the range of nearby SN Ia properties and not just those in the Hubble
flow so we do not impose any cut on redshift.  We are not trying to ensure
high-precision distances in these non-residuals panels so we also relax the
time of first observation to $+20$ days and allow for a slightly higher
\xx~(roughly $30\%$ higher than in the minimal cuts).  These are still
stringent enough to ensure that the light-curve parameters are useful.  For
example, J07 use $+20$ days as their cut on $t_{1st}$.  We remind the reader
that PGCD is a lower limit on the actual distance from the host center.
Several interesting properties are seen.

\subsubsection{Light-Curve Shape}

First, we focus on the light-curve shape plots in the lower panels of
Figures \ref{fig_morph_salt2} and \ref{fig_morph_mlcs17}.

There are several negative-\dd~(broad light-curve shape) SN Ia in elliptical
and S0 hosts.  Due to the fact that stretch ($s$ and $x_1$) and \dd~are not
linearly correlated, there is a difference in their respective distributions
versus morphology.  The largest-stretch SN Ia are clearly found in the Sb-Sc
hosts and not in the Sd/Irr or E/S0 hosts while the distribution of \dd~is much
flatter across host morphology.  Close examination does show a gentle maximum
of negative-\dd~in the Sb-Sc galaxies but with negative-\dd~objects in E/S0 and
Sd/Irr hosts as well.  An example of this is the two most-negative-\dd~ SN Ia
in the E hosts: SN 2002dj and 2008bf.  SN 2002dj has \dd$=-0.20$, $s=0.93$ (but
with a poorer SALT fit) and $x_1=-0.114$.  SN 2008bf has \dd$=-0.18$, $s=1.04$
and $x_1=0.17$, with the highest $s$ and $x_1$ for SN Ia within E hosts.  Both
of these are at the high end of the range of $s$ and $x_1$ within E hosts but
they do not seem particularly high when compared to the highest values seen in
the Sb hosts.  Nonetheless, they do extend the range of light curve shapes seen
in E hosts.

Additionally, the CfA3 sample adds many slow decliners.  This is very important
since high redshift searches are often biased towards finding these broad and
bright SN Ia.  Many of these are found at high PGCD, something that was not
seen often in previous nearby samples.  There is an absence of very-fast
decliners ($x_1 < -3$; \dd$>0.8$) in Sbc-Irr galaxies but the CfA3 sample
does extend the range in these galaxy types to include faster decliners than
typically seen in the OLD sample.  As was seen in the OLD sample, most of the
CfA3 high-\dd~(low-luminosity) SN Ia are found closer to their host centers.

\subsubsection{Reddening and Extinction}

\begin{figure}
\scalebox{0.8}[0.80]{
\plotone{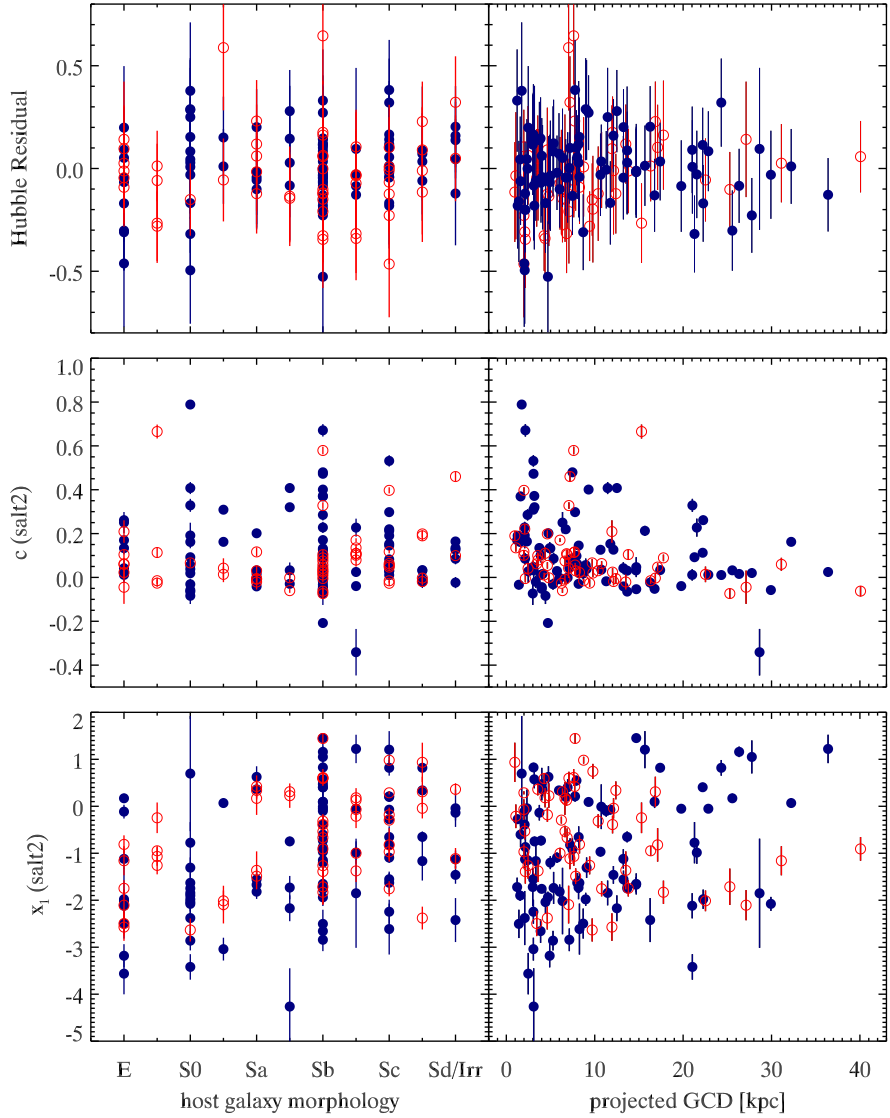}
}
\caption{
The top panels shows the SALT2 Hubble residuals versus host galaxy morphology
and projected galactocentric distance.  The middle panels plot SALT2 $c$ and
the bottom panels plot stretch, $x_1$, versus morphology and PGCD. 
The Hubble residuals are more negative on average in the E/S0 hosts
than in the Scd/Sd/Irr hosts by roughly $2\sigma$.  
The Scd/Sd/Irr SN Ia show the smallest dispersion, although we caution
that this may be due to small numbers of objects.
The CfA3 sample adds many slow decliners, of great 
importance in matching the High-z sample as much as possible. 
Red empty circles are OLD and blue filled circles are CfA3.  
}
\label{fig_morph_salt2}
\end{figure}

Looking at the plots of \av~and $c$ in the middle panels of Figures
\ref{fig_morph_salt2} and
\ref{fig_morph_mlcs17} shows the striking feature that the estimated extinction
or color is roughly as low in the Scd/Sd/Irr galaxies as it is in the E/S0
hosts while the higher extinction and colors are seen in the middle-type
spirals.  The MLCS17 plot shows this best but the trend is there with all four
fitters.  With a larger sample it is also interesting to see that most of the
reddened SN Ia occur within 10-12 kpc of the host center.  

\begin{figure}
\scalebox{0.8}[0.80]{
\plotone{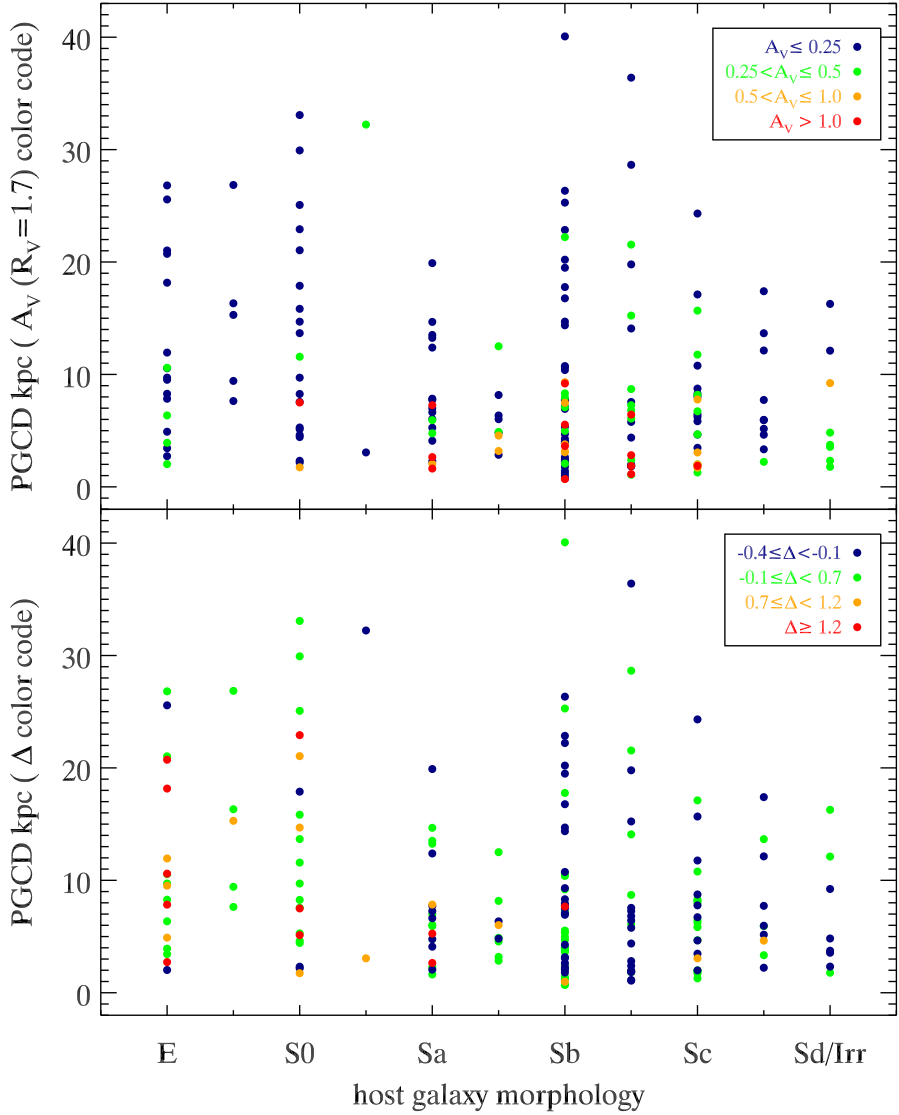}
}
\caption{
MLCS17 PGCD versus host galaxy morphology with dust extinction color coded 
in the top panel and \dd~in the bottom panel.  
Top panel:  blue ($A_V\leq 0.25$), green ($0.25<A_V\leq 0.5$), orange 
($0.5<A_V\leq 1$), red ($A_V>1$).  Bottom panel:  blue ($-0.4\leq\Delta<-0.1$),
green ($-0.1\leq \Delta < 0.7$), orange ($0.7\leq \Delta < 1.2$), red 
($\Delta~\geq 1.2$).  Light extinction in the
inner regions of Scd/Sd/Irr hosts.  High extinction is more prevalent in the 
Sa/Sb/Sc hosts, extending to highest PGCD in Sb hosts.  Extinction is low
in E hosts with ``green" extinction SN Ia in the lower range of PGCD.  
}
\label{fig_pgcd_morph}
\end{figure}

To get a three-dimensional perspective we plot PGCD versus morphology and
color-code \av~and \dd~for MLCS17 in Figure \ref{fig_pgcd_morph}.  In the top
panel, blue means low reddening and red means high reddening and we see that
the SN Ia found in the Scd and Sd/Irr hosts are lowly-reddened (mostly blue and
green points) and only extend out to $\rm\sim18kpc$.  This suggests that although
Scd and Sd/Irr galaxies have a lot of gas that can form new stars they do not
have a lot of dust that reddens SN Ia.  As we move into the Sc and Sb spirals we
see reddening start to occur near the galaxy centers and move outwards, perhaps
suggesting a change in the amount or type of dust in the inner regions.  The
projected distance at which SN Ia are found also increases.  In the lower
panel, blue means bright and red means faint and there does not seem to be much
of a trend in the distribution of \dd~along PGCD in the Sb-Irr galaxies
but it is very interesting to start to see faint, fast decliners (red) in the
central regions of the Sa galaxies, where the star populations are older.  

Then, as we move into the S0 and E galaxies in the top panel of Figure 
\ref{fig_pgcd_morph}, we again see a change in the SN Ia
reddening.  It is not completely gone, as evidenced by the green symbols!  But
it is once again at levels seen in the Scd/Sd/Irr spirals and irregulars.
The distribution of \dd~throughout the S0 and E hosts in the lower panel
seems fairly uniform.
Especially of note is the fast decliners occuring at both small and large PGCD,
in agreement with a relatively uniform and old stellar population 
throughout the E/S0 galaxies.  

We have just narrated the change of both supernova and dust properties as we
move from Sd/Irr to E/S0 galaxies and their presumed star populations.
\citet{scannapieco05}, \citet{mannucci06} and \citet{sullivan06} find two
populations of SN Ia progenitors, one ``prompt", that follows the
star-formation rate, and one ``tardy", that follows the mass distribution.  The
fast-declining SN Ia occur in regions of older star populations while the
slow-declining SN Ia predominantly occur in regions of intermediate star
populations and star formation (Sa-Sc galaxies for SALT and Sb-Sc for SALT2).
This is slightly different than the mantra that the brightest, slowest SN Ia
are found in the youngest galaxies.  Bright SN Ia (negative \dd) are certainly
found in the Sd/Irr hosts but the slowest decliners (largest stretch) are not.
Meanwhile, the ``normal" SN Ia occur throughout all morphologies and PGCD.
\citet{sullivan06} show that at high redshift the high-stretch SN Ia are found
in star-forming galaxies while low-stretch SN Ia are found more predominantly
in passive hosts and a mix is found in weakly star-forming hosts.
\citet{gallagher08} find a similar trend, where high star-formation rate
galaxies have a flat distribution of of bright and normal SN Ia while moderate
star-formation rate galaxies host a wide range of SN Ia and the very-low
star-formation rate galaxies do not host the brightest SN Ia but do host faint
SN Ia.
There is a definite uptrend in stretch from Sd/Irr to Sb (Figure 
\ref{fig_morph_salt2}).  

It is
tempting to speculate that the prompt component is related to the normal
decliners of the type found in the Sd/Irr hosts while the tardy component is
related to the fast decliners.  Normal decliners in older hosts may be slightly
different beasts from those in young hosts.  Whether the slowest decliners can
be assigned to either prompt or tardy, or split between the two somehow is
unclear.  Perhaps there is a third population that is neither prompt nor tardy
and gives rise to the slowest decliners.  Or, perhaps there is more of a 
continuum of progenitors.

\subsubsection{Hubble Residuals and Three Morphology Bins}

We now focus on the Hubble Residual plots in the upper panels of Figures
\ref{fig_morph_salt2} and \ref{fig_morph_mlcs17}.  The SN Ia in the problematic
regions of $A_V>0.5$ and $0.7\leq \Delta \leq1.2$ have been excluded from these
plots and the calculations that follow.  We notice that the residuals of the SN
Ia in ellipticals have more negative residuals with all four fitting methods
while the SN Ia in Scd and Sd/Irr hosts have more positive residuals.  We
divide the SN Ia in these plots into three groups: E-S0, S0a-Sc, and
Scd/Sd/Irr.  We calculate the standard deviation and the weighted mean of the
residuals in each subsample, for each fitter.  The results are shown in Table
\ref{table_res_morph}.  For each fitter's results, we find that SN Ia in E-S0
hosts have negative mean residuals beyond the $1\sigma$ level and that the SN
Ia in Scd/Sd/Irr hosts have positive mean residuals beyond the $1\sigma$ level,
except for the SALT Scd/Sd/Irr subsample that is nearly at the $1\sigma$ level.
The negative residuals in the E-S0 hosts is somewhat similar to the $1\sigma$
negative mean residual found by J07 in E hosts.  We find the average residual
in the Scd/Sd/Irr hosts is $0.091\pm0.057$ mag while in the E/S0 hosts it is
$-0.054\pm0.041$ mag, leading to the average residual in the Scd/Sd/Irr hosts
being $0.144\pm0.070$ mag larger than the average E/S0 residual, a $2\sigma$
measurement.  This is not an effect of host reddening since neither the E/S0
nor the Scd/Sd/Irr hosts produce highly-reddened SN Ia.  It is not the effect
of the negative MLCS2k2 residuals in the range $0.7\leq \Delta \leq 1.2$ since
those were removed.  We examine some of the light curve fits of the most severe
outliers for irregularities but find no good reason to remove them.  The
residuals in the intermediate group of S0a-Sc spirals are mostly consistent
with zero mean residuals.  The fact that each fitter finds the same trend is
reassuring.  These results suggest that SN Ia are different, after light-curve
and color/reddening correction, in the E/S0 and Scd/Sd/Irr galaxies.
It may be fruitful to train the various light curve fitters separately on the
three subsets of host galaxy types.  

\begin{figure}
\scalebox{0.8}[0.80]{
\plotone{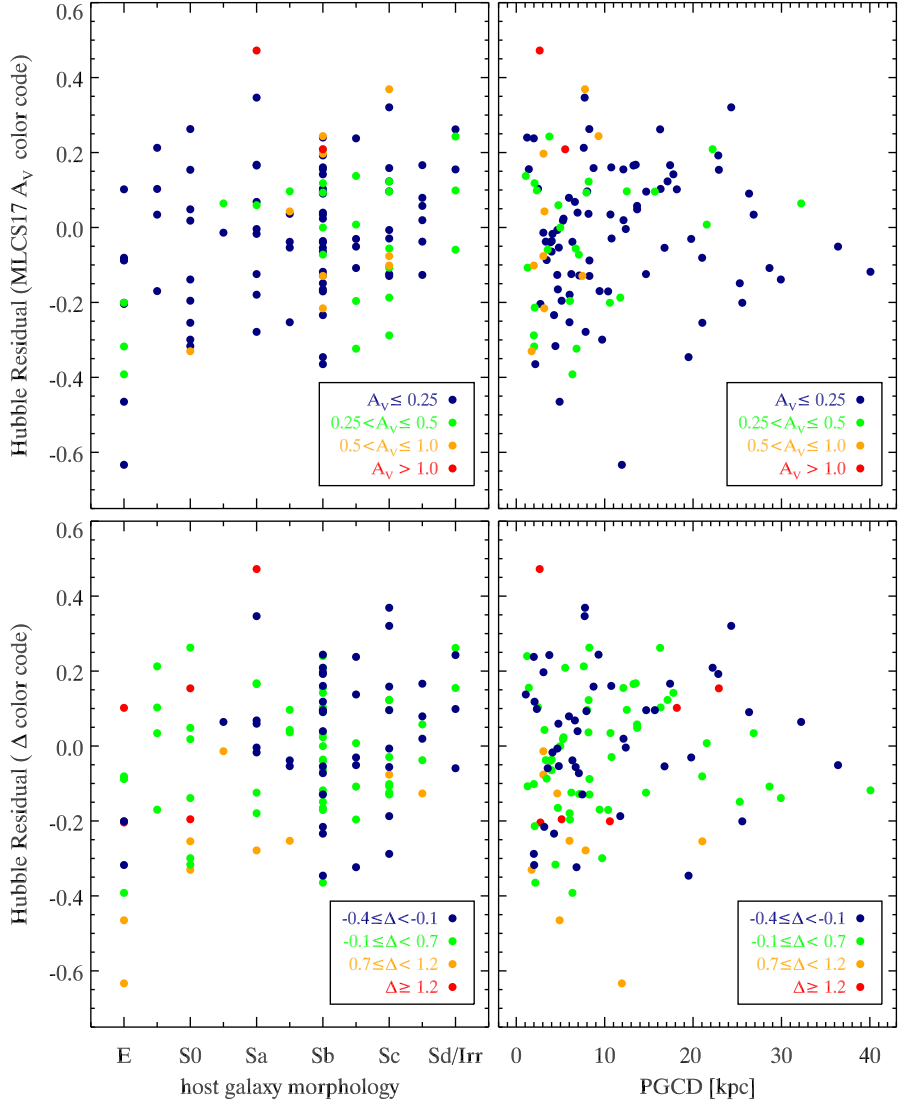}
}
\caption{
MLCS17 Hubble Residual versus morphology and PGCD.  Same color
coding as \ref{fig_pgcd_morph}.  Notice especially the bottom-left
panel where the most-negative residual SN Ia in the E and S0 hosts
are fast (orange), but not very-fast, decliners.  These objects
($0.7\leq \Delta < 1.2$) have been included in this plot to show that
they are found across most morphological types and all have negative
residuals.  These orange points are excluded for our mean Hubble residual
calculations of the three binned morphologies (E-S0, S0a-Sc, and Scd-Irr)
and there is still a $\sim 2\sigma$ difference between the E-S0 and Scd-Irr
mean Hubble residuals.
}
\label{fig_res_morph17_colorcode}
\end{figure}

In Figure \ref{fig_res_morph17_colorcode}, in the bottom left panel, we plot
the MLCS17 Hubble residuals versus morphology and color code \dd~so that the
very faint SN Ia are red ($\Delta>1.2$), faint ones are orange ($0.7\leq \Delta
< 1.2$), normal and moderately faint ones are green ($-0.1\leq \Delta < 0.7$),
and bright ones are blue ($\Delta<-0.1$).  The most-negative residual SN Ia in
the E and S0 hosts are faint (orange), but not very faint (red).  We reiterate
that the $A_V>0.5$ and $0.7\leq \Delta \leq1.2$ objects were not included in
calculating the mean residuals and scatter in the three morphology bins but
we show them in \ref{fig_res_morph17_colorcode} so that their distribution
can be seen.



We also find that the Scd/Sd/Irr subsample has the lowest average scatter
across the four fitters (0.100 mag) while the E-S0 subsample scatter (0.182
mag) is slightly larger than that of the S0a-Sc subsample (0.166 mag) (see
Table \ref{table_res_morph}).  At face value this seems to disagree with the
findings by \citet{sullivan03}, where the three partitions of SN Ia we have
used have $\sigma=0.16$, $\sigma=0.20$, and $\sigma=0.27$ mag, respectively.
However, they use the stretch method that attempts no color or reddening
correction and many of their objects are at high redshift.  We find that two of
the arguments that are used to promote the search for SN Ia in elliptical
galaxies at high redshift would also apply to the Scd/Sd/Irr hosts: namely,
that the extinction is low and the intrinsic dispersion is small.  Granted, our
study only looks at the low redshift sample so we cannot say if the high
redshift SN Ia in Scd/Sd/Irr galaxies have this same low reddening and
dispersion but it is worth keeping in mind in designing future SN Ia surveys.
Our finding of high dispersion in the early-type hosts, if not just an artifact
of the four fitters we use, would weaken but not kill the argument for
searching for high-redshift SN Ia in early-type hosts.  The reddening would
still be low but the scatter in the Hubble residuals may not be as low as
expected.

Finally, we divide the nearby sample into inner and outer subsamples at $\rm
PGCD=12kpc$, where the distribution of highly reddened objects ends.  We find
that the scatter is about 0.05 mag less in the outer subsample with three of
the four fitters and about the same in the fourth.  This may be informative to
high-redshift searches in that higher-dispersion and highly-reddened SN Ia can
be avoided by not including SN Ia within $\rm\sim10kpc$ of the host center.

\subsection{Systematics}

WV07 present a thorough list of potential sources of systematic error on the
measurment of $1+w$, leading to an estimate of $\Delta_w(syst) = 0.13$.  This
is larger than what is presented by A06 and K08:  $\Delta_w(syst) = 0.054$ and
$\Delta_w(syst) \approx 0.085$, respectively for their SN+BAO fits.  A06, WV07
and K08 do not explicitly explore the effects of different light-curve/distance
fitters on $1+w$, although some of the things they explore are components of an
individual fitter's systematic uncertainty.  Since probably the largest
component of systematic uncertainty is the reddening/color correction, the WV07
estimate implicitly includes much of the systematic uncertainty due to choice
of light-curve/distance fitter.  Because of this, and as our findings confirm,
the WV07 estimate of systematic uncertainty is probably closer to reality than
K08 and certainly closer than A06.  The majority of the WV07 value comes from
three sources:  0.08 from the host galaxy extinction treatment, 0.06 from the
uncertainty of the intrinsic color of SN Ia and 0.06 from a potential Hubble
bubble.  If all three of these are removed then the WV07 uncertainty would be
$\Delta_w(syst) = 0.057$, in line with the A06 value, where these effects are
not considered.  

The choice of which potential sources of systematic error to include and how
much of an effect they may have is a somewhat subjective decision but it can be
informed by looking at how different samples are affected by different fitters.
One measure of the systematic effects is to look at the differences in $1+w$
between the different fitters and samples under the minimal cuts, and again,
with best cuts.  Our results show the WV07 value seems plausible.  The largest
difference in $1+w$ for the minimal-cuts OLD+CfA3 sample is 0.12, between
MLCS31 and MLCS17.  The host galaxy extinction and the Hubble bubble change
significantly by using \rvone~instead of \rvth.  The uncertainties in these
individual effects in WV07 is in rough agreement with our difference in $1+w$
of 0.12.  The largest difference in $1+w$ for the best-cuts OLD+CfA3 sample is
0.09, between MLCS17 and SALT2.  The choice of fitter (and the underlying
assumptions) is very important.  The choice of which SN Ia samples to use is
also important.  SALT2 and MLCS31 show large differences in $1+w$ between the
OLD and the CfA3 samples (-0.15 and -0.13, respectively) while SALT and MLCS17
do not.  


By making cuts on \dd, color and host extinction, a slightly better agreement
in $1+w$ is achieved in the OLD+CfA3 samples between MLCS17 and MLCS31 and
between SALT and SALT2 but not between either of the SALT/2 cases and either of
the MLCS17/31 cases.  We note that the best agreement between all four fitters
occurs when the color and host-extinction cuts are made without any cuts on
\dd.  At lower values of host reddening, the influence of the particular
MLCS2k2 prior and the value of \rv~are less important.  Similarly, in SALT and
SALT2, for the narrower range of color considered, the value of $\beta$ will be
more accurate.  The difference in $1+w$ between MLCS31 and MLCS17 with
\av$\leq0.5$ is 0.05, with the SALT and SALT2 values agreeing extremely well
with MLCS31, and so we adjust the WV07 uncertainty on $1+w$ due to host-galaxy
extinction from 0.08 to 0.05.  The choice of OLD or CfA3 as the nearby sample
leads to an average difference in $1+w$ of 0.06-0.07.  By using both, the
systematic on $1+w$ should be less and we adopt 0.03 as our value for choice of
nearby sample.  We also saw in $\S$3.8 that making cuts on \av~made our results
even more consistent with no Hubble Bubble and so we reduce our Hubble bubble
amplitude uncertainty from 0.02 to 0.01, reducing the corresponding systematic
uncertainty in $1+w$ from 0.06 to 0.03.  Making the cuts on \dd~raises the
values of $1+w$ by about 0.04 for both MLCS17 and MLCS31 and so we add this as
another component of systematic uncertainty.  These three changes result in a
systematic uncertainty in $1+w$ of 0.11 for our best-cuts samples.  This is
roughly 1.4-1.7 times the typical statistical uncertainty of the best-cuts
values of $1+w$.

\section{Conclusion}

For people that have used the Union compilation from K08 to perform their own
cosmology fits, we recommend that they use our Constitution set (Union+CfA3) in
Table \ref{table_salt}.  The CfA3 sample adds 90 nearby objects to the 57 Union
objects.  The Constitution set of SN Ia (using the SALT output and a linear
luminosity dependence on color and stretch for the cosmology calculation),
combined with the BAO prior, give $1+w=0.013^{+0.066}_{-0.068}$, consistent
with a cosmological constant.  We estimate a systematic uncertainty of 0.11.
Further progress will mostly come by reducing systematic errors and
uncertainty.  We also invite people to use our SALT2, MLCS31, and MLCS17 light
curve fits in Tables \ref{table_salt2}, \ref{table_mlcs31}, and
\ref{table_mlcs17}.  We recommend a cut on $c$, $\Delta$ and $A_V$ as we have
employed in $\S$3.11.  

The four fitters are seen to be relatively consistent in the
light-curve-shape and color parameters they measure but that there are areas
of concern.  MLCS2k2 produces negative residuals in the range $0.7<\Delta<1.2$
and when \rvth~is used the host extinction is overestimated.  SALT and SALT2
give rise to different light-curve-shape/luminosity coefficients, $\alpha$, at
high and low redshift.  Both SALT and SALT2 have higher scatter at high
redshift.  SALT gives rise to a strong trend in residuals versus color at high
redshift.  SALT2 poorly fits the nearby \emph{U}-band light curves.  This shows
that light curve and distance fitters still have room for improvement and
provide a considerable amount of systematic uncertainty to any analysis.
However, by addressing these issues, future iterations of these fitters should
be much more robust and agree better with each other.

We have shown by using multiple fitters on the OLD+CfA3 sample, combined with
reasonable cuts on color and host reddening, that a consistent value of $1+w$
emerges.  However, when we make the additional cut on \dd~in MLCS2k2, which we
believe is warranted, the MLCS17 and MLCS31 values of $1+w$ still agree equally
well with each other but less well with the SALT and SALT2 values.  We
consider our best cuts to be $-0.1<c<0.2$ for SALT and SALT2, and $A_V\leq0.5$
and $\Delta < 0.7$ for MLCS31 and MLCS17.  

Our investigation of the Hubble bubble is consistent with no Hubble bubble to
about the 1-2$\%$ level.  We also find that SN Ia in E/S0 hosts have negative
mean Hubble residuals and low reddening while SN Ia in Scd/Sd/Irr hosts have
positive mean Hubble residuals, low scatter and low reddening.  After
correction for color and light-curve shape, the SN Ia in Scd/Sd/Irr
hosts are fainter than those in E/S0 hosts by $2\sigma$, suggesting that they
may come from different populations.  It may be worthwhile to form two or more
separate samples, based on host-galaxy morphology or color, to train
light-curve fitters.

Systematic uncertainties are now the largest obstacle for progress in supernova
cosmology, both in better measuring constant $1+w$ and in measuring any time
dependence.  Three large potential sources of systematic error are SN Ia
photometric accuracy, host-galaxy reddening, and SN Ia population differences
or evolution.  Regarding photometric accuracy, we have broken the nearby sample
into subsamples based on the main groups or instruments involved and find that
they agree with each other at about the 0.03 mag level.  This is an issue which
will be improved as the KAIT and Carnegie groups publish their photometry, a
good portion of which overlaps with CfA2 and CfA3 photometry.  This will allow
comparisons and possibly corrections so that the various nearby samples can be
combined more consistently.  Regarding reddening, the forthcoming sets of
near-infrared SN Ia light curves from the CfA and Carnegie groups, combined
with optical measurements, should improve our understanding of host-galaxy dust
and the intrinsic color variation in SN Ia.  As for SN Ia population difference
and evolution, continued investigation of the possible two-or-more SN Ia
populations should help reduce the systematic uncertainty that may arise.  We
encourage further efforts in SN Ia data gathering, distance estimation methods
and theoretical understanding so that the mystery of the dark energy may be
more fully illuminated.

\acknowledgments

This work has been supported, in part, by NSF grant AST0606772 to Harvard 
University.  AR thanks the Goldberg Fellowship Program for its support.

\begin{deluxetable}{lrrrrrrrrr}
\tablecolumns{10}
\tabletypesize{\scriptsize}
\tablewidth{0pc}
\tablecaption{SALT Light Curve Fit Parameters and Distances.  Constitution Set.}
\tablehead{
\colhead{SN Ia} & \colhead{$z_{CMB}$} &\colhead{$m_B$ \tablenotemark{1}} &  \colhead{$s$} & 
\colhead{$c$} & \colhead{$\mu$ \tablenotemark{2}} 
}
\startdata
1993ag & 0.050(0.001) & 17.790(0.108) & 0.910(0.020) & 0.090(0.020) & 36.910(0.135)\\
1993o & 0.053(0.001) & 17.610(0.111) & 0.900(0.010) & -0.010(0.020) & 36.977(0.134)\\
1993h & 0.025(0.001) & 16.740(0.145) & 0.680(0.010) & 0.210(0.010) & 35.264(0.186)\\
1993b & 0.070(0.001) & 18.380(0.040) & 0.990(0.030) & 0.040(0.040) & 37.729(0.124)\\
1992bs & 0.063(0.001) & 18.180(0.108) & 1.000(0.020) & -0.030(0.020) & 37.722(0.131)\\
1992br & 0.088(0.001) & 19.400(0.001) & 0.650(0.040) & 0.030(0.050) & 38.353(0.143)\\
1992bp & 0.079(0.001) & 18.280(0.108) & 0.870(0.020) & -0.040(0.020) & 37.687(0.128)\\
1992bo & 0.017(0.001) & 15.750(0.168) & 0.740(0.010) & 0.030(0.010) & 34.815(0.235)\\
\enddata
\tablenotetext{1}{The uncertainty of $m_B$ is a modified value for the Union set SN Ia, 
calculated by subtracting the color and stretch uncertainties
from $\sigma_\mu$ of K08 so as to include the sample-dependent systematic uncertainty
of K08.  For the CfA3 sample, 0.138 mag is added in quadrature to the actual uncertainty
of $m_B$.}
\tablenotetext{2}{Distance moduli were calculated with $\rm H_o=65 km s^{-1}$, $M_B=-19.46, \alpha = 1.34^{+0.08}_{-0.08}$ and $\beta = 2.59^{+0.12}_{-0.08}$.  Uncertainty of $\mu$ includes uncertainties of $m_B$, $s$, $c$, $\sigma_{v-pec}(400$ \kms$)$, and $\sigma_z$.}
\label{table_salt}
\end{deluxetable}

\begin{deluxetable}{lrrrrrrr}
\tabletypesize{\scriptsize}
\tablecolumns{8}
\tablewidth{0pc}
\tablecaption{SALT2 Light Curve Fit Parameters and Distances.}
\tablehead{
\colhead{SN Ia} & \colhead{$z_{CMB}$} &\colhead{$m_B$} & \colhead{$x_1$} & 
\colhead{$c$} & \colhead{$\mu$ \tablenotemark{1}}
}
\startdata
03D1au & 0.504(0.001) & 22.990(0.013) & 1.058(0.209) & 0.018(0.022) & 42.498(0.169)\\
03D1aw & 0.582(0.001) & 23.557(0.018) & -0.039(0.240) & 0.055(0.031) & 42.858(0.178)\\
03D1ax & 0.496(0.001) & 22.956(0.015) & -0.931(0.167) & -0.033(0.026) & 42.384(0.172)\\
03D1bp & 0.346(0.001) & 22.412(0.036) & -1.153(0.224) & 0.081(0.039) & 41.534(0.190)\\
03D1co & 0.679(0.001) & 24.102(0.046) & 0.666(0.587) & -0.008(0.055) & 43.632(0.222)\\
03D1ew & 0.868(0.001) & 24.361(0.056) & 0.297(0.411) & -0.072(0.180) & 44.012(0.479)\\
03D1fc & 0.331(0.001) & 21.813(0.020) & 0.218(0.186) & 0.024(0.017) & 41.220(0.166)\\
03D1fl & 0.688(0.001) & 23.599(0.017) & -0.262(0.255) & -0.070(0.028) & 43.187(0.176)\\
\enddata
\tablenotetext{1}{Distance moduli were calculated with $\rm H_o=65 km s^{-1}$, $M_B=-19.44, \alpha = 0.104^{+0.018}_{-0.018}$ and $\beta = 2.48^{+0.10}_{-0.12}$.  Uncertainty of $\mu$ includes 0.158 mag ``intrinsic", $\sigma_{v-pec}(400$ \kms$)$, and $\sigma_z$.}
\label{table_salt2}
\end{deluxetable}

\begin{deluxetable}{lrrrrrrr}
\tabletypesize{\scriptsize}
\tablecolumns{8}
\tablewidth{0pc}
\tablecaption{MLCS31 Light Curve Fit Parameters.}
\tablehead{
\colhead{SN Ia} & \colhead{$z_{CMB}$} & \colhead{$\Delta$} & 
\colhead{\av} & \colhead{$\mu$ \tablenotemark{1}}
}
\startdata
03D1au & 0.504(0.001) & -0.296(0.065) & 0.147(0.127) & 42.540(0.179)\\
03D1aw & 0.582(0.001) & -0.262(0.141) & 0.144(0.113) & 43.098(0.207)\\
03D1ax & 0.496(0.001) & 0.178(0.118) & 0.117(0.115) & 42.322(0.194)\\
03D1bp & 0.346(0.001) & 0.157(0.121) & 0.268(0.224) & 41.516(0.255)\\
03D1co & 0.679(0.001) & -0.244(0.112) & 0.172(0.162) & 43.589(0.261)\\
03D1ew & 0.868(0.001) & -0.306(0.099) & 0.095(0.071) & 43.951(0.159)\\
03D1fc & 0.331(0.001) & -0.123(0.153) & 0.136(0.123) & 41.313(0.206)\\
03D1fl & 0.688(0.001) & -0.225(0.119) & 0.136(0.138) & 43.123(0.226)\\
\enddata
\tablenotetext{1}{Distance moduli were calculated with $\rm H_o=65 km s^{-1}$ and $M_V=-19.504$.  Uncertainty of $\mu$ includes 0.078 mag ``intrinsic", $\sigma_{v-pec}(400$ \kms$)$, and $\sigma_z$.}
\label{table_mlcs31}
\end{deluxetable}

\begin{deluxetable}{lrrrrrrr}
\tabletypesize{\scriptsize}
\tablecolumns{8}
\tablewidth{0pc}
\tablecaption{MLCS17 Light Curve Fit Parameters.}
\tablehead{
\colhead{SN Ia} & \colhead{$z_{CMB}$} & \colhead{$\Delta$} & 
\colhead{\av} & \colhead{$\mu$ \tablenotemark{1}}
}
\startdata
03D1au & 0.504(0.001) & -0.346(0.076) & 0.108(0.089) & 42.514(0.173)\\
03D1aw & 0.582(0.001) & -0.260(0.139) & 0.098(0.082) & 43.054(0.194)\\
03D1ax & 0.496(0.001) & 0.170(0.119) & 0.065(0.056) & 42.194(0.180)\\
03D1bp & 0.346(0.001) & 0.163(0.152) & 0.158(0.104) & 41.459(0.208)\\
03D1co & 0.679(0.001) & -0.313(0.121) & 0.151(0.126) & 43.575(0.256)\\
03D1ew & 0.868(0.001) & -0.255(0.120) & 0.111(0.084) & 43.794(0.207)\\
03D1fc & 0.331(0.001) & -0.170(0.147) & 0.123(0.078) & 41.242(0.181)\\
03D1fl & 0.688(0.001) & -0.264(0.132) & 0.130(0.101) & 43.075(0.211)\\
\enddata
\tablenotetext{1}{Distance moduli were calculated with $\rm H_o=65 km s^{-1}$ and $M_V=-19.504$.  Uncertainty of $\mu$ includes 0.078 mag ``intrinsic", $\sigma_{v-pec}(400$ \kms$)$, and $\sigma_z$.}
\label{table_mlcs17}
\end{deluxetable}

\begin{deluxetable}{l|ccc|r}
\tablecolumns{5}
\tablewidth{0pc}
\tabletypesize{\footnotesize}
\tablecaption{SN+BAO Values of $1+w$ for Each Fitter.}
\tablecomments{The last column is
the difference in $1+w$ between the OLD and the CfA3 samples.  Below the four
rows are the mean and standard deviation of the values from the four fitters. 
The top section shows $1+w$ for the uncut samples (no cuts beyond the original 
cuts on $t_{1st}$, \xx, $z$, and $A_V$) and the second section
shows $1+w$ after the ``best" cuts ($-0.1<c<0.2$,$A_V\leq0.5$).  Below are 
results from other cuts and fits as described in the text.}
\tablehead{
\colhead{Fitter} & \colhead{OLD} & \colhead{CfA3} &\colhead{OLD+CfA3} &  \colhead{OLD$-$CfA3}
}
\startdata
Best cuts:  \dd,$A_V$, $c$& $1+w$ &     $1+w$       &     $1+w$       &   $\Delta (1+w)$\\
\hline
SALT  &  0.036$^{+0.080}_{-0.082}$    &    -0.016$^{+0.078}_{-0.080}$    &     0.026$^{+0.069}_{-0.071}$    &    +0.052 \\
SALT2 & -0.075$^{+0.092}_{-0.097}$    &     0.071$^{+0.083}_{-0.087}$    &     0.026$^{+0.078}_{-0.082}$    &    -0.147 \\
MLCS31&  0.006$^{+0.077}_{-0.080}$    &     0.099$^{+0.066}_{-0.067}$    &     0.074$^{+0.063}_{-0.065}$    &    -0.093 \\
MLCS17&  0.083$^{+0.076}_{-0.079}$    &     0.134$^{+0.065}_{-0.067}$    &     0.118$^{+0.063}_{-0.065}$    &    -0.051  \\
\hline
Mean  &  0.012$^{+0.081}_{-0.085}$    &     0.072$^{+0.073}_{-0.075}$    &     0.061$^{+0.068}_{-0.071}$    &    -0.068 \\
Stddev $1+w$&  0.066    &     0.064    &     0.044    &\\
\hline

Minimal cuts & $1+w$ &     $1+w$       &     $1+w$       &   $\Delta (1+w)$\\
\enddata
\label{table_w}
\end{deluxetable}

\begin{deluxetable}{l|rrrc|rr}
\tablecolumns{7}
\tabletypesize{\footnotesize}
\tablewidth{0pc}
\tablecaption{Mean of the Minimal-Cuts Hubble Residuals by Sample and Fitter.}
\tablecomments{Mean and weighted mean of the Hubble residuals, relative to 
the best-fit cosmology, for each fitter and sample.  
$N$ is the number of SN Ia in each
sample for each fitter.  The standard deviation of the Hubble residuals 
and the uncertainties of the mean and the weighted mean are shown.  The
bottom row for each fitter contains the difference in the means and the weighted
means of the OLD sample minus the CfA3 sample, and their uncertainties.}
\tablehead{
\colhead{Sample} & \colhead{$N$} & \colhead{Mean} & \colhead{StdDev} & 
\colhead{StdDev/$\sqrt{N}$} & \colhead{WMEAN} & \colhead{$\sigma_{WM}$}  
}
\startdata
MLCS31\\
\hline
OLD 	& 72 & -0.0385  &   0.209  &   0.0246 &  -0.0233  &  0.0221 \\
CfA3  	& 129& -0.0099   &   0.215  &   0.0189 &    0.0150 &  0.0159 \\
OLD+CfA3& 201 & -0.0202  &    0.213   &  0.0150 &   0.0019  &  0.0129 \\
High-z  & 165& -0.0099  &    0.236   &  0.0183 & -0.0009   &  0.0166 \\
OLD$-$CfA3 & &   -0.0286  &   		& 0.0310  &  -0.0383 &    0.0272 \\
\hline
MLCS17\\
\hline
OLD  &70 &   -0.0132   &   0.200   &  0.0239  &  -0.0121   &  0.0223 \\
\enddata
\label{table_meanres}
\end{deluxetable}

\begin{deluxetable}{l|rr|rr|rr|rr}
\tablecolumns{5}
\tabletypesize{\footnotesize}
\tablewidth{0pc}
\tablecaption{Weighted Mean of the Hubble Residuals of Survey Subsamples}
\tablecomments{The nearby sample is broken up into subsamples.
The top half is for the minimal cuts while 
$0.7\leq\Delta\leq1.2$ and $A_V>0.5$ are excluded from MLCS2k2
in the bottom half.  The weighted mean of the residuals is presented for 
each subsample.  The standard deviation of the weighted means
is presented below.  The difference between the weighted means of 
the residuals, OLD$-$CfA3, is presented.  * means the 
most-discrepant subsample is excluded from OLD.  
} 
\tablehead{
\colhead{Sample} & \colhead{} & \colhead{MLCS17} & \colhead{} &\colhead{MLCS31} & \colhead{} & \colhead{SALT} & \colhead{} & \colhead{SALT2}
}
\startdata
Minimal Cuts & N & WMEAN & N & WMEAN & N & WMEAN & N & WMEAN \\
\hline
CalanTololo&21&-0.022(0.039)&22&-0.034(0.038)&17& 0.043(0.036)&20&-0.077(0.044)\\
CfA1	 &13&-0.064(0.051)&12&-0.115(0.056)&11&-0.031(0.057)&12&-0.068(0.059)\\
CfA2	 &25& 0.028(0.038)&26&-0.003(0.038)&16&-0.010(0.066)&22& 0.054(0.046)\\
CfA3(4Sh)&45&-0.008(0.028)&43&-0.007(0.029)&29&-0.038(0.033)&40&-0.020(0.035)\\
CfA3(Kep)&88& 0.013(0.018)&86& 0.024(0.019)&61& 0.029(0.022)&75& 0.038(0.023)\\
Other	 &11&-0.015(0.061)&11& 0.032(0.056)&5 &-0.117(0.063)&11&-0.083(0.068)\\
K08	 &  &             &  &             &8 &-0.015(0.030)&  &             \\
\enddata
\label{table_res_6samples}
\end{deluxetable}

\begin{deluxetable}{l|l|llrrr}
\tablecolumns{7}
\tabletypesize{\footnotesize}
\tablewidth{0pc}
\tablecaption{Hubble Bubble Results.}
\tablecomments{The Hubble bubble calculations are listed for the four fitters and
three SN Ia samples.  The partition redshift, void amplitude, $\delta_H$, 
void amplitude uncertainty, $\sigma$, and void significance, $\delta_H/\sigma$,
are presented.  The MLCS2k2 Hubble bubble results with $A_V\leq0.5$
are at the bottom.  The most robust
results are from the OLD+CfA3 sample for SALT and SALT2 and for MLCS31 and
MLCS17 with $A_V\leq0.5$, all with absolute-value void significances 
just greater than one.  The low significance and the different partition 
redshifts of the most significant voids weakens the case for a Hubble bubble.}
\tablehead{
\colhead{Fitter} & \colhead{Bubble Sample} &\colhead{       } & \colhead{$z$ (partition)} 
&\colhead{$\delta_H$} &  \colhead{$\sigma$} & \colhead{$\delta_H/\sigma$}
}
\startdata
MLCS31&Old+CfA3&  &  0.028 &  0.029 &  0.005  &   5.556\\
      &Old     &  &  0.025 &  0.036 &  0.012  &   2.933\\
      &CfA3    &  &  0.040 &  0.033 &  0.009  &   3.660\\
\hline
MLCS17&Old+CfA3&  &  0.034 & -0.020 &  0.007  &   2.752 \\
      &Old     &  &  0.046 & -0.027 &  0.021  &   1.301 \\
      &CfA3    &  &  0.034 & -0.027 &  0.008  &   3.467 \\
\hline
SALT&Old+CfA3  &  &  0.035 & -0.014 &  0.010  &   1.306 \\
    &Old       &  &  0.025 &  0.037 &  0.018  &   2.089 \\
\enddata
\label{table_hubble_bubble}
\end{deluxetable}

\begin{deluxetable}{l|l|r|rr|rr}
\tablecolumns{5}
\tablewidth{0pc}
\tablecaption{Std. Dev. and Weighted Means of Hubble Residuals by Host-Galaxy Morpology.}
\tablecomments{For these results, SN Ia with $0.7\leq\Delta\leq1.2$ and $A_V>0.5$ have been 
excluded from MLCS31 and MLCS17.  SN Ia with $-0.1\leq c \leq0.2$ from SALT/2 
have been included.  N is the number in each bin.  RelativeWM is the weighted 
mean relative to the S0a-Sc bin.  The Scd/Sd/Irr bin has the smallest average 
standard deviation while the E-S0 bin has the largest.  The mean of Hubble 
residuals of the Scd/Sd/Irr hosts is greater by 2-$\sigma$ than that of 
the E-S0 bin.}
\tablehead{
\colhead{Fitter} & \colhead{Morph} & \colhead{N} & \colhead{StdDev} &\colhead{WMEAN} & \colhead{RelativeWM} & \colhead{$\sigma_{WM}$}
}
\startdata
SALT&E-S0     &21& 0.170 & -0.0641 & -0.0712& 0.0366 \\
&S0a-Sc       &63& 0.173 &  0.0071 &  0.0   & 0.0224 \\
&Scd/Sd/Irr   &9 & 0.064 &  0.0506 &  0.0435& 0.0602 \\
\hline
SALT2&E-S0    &26& 0.208 & -0.0686 & -0.0490& 0.0411 \\
&S0a-Sc	      &72& 0.163 & -0.0196 &  0.0   & 0.0242 \\
&Scd/Sd/Irr   &14& 0.108 &  0.0717 &  0.0913& 0.0560 \\
\hline
MLCS31&E-S0   &19& 0.166 & -0.0187 & -0.0405& 0.0432 \\
&S0a-Sc       &59& 0.172 &  0.0218 &  0.0   & 0.0234 \\
&Scd/Sd/Irr   &9 & 0.118 &  0.1367 &  0.1149& 0.0583 \\
\hline
MLCS17 & E-S0 &21& 0.186 & -0.0521 & -0.0537& 0.0416 \\
&S0a-Sc       &64& 0.155 &  0.0016 &  0.0   & 0.0221 \\
&Scd/Sd/Irr   &10& 0.109 &  0.1138 &  0.1122& 0.0538 \\
\hline
Average   & E-S0 &  &0.182        &         & -0.0536& 0.0406 \\
&S0a-Sc       &  &0.166        &         &  0.0   & 0.0230 \\
&Scd/Sd/Irr   &  &0.100        &         &  0.0905& 0.0571 \\
\hline
Diff.&Scd-Irr$-$E-S0& &        &         &  0.1441& 0.0701 \\
\enddata
\label{table_res_morph}
\end{deluxetable}

\clearpage

\clearpage

\clearpage

\clearpage

\clearpage

\clearpage

\clearpage

\clearpage

%

\clearpage



\clearpage



\clearpage

%
%
%
%
\end{document}